\documentclass[instruments,article,submit,pdftex,moreauthors]{Definitions/mdpi}

\firstpage{1} 
\pubvolume{1}
\issuenum{1}
\articlenumber{0}
\pubyear{2024}
\copyrightyear{2024}
\datereceived{ } 
\daterevised{ }
\dateaccepted{ } 
\datepublished{ } 
\hreflink{https://doi.org/} 

\usepackage[utf8]{inputenc}
\usepackage{amsmath}
\usepackage{tensor}
\usepackage{graphicx}
\usepackage{dcolumn}
\usepackage{bm}
\usepackage{xcolor}
\usepackage{subfigure}
\usepackage{amsmath}
\usepackage{placeins}
\usepackage{booktabs}
\usepackage{xcolor}
\usepackage{siunitx}
\usepackage[normalem]{ulem}

\Title{Performance of a modular ton-scale pixel-readout liquid argon time projection chamber}
\TitleCitation{Performance of a modular ton-scale pixel-readout liquid argon time projection chamber}

\Author{%
A.~Abed Abud$^{1}$, B.~Abi$^{2}$, R.~Acciarri$^{3}$, M.~A.~Acero$^{4}$, M.~R.~Adames$^{5}$, G.~Adamov$^{6}$, M.~Adamowski$^{3}$, D.~Adams$^{7}$, M.~Adinolfi$^{8}$, C.~Adriano$^{9}$
% ...
and On behalf of the DUNE Collaboration
}

\address{%
$^{1}$ \quad CERN, The European Organization for Nuclear Research, 1211 Meyrin, Switzerland \\
$^{2}$ \quad University of Oxford, Oxford, OX1 3RH, United Kingdom \\
$^{3}$ \quad Fermi National Accelerator Laboratory, Batavia, IL 60510, USA \\
$^{4}$ \quad Universidad del Atl\'antico, Barranquilla, Atl\'antico, Colombia \\
$^{5}$ \quad Universidade Tecnol\'ogica Federal do Paran\'a, Curitiba, Brazil \\
$^{6}$ \quad Georgian Technical University, Tbilisi, Georgia \\
$^{7}$ \quad Brookhaven National Laboratory, Upton, NY 11973, USA \\
$^{8}$ \quad University of Bristol, Bristol BS8 1TL, United Kingdom \\
$^{9}$ \quad Universidade Estadual de Campinas, Campinas - SP, 13083-970, Brazil \\
% ...
}

\longauthorlist{yes}
\preto{\abstractkeywords}{\nolinenumbers}

\firstnote{In memory of our colleague, Dr. Davide Salvatore Porzio, who is no longer with us.}

\abstract{%
The Module-0 Demonstrator is a single-phase 600 kg
liquid argon time projection chamber operated as a prototype for the DUNE liquid argon near detector. Based on the ArgonCube design concept, Module-0 features a novel 80k-channel pixelated charge readout and advanced high-coverage photon detection system. In this paper, we present an analysis of an eight-day data set consisting of 25 million cosmic ray events collected in the spring of 2021. We use this sample to demonstrate the imaging performance of the charge and light readout systems as well as the signal correlations between the two. We also report argon purity and detector uniformity measurements, and provide comparisons to detector simulations.}

\begin{document}

\maketitle
\flushbottom

\section{Introduction}
\label{sec:overview}

Charge readout in liquid argon time projection chambers (LArTPCs) has traditionally been accomplished via a set of projective wire planes, as successfully demonstrated e.g. in the ICARUS \cite{ICARUS:2004wqc}, ArgoNeuT~\cite{Anderson:2012vc}, MicroBooNE \cite{MicroBooNE:2016pwy} and ProtoDUNE-SP~\cite{DUNE:2020cqd,DUNE:2021hwx} experiments, and as planned for the first large detector module of the DUNE experiment currently in preparation at the Sanford Underground Research Facility (SURF) underground laboratory in South Dakota \cite{DUNE:2020lwj}. However, this approach leads to inherent ambiguities in the 3D reconstruction of charge information that present serious challenges for LArTPC-based near detectors, where a high rate of neutrino interactions and an associated high-intensity muon flux cannot be avoided. In particular, 3D reconstruction becomes limited by overlap of charge clusters in one or more projections, and the unique association of deposited charge to single interactions becomes intractable.

To overcome event pile-up, a novel approach has been proposed and is being developed for the LArTPC of the Near Detector (ND) complex of the DUNE experiment, close to the neutrino source at Fermilab. This technology implements three main innovations compared to traditional wire-based LArTPCs: a pixelated charge readout enabling true 3D reconstruction, a high-performance light readout system providing fast and efficient detection of scintillation light, and segmentation into optically isolated regions. By achieving a low signal occupancy in both readout systems, the segmentation enables efficient reconstruction and unambiguous matching of charge and light signals.

This paper describes the first tonne-scale prototype of this technology, referred to as Module-0, and its performance as evaluated with a large cosmic ray data set acquired over a period of several days at the University of Bern. Section~\ref{sec:m0} provides an overview of the detector, as well as of its charge and light readout systems.  Section~\ref{sec:charge} discusses the performance of the charge readout system in detail, and Section~\ref{sec:light} does the same for the light readout system. Section~\ref{sec:cosmic-analysis} then reviews several analyses performed with reconstructed tracks from the cosmic ray data set collected during the Module-0 that allow to assess the performance of the fully-integrated system. Important metrics for successful operation are addressed, such as electron lifetime, electric field uniformity, and the ability to match charge and light signals, among others. Section~\ref{sec:conclusions} offers some concluding thoughts. 

\section{The Module-0 Demonstrator}
\label{sec:m0}

\subsection{Detector Description} 
\begin{figure}[htbp!]
 \includegraphics[width=1.0\linewidth]{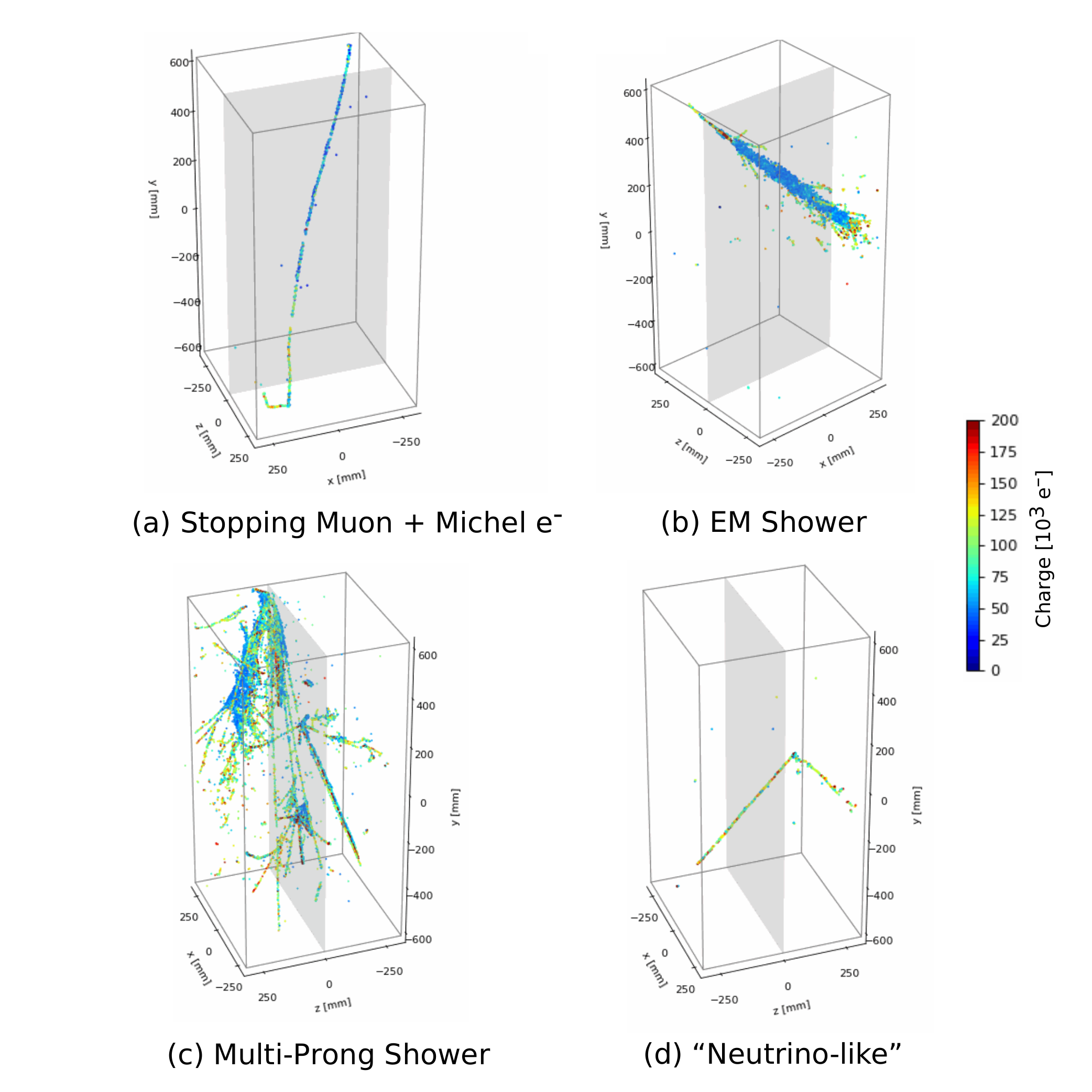}
 \caption{\label{fig:fig_eventgallery} Gallery of four representative cosmic ray-induced events collected with \mbox{Module-0}, as recorded in the raw event data, with collected charge converted to units of thousands of electrons. In all cases, the central plane in grey denotes the cathode, and the color scale denotes the collected charge. (a) shows a stopping muon and the subsequent Michel electron decay, (b) denotes an electromagnetic (EM) shower, (c) is a multi-prong shower, and (d) is ``neutrino-like'' in that the vertex of this interaction appears to be inside the active volume.}
\end{figure}

The \mbox{Module-0} demonstrator is the first fully integrated, tonne-scale prototype of the DUNE Liquid Argon Near Detector (ND-LAr) design. That detector will consist of a $7\times5$ array of $1\times1\times3$~m$^3$ detector modules \cite{ndcdr} based on the ArgonCube detector concept \cite{argoncube}, each housing two 50~cm--drift TPC volumes with 24.9\% optical detector coverage of the interior area. \mbox{Module-0} has dimensions of $0.7~\mathrm{m}\times0.7~\mathrm{m}\times1.4~\mathrm{m}$, and brings together the innovative features of LArPix \cite{larpix,larpixv2} pixelated 3D charge readout, advanced ArCLight \cite{arclight} and Light Collection Module (LCM) \cite{lcm} optical detectors, and field shaping provided by a low-profile resistive shell \cite{fieldcage}.
This integrated prototype also tests the charge and light system control interfaces, data acquisition, triggering, and timing. \mbox{Module-0} is the first of four functionally-identical modules that together will comprise an upcoming $2\times2$ ND-LAr prototype, known as ProtoDUNE-ND. Following construction and initial tests with cosmic ray event samples, this larger detector will be deployed underground in the NuMI neutrino beam at Fermilab \cite{numi2016} to demonstrate the physics performance of the technology in a similar neutrino beam environment to the DUNE ND. 
The work presented here describes the analysis of a data set of cosmic ray events obtained with the \mbox{Module-0} detector, installed in a liquid argon cryostat at the Laboratory for High-Energy Physics of the University of Bern. Over a period of eight days, the detector collected a sample of approximately 25 million self-triggered cosmic ray--induced events along with sets of diagnostic and calibration data.
The data collection period included an array of
characterization tests and data collection with changes to detector trigger conditions, thresholds, and with the TPC drift field as high as 1~kV/cm. For a brief second running period, the cryostat was emptied and refilled following a series of gas purges rather than complete evacuation, to assess the purity impact; this is discussed further in Section~\ref{sec:electron-lifetime}.
A gallery of events of different types is shown in Fig.~\ref{fig:fig_eventgallery}.
These images illustrate the rich 3D raw data from the pixelated charge readout system, the imaging capabilities for complex event topologies, and the low noise levels.

\begin{figure}[h]
 \includegraphics[width=0.99\textwidth]{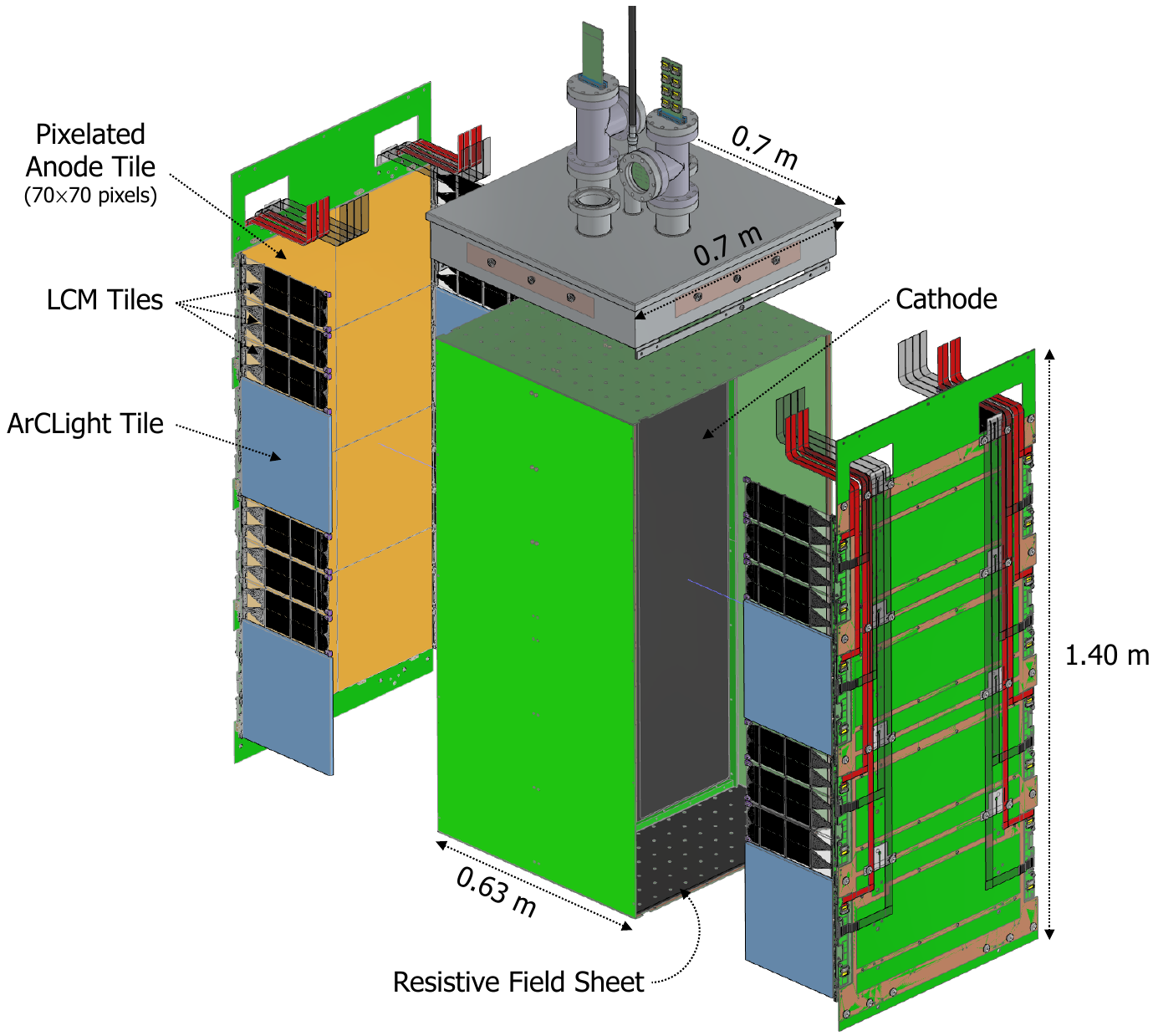}
 \caption{Schematic of the $0.7~\mathrm{m}\times0.7~\mathrm{m}\times1.4~\mathrm{m}$ Module-0 detector with annotations of the key components.
 \label{fig:module0-schem}}
\end{figure}

\begin{figure}[h]
 \includegraphics[width=0.79\textwidth]{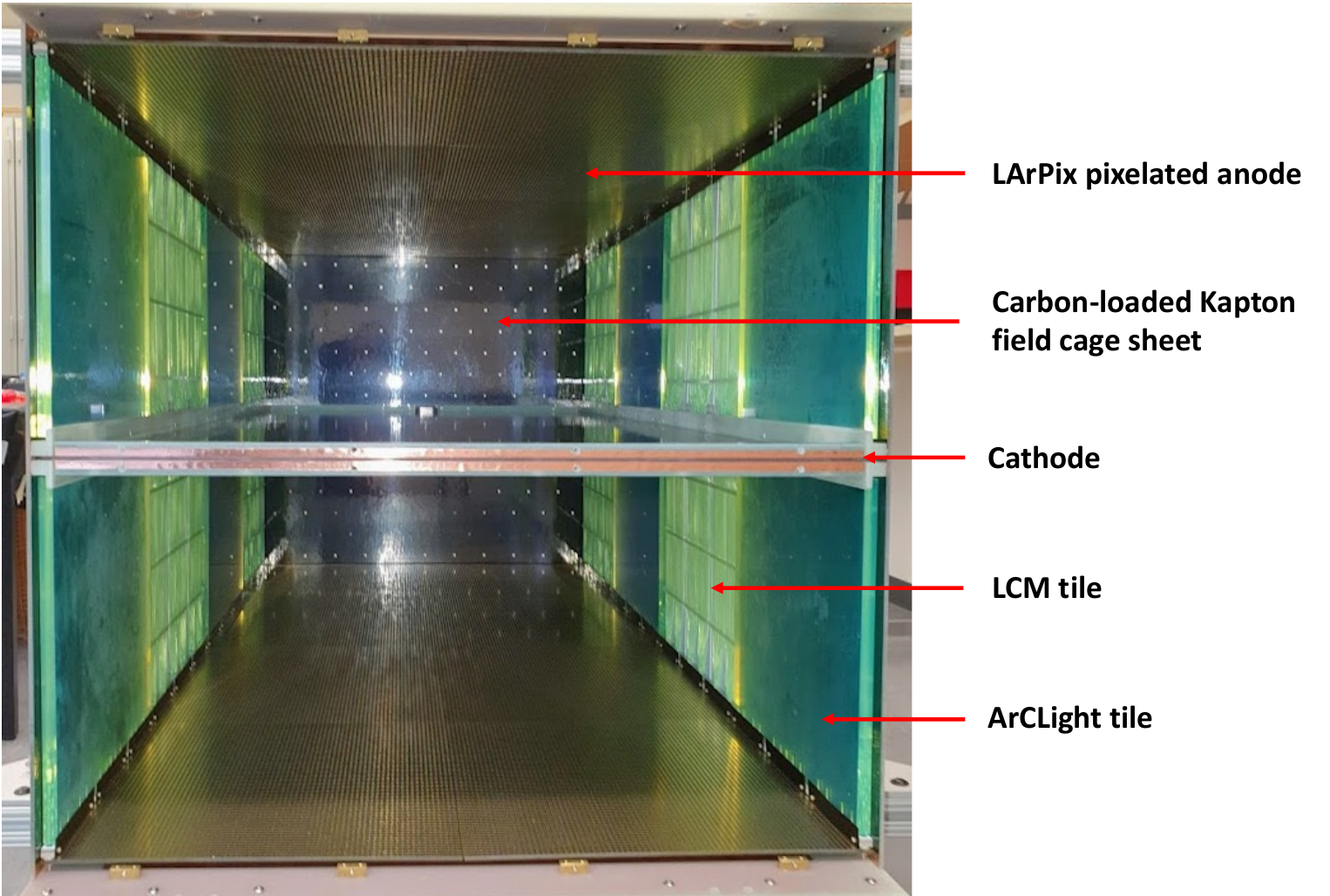}
 \caption{Photograph of the Module-0 detector interior as seen from the bottom, with annotations of the key components. 
 \label{fig:module0-overview}}
\end{figure}

A schematic showing an exploded view of \mbox{Module-0} with annotations of the key components is provided in Fig.~\ref{fig:module0-schem}, and a photograph of the interior of the \mbox{Module-0} detector as seen from the bottom
prior to final assembly in Fig.~\ref{fig:module0-overview}. The module is divided into two identical TPC drift regions sharing a central high-voltage cathode that provides the drift electric field.
Opposite the cathode at a distance of 30~cm are the anode planes, pixelated with charge-sensitive gold-plated pads where drifting ionization electrons are collected. The sides of the module are covered with photon detectors --- alternating ArCLight and LCM tiles. The TPC drift region is surrounded by a resistive field shell made of carbon-loaded Kapton films. This low-profile field cage provides field shaping to ensure a uniform electric field throughout the TPC volumes.

\subsection{The Charge Readout System}

\begin{figure}[htb]
 \includegraphics[width=0.4\linewidth]{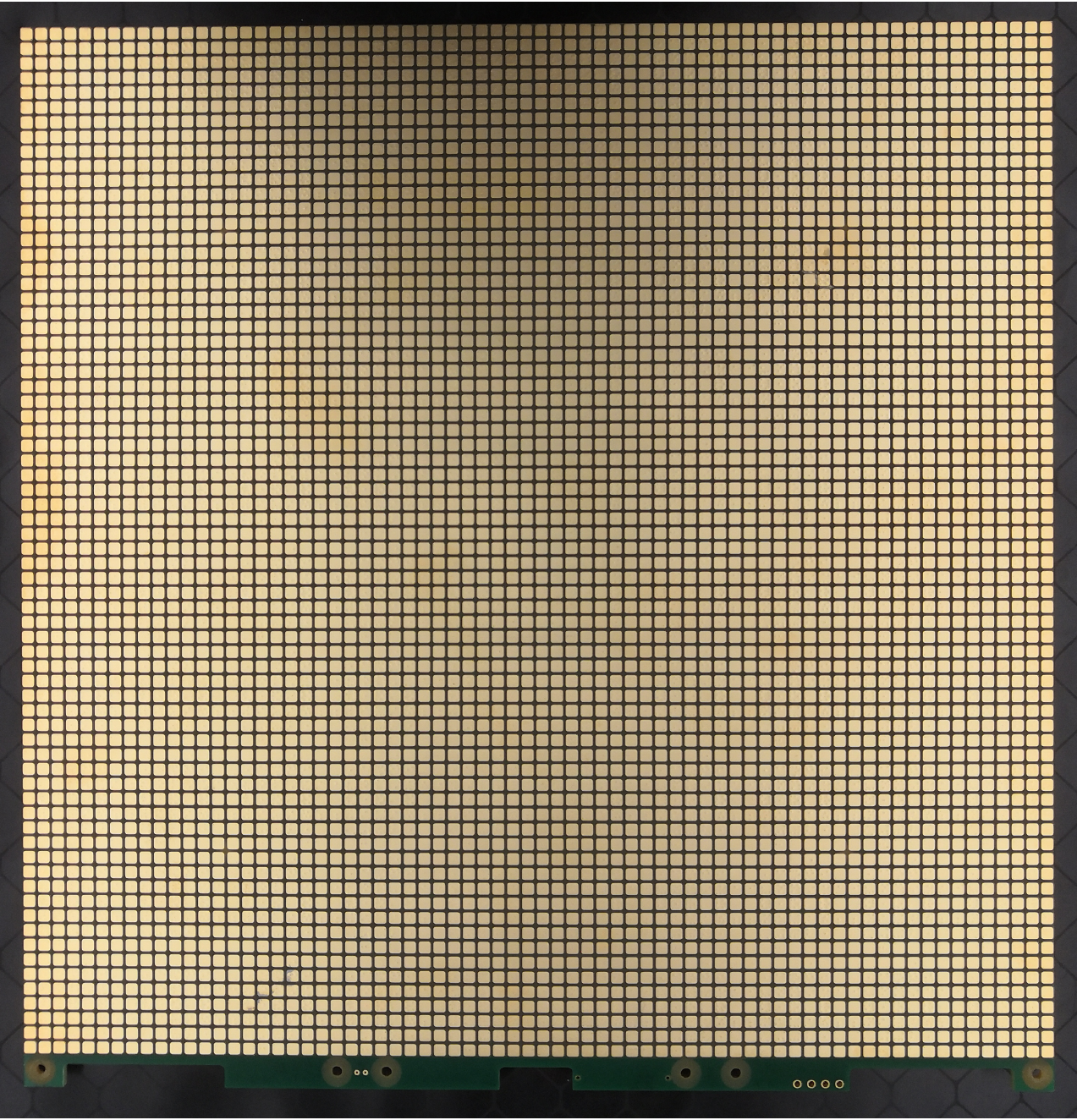}
 \includegraphics[width=0.4\linewidth]{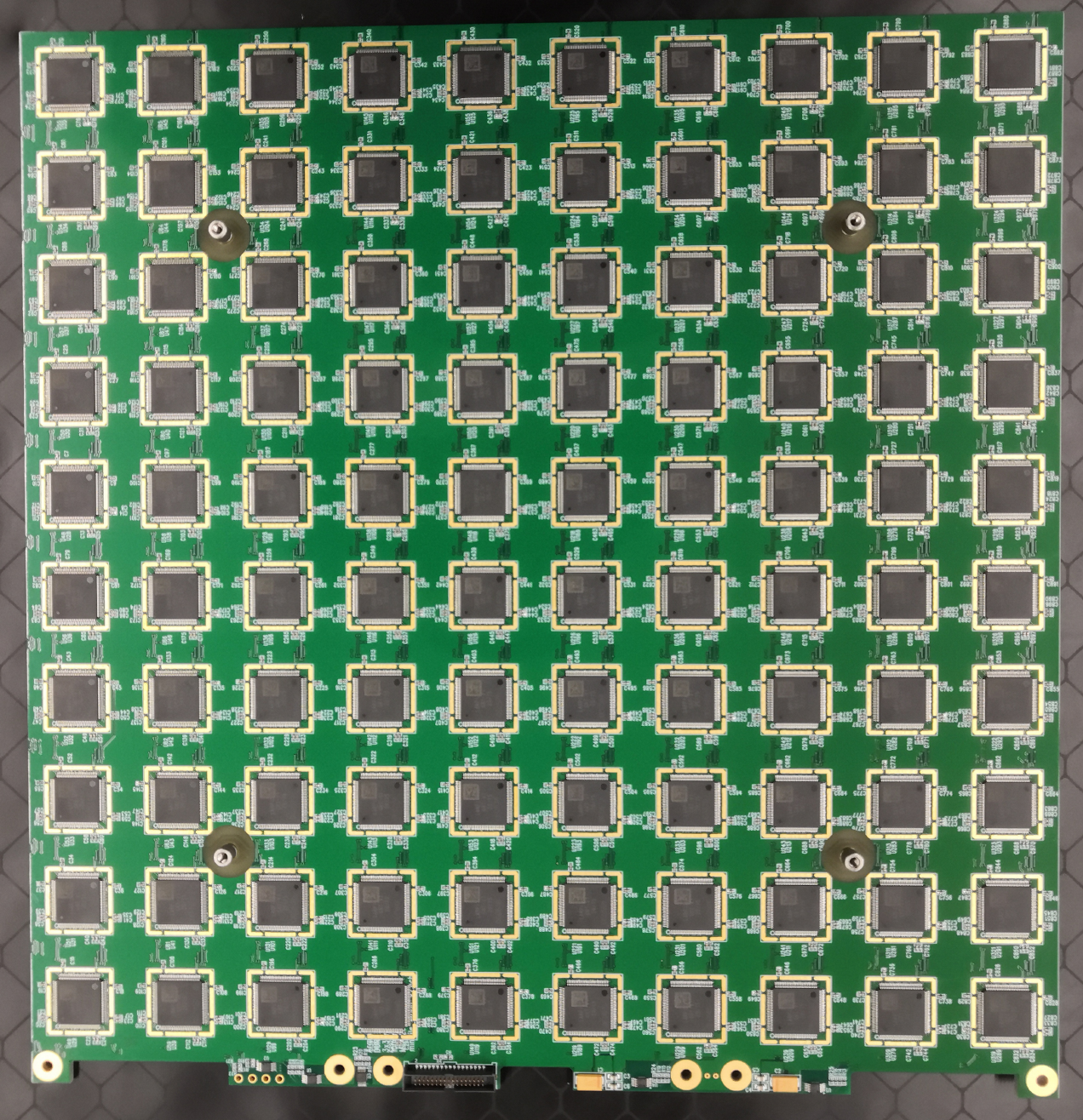}
 \qquad
 \caption{\label{fig:larpix_tile} Front (left) and back (right) of a TPC anode tile. The front contains 4,900 charge-sensitive pixels with 4.43~mm pitch that face the cathode, and the back contains a $10\times10$ array of LArPix ASICs. The dimensions are $31~\mathrm{cm}\times 32~\mathrm{cm}$, with the extra centimeter providing space for the light system attachment points.}
\end{figure}

The charge readout is accomplished using a two-dimensional array of charge-sensitive pads on the two anode planes parallel to the cathode.
While pixel-based charge readout has already been implemented in gaseous TPCs, LArTPCs have additional challenges due to restrictions on power dissipation. 
A proof of principle for pixelated charge readout in a single-phase LArTPC is described in Ref. \cite{Asaadi:2018oxk}, where a test device was exposed to cosmic ray muons. Readout electronics were also developed \cite{Asaadi:2018xfh,larpix} and successfully applied in a pixel-readout LArTPC. 
Each of the anode planes on opposite sides of the central cathode is comprised of a $2\times4$ array of anode tiles. Each tile is a large-area printed circuit board (PCB) containing a $70\times70$ grid of 4,900 charge-sensitive pixel pads with a 4.43 mm pitch. On the back of each PCB is a $10\times10$ grid of custom low-power, low-noise cryogenic-compatible LArPix-v2 application-specific integrated circuits (ASICs) \cite{larpixv2}, as shown in Fig.~\ref{fig:larpix_tile}. Each ASIC is a mixed-signal chip consisting of 64 analog front-end amplifiers, 64 analog-to-digital converters, and a shared digital core that manages configuration and data I/O. Each pixel channel functions as an independent self-triggering detector with nearly 100$\%$ uptime, and is only unresponsive to charge for 100~ns while the frontend resets. The LArPix ASIC leverages the sparsity of LArTPC signals. The chip is in a quiescent mode when not self-triggering on ionization activity higher than $\mathcal{O}(100)$ keV. Thus, it avoids digitization and readout of mostly-quiescent data. At liquid argon temperatures, the rate of accumulation of spurious charge (leakage current) is about 500 electrons/second. Each channel periodically resets to discard spurious charge that has collected at the input. In total, \mbox{Module-0} comprises 78,400 instrumented LArTPC pixels.

\begin{figure}[htb]
 \includegraphics[angle=90,width=0.70\textwidth]{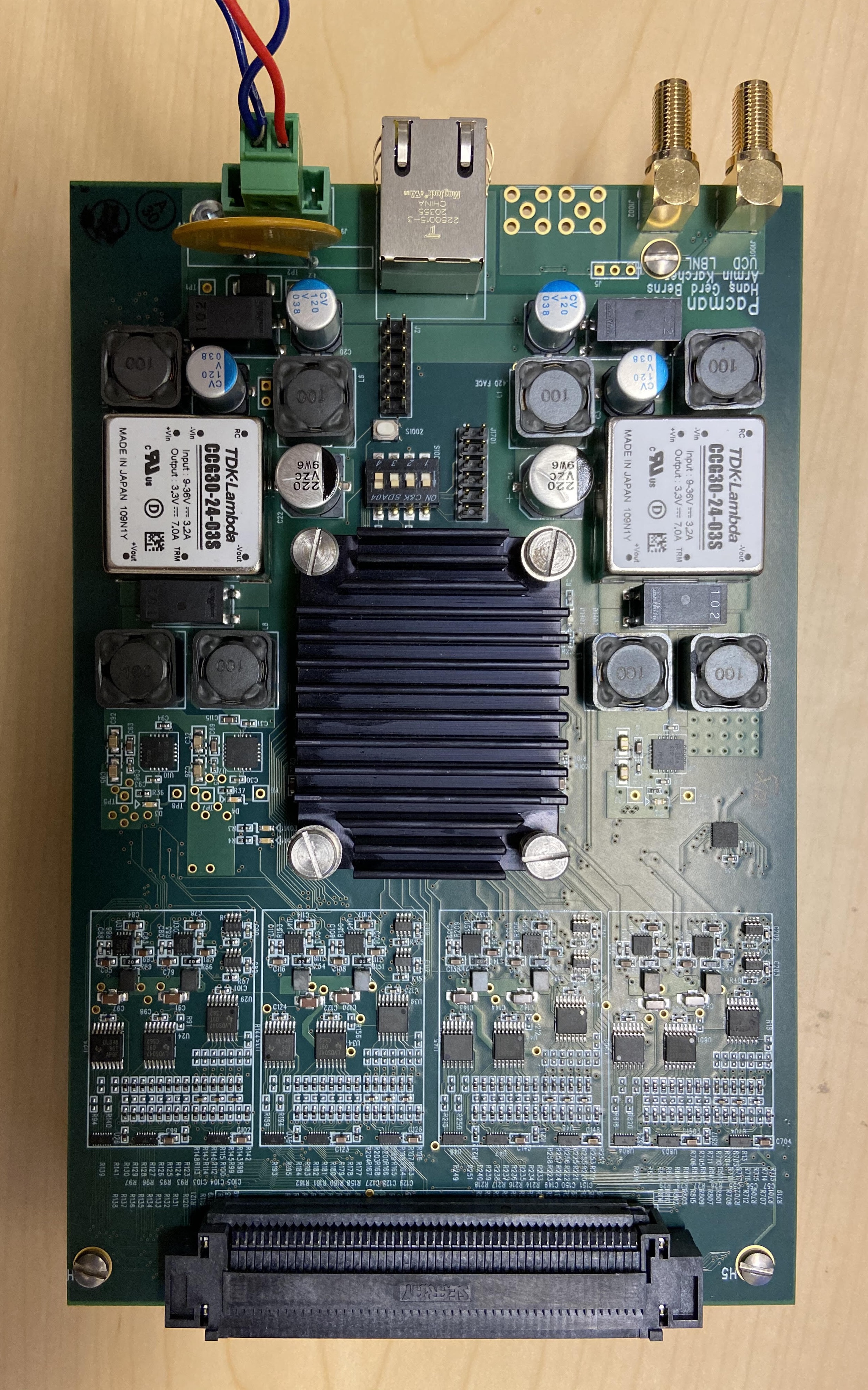}
 \caption{The Pixel Array Controller and Network card (PACMAN), which controls the data acquisition and power for the charge readout system.}
 \label{fig:pacman}
\end{figure}

Power and data I/O is provided to each tile by a single 34-pin twisted-pair ribbon cable. These cables are connected at the cryostat flange to a custom feedthrough PCB mounted on the cryostat lid. Data acquisition is controlled by the Pixel Array Controller and Network (PACMAN) card (Fig.~\ref{fig:pacman}), which provides filtered power and noise-isolated data I/O to eight tiles. Two PACMAN controllers are mounted in metal enclosures attached to the outer surface of each feedthrough. During \mbox{Module-0} operation, the PACMAN controller received a pulse-per-second timing signal for data synchronization between charge readout and light readout systems, and external trigger signals from the light readout system were embedded as markers into the charge readout data stream. Data are carried over a standard copper ethernet cable connected at each PACMAN to a network switch. Subsequently, data are transferred to and from the DAQ system via an optical fiber connection.

\begin{figure}[htb]
 \includegraphics[width=0.95\linewidth]{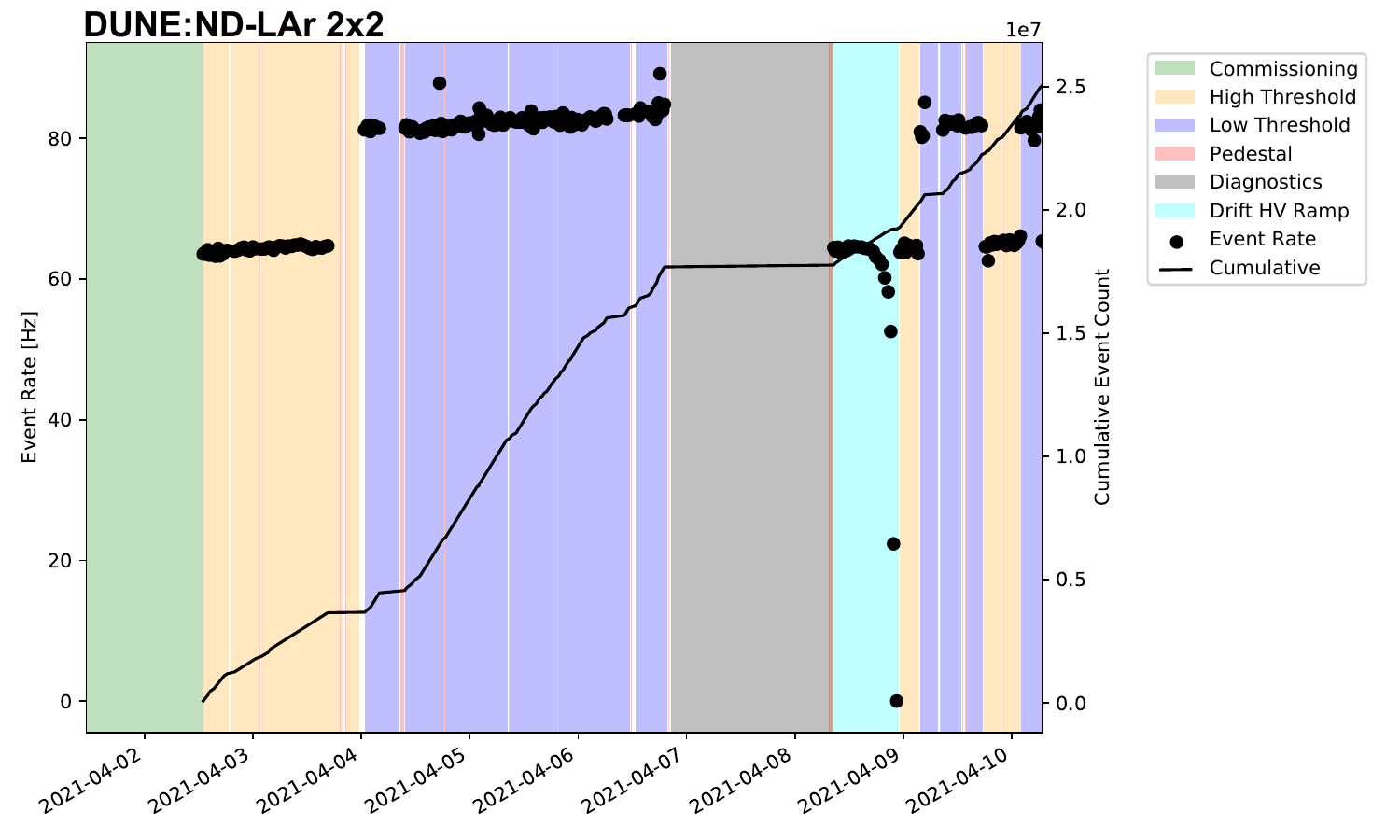}
 \caption{\label{fig:charge-configuration-events} Run event rate and cumulative events as a function of time with respect to charge readout operating condition.}
\end{figure}

For the LArTPC ionization charge measurement, LArPix ASICs mainly operate in self-trigger mode, where a trigger is initiated on a per-channel basis when a channel-level charge threshold is exceeded. In this mode of operation LArPix incurs negligible dead time and produces only modest data volumes, due to the sparsity of ionization signals in 3D, even for high-energy events. Serial data packets stream out of the system continuously via the PACMAN boards and are processed offline for analysis. A programmable channel-level threshold is set using internal digital to analog converters (DACs), which are tuned so that the spurious (i.e. noise-related) trigger rate is less than 2 Hz for each channel.
For \mbox{Module-0}, channel thresholds were operated in two regimes: low and high threshold (see Fig.~\ref{fig:charge-configuration-events}). Low threshold (${\sim5.8}$~ke$^-$/pixel or ${\sim\frac{1}{4}}$~MIP/pixel) operation optimized charge signal sensitivity at the expense of incurring additional triggers due to e.g. digital pickup, whereas high threshold (${\sim10.7}$~ke$^-$/pixel or ${\sim\frac{1}{2}}$~MIP/pixel) operation benefited from improved trigger stability at the expense of charge sensitivity. Updated revisions of the LArPix ASIC include additional pickup mitigation that will allow channel thresholds to be lowered further. Also, a slight rising trend in event rate can be seen over some of the different periods, most likely due to the emergence during data taking of pixels with a high data rate. It is believed that this small effect, which has no impact on the physics performance, can be mitigated by improving the procedure used to set the thresholds. 

ASICs within an anode tile are routed out to the DAQ through a configurable ``hydra'' network, wherein each ASIC has the ability to pass data packets to and from any adjacent neighbor. The scheme allows for system robustness in the event that an ASIC along the signal path becomes nonfunctional, though none of the 1600 ASICs failed during \mbox{Module-0} operation. A few-millisecond delay is incurred for data packets produced deeper in the network to reach the PACMAN controller relative to data packets produced closer to it. This is accounted for during hit digitization: each data packet carries a timestamp at creation when the hit signal is digitized, and when packets reach the PACMAN controller, a receipt timestamp is also assigned. Time ordering and filtering on packet trigger type is performed offline.
In order to monitor the integrity of the data in near--real time, a dedicated nearline monitoring system was developed and operated during the \mbox{Module-0} run. An automated analysis was performed on each run's raw data once the run ended and provided metrics including system trigger rates, trigger timing and offsets, channel occupancy and trigger rates, and data corruption checks.
Cosmic rays produced a self-trigger rate of ${\sim0.25}$~Hz per pixel.  This resulted in a total pixel hit rate of ${\sim20}$~kHz for the entire \mbox{Module-0} detector, yielding a modest data rate of 2.5~Mb/s.

\subsection{The Light Readout System} 

\begin{figure}[!ht]
 \includegraphics[width=0.8\linewidth]{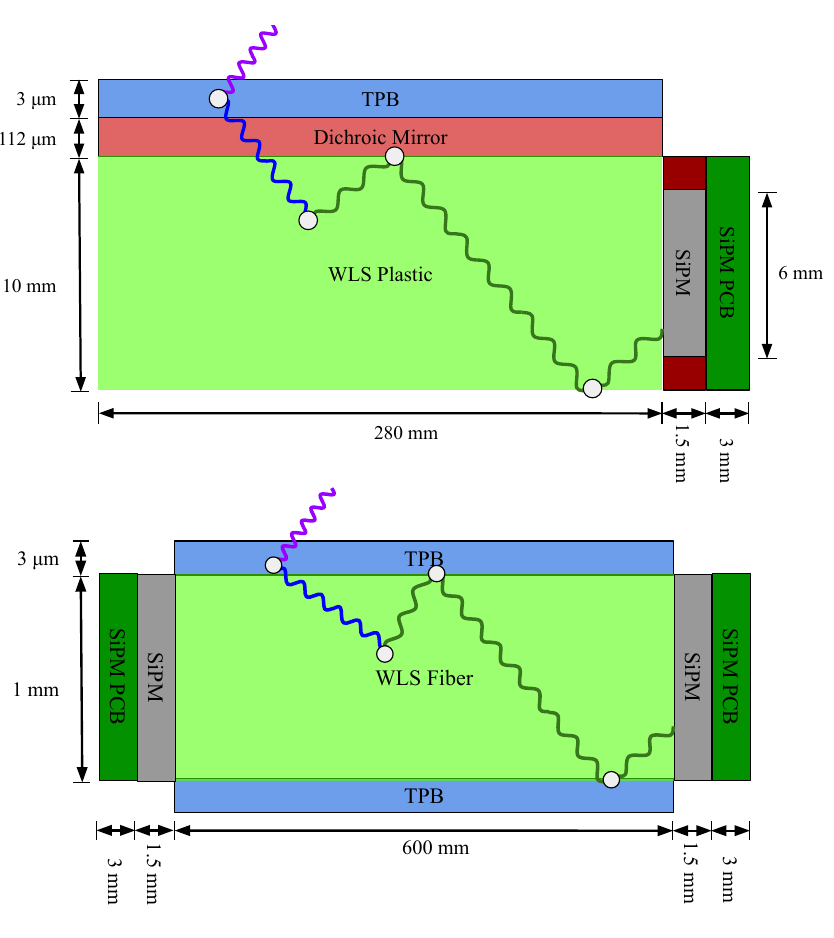}
 \caption{\label{fig:fig_modules} Detection principle of the two types of modules comprising the LRS: a segment of an ArCLight tile (top) and a single LCM optical fiber (bottom). The wave-like lines indicate example photon trajectories, where the white points indicate interactions. Drawings are not to scale.}
\end{figure}

The Light Readout System (LRS) provides fast timing information using the prompt ${\sim128}$~nm scintillation light induced by charged particles in LAr. The detection of scintillation photons provides absolute reference for event timing ($t_0$) and, when operated in an intense neutrino beam, will allow for unambiguous association of charge signals from the specific neutrino interactions of interest (i.e. pile-up mitigation). The LRS uses a novel dielectric light detection technique capable of being placed inside the field-shaping structure to increase light yield and localization of light signals. The LRS consists of two functionally-similar silicon photomultiplier (SiPM)-based detectors for efficient collection of single UV photons with large surface coverage: the Light Collection Module (LCM) and the ArCLight module. The full LRS system includes these modules together with the ancillary readout, front-end electronics, DAQ (ADCs, synchronization, and trigger), feedthrough flanges, SiPM power supply subsystem, and slow controls, as well as cabling and interconnection between different elements. LCM and ArCLight modules share the same basic operation principle. The vacuum ultraviolet (VUV) scintillation light produced by LAr is shifted from 128 nm to visible light by a wavelength shifter (WLS). Tetraphenyl butadiene (TPB) coated on the surface of the light collection systems provides an efficient WLS, and the emission spectrum of TPB is quite broad with a peak intensity of around 425 nm (violet light). Part of the light emitted at the surface of the light detection system eventually enters the bulk structure of the detector and is shifted to green light by a dopant (coumarin) in a bulk material, which also acts as a light trap (see Fig.~\ref{fig:fig_modules}).

\begin{figure}[htbp]
 \includegraphics[width = 0.8\linewidth]{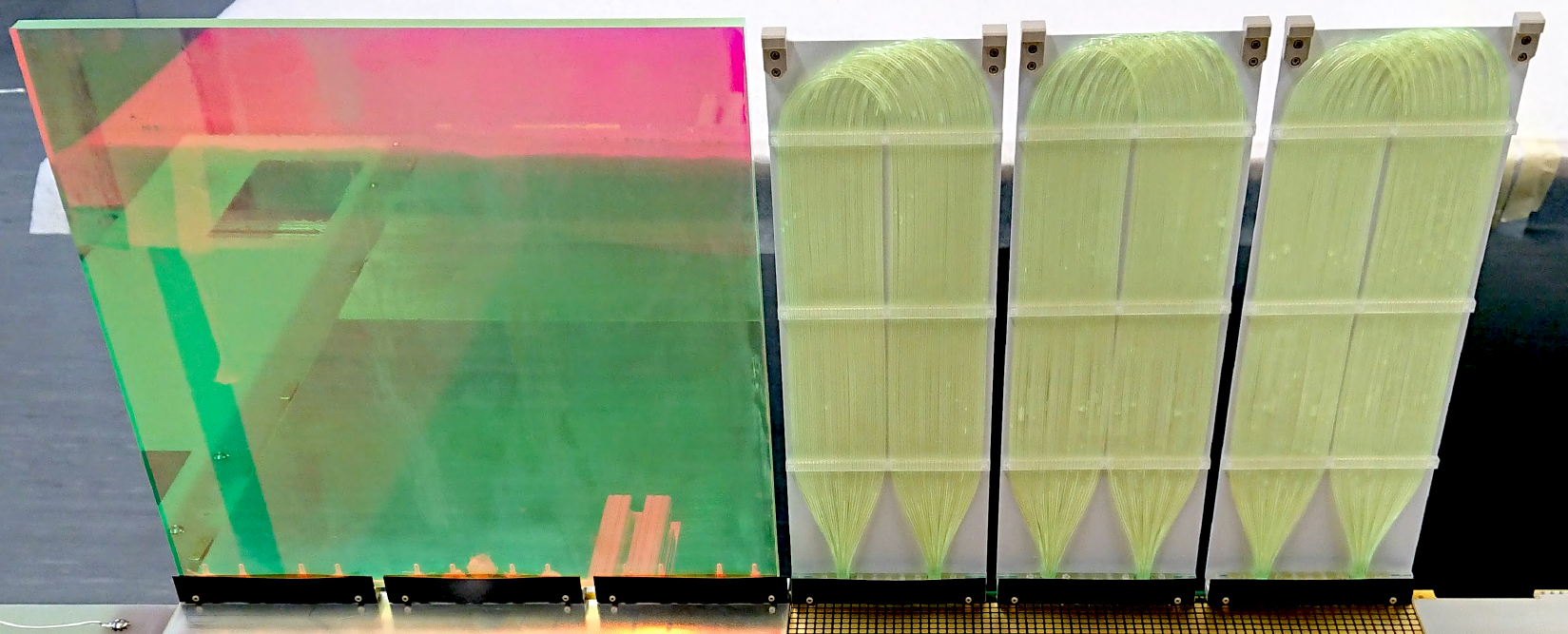}
 \caption{\label{fig:fig_ACL_LCM} An ArCLight tile (left) and three LCM tiles (right), as assembled within the Module-0 structure.}
\end{figure}

The ArCLight module has been developed by Bern University \cite{arclight} and uses the ARAPUCA \cite{arapuca} principle of light trapping. The general concept, illustrated in Fig.~\ref{fig:fig_modules} (top), is that violet light enters a bulk WLS volume and is re-emitted as green light, and the volume has a coating reflective to green light on all sides except on the SiPM photosensor window. A dichroic filter transparent to the violet light and reflective for the green is used on the WLS (tetraphenyl butadiene, TPB) side. The overall module dimensions are 300~mm~$\times$~300~mm~$\times$~10~mm.
A photograph of an ArCLight module is shown in Fig.~\ref{fig:fig_ACL_LCM} (left).

The LCM prototype is a frame cantilevered by a PVC plate that holds 25 WLS fibers bent into a bundle whose both ends are readout by a SiPM light sensor. Fibers are grouped and held by spacer bars with holes fixed on the PVC plate by means of polycarbonate screws to provide matching of thermal contraction. The PVC plate with the WLS fibers is coated with TPB, which re-emits the absorbed VUV light to the violet (${\sim425}$~nm). This light is then shifted inside multi-cladding $\varnothing$=1.2 mm Kuraray Y-11 fibers to green (${\sim 510}$~nm), and hence is trapped by total internal reflection guiding it to the SiPM readout at the fiber end, as depicted in Fig.~\ref{fig:fig_modules} (bottom). For each group of LCMs, the center module uses bis-MSB as a WLS rather than TPB to evaluate this alternative option; the photon detection efficiency performance is discussed in Section~\ref{sec:light} and the relative performance can be observed in Fig.~\ref{fig: PDE}. The LCM dimensions are 100~mm~$\times~$300~mm~$\times$~10~mm.
Fig.~\ref{fig:fig_ACL_LCM} (right) shows three LCMs.

\begin{figure}[htb]
 \includegraphics[width=0.35\linewidth]{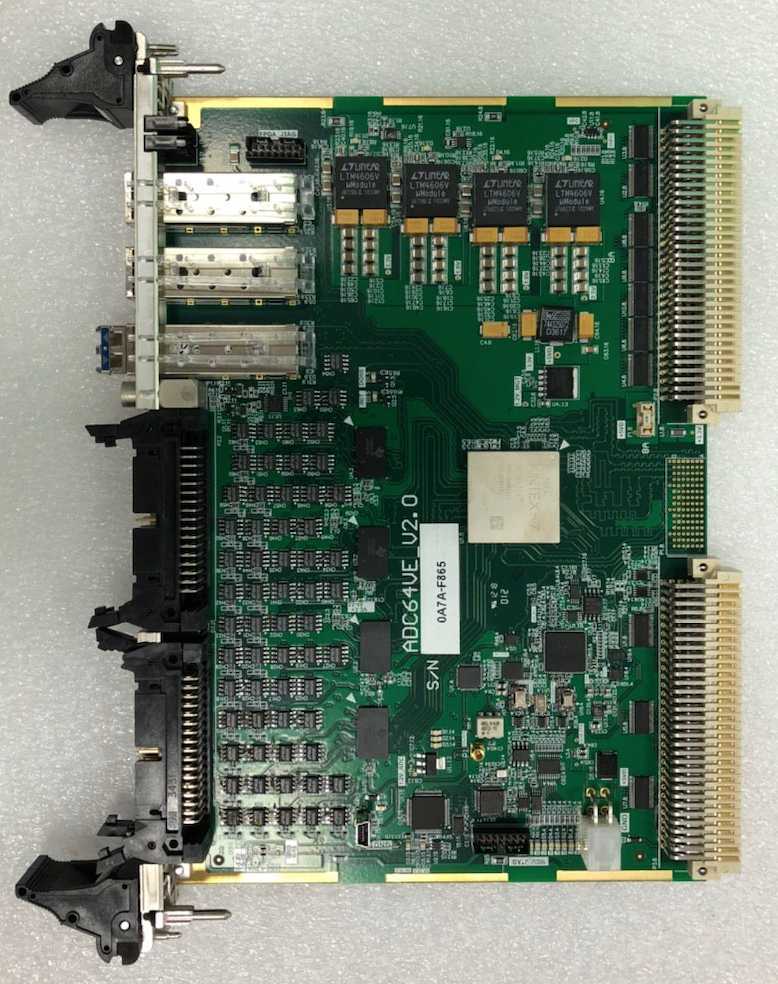}
 \qquad
 \includegraphics[width=0.55\linewidth]{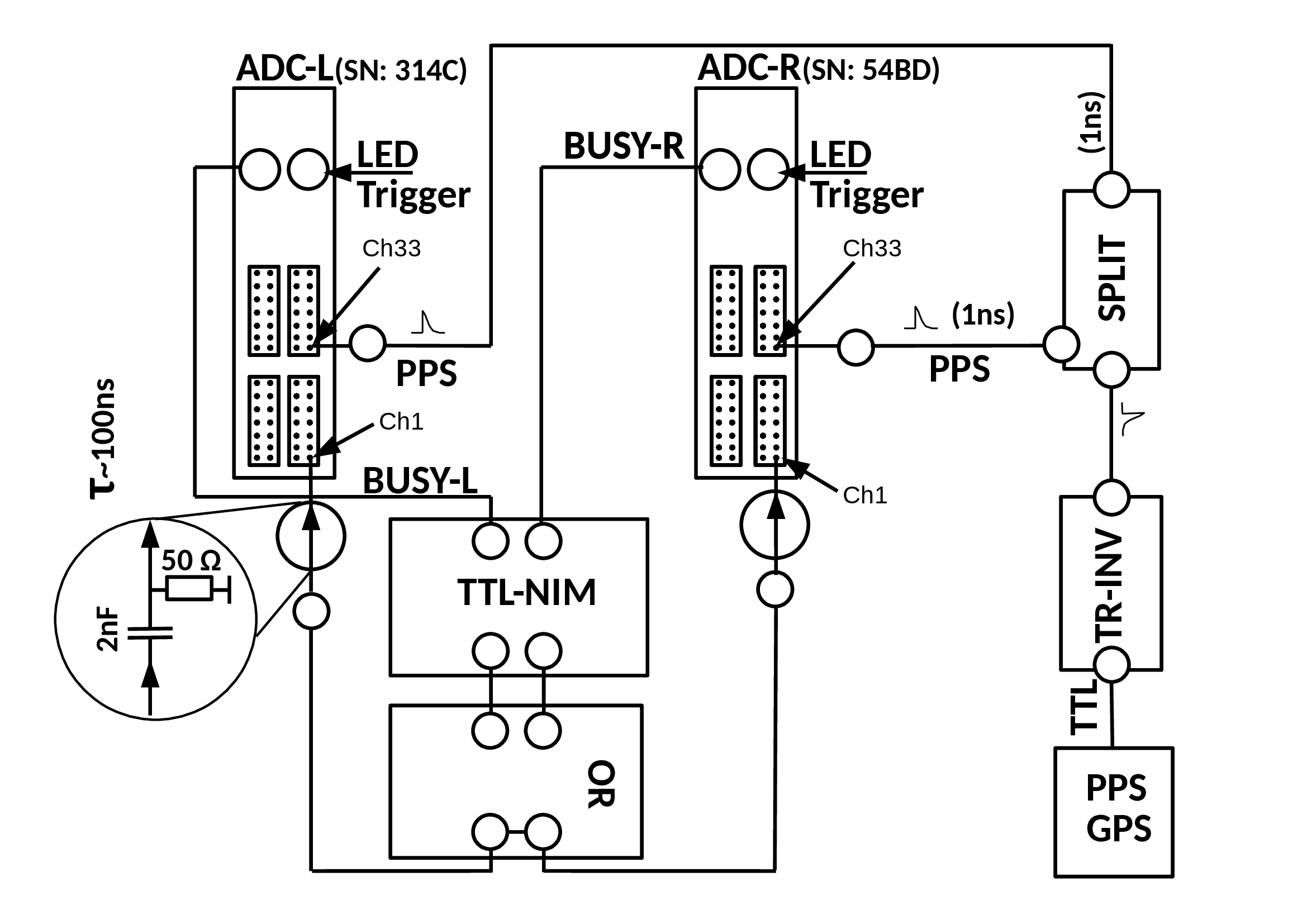}
\caption{\label{fig:adc} LRS data acquisition components: JINR ADC board (left), synchronization and trigger scheme (right).}
\end{figure}

In order to digitize analog signals from SiPMs, a 100 MHz, 10-bit, 64-channel (differential signals, full range $\pm1.6$~V) ADC prototype module in VME standard produced at the Joint Institute for Nuclear Research (JINR) was used (see Fig.~\ref{fig:adc}~left). This ADC module streams UDP/TCP data packets via M-link MStream protocol using a 10 Gbps optical link. The ADC boards have the capability to be synchronized via a White Rabbit system \cite{white-rabbit}. This was not available for \mbox{Module-0} run, for which timing synchronization between the charge light systems was provided by a dedicated system shown in Fig.~\ref{fig:adc} (right). To merge data between light and charge systems, a trigger signal generated by the LRS is written out to the charge readout data stream.
This trigger signal is also fed to the analog input of both ADCs to allow for precise time matching between ADC boards for further LRS data analysis. Additionally, a pulse-per-second from a stable GPS source was used for both detection systems to provide accurate synchronization. For the LRS, the pulse-per-second signal was fed to the analog input of each ADC. During the \mbox{Module-0} run, the LRS operated in a self-triggered mode with adjustable threshold settings.
The thresholds for the LCMs are approximately 30 photoelectrons, as discussed in Section~\ref{sec:light}.

\section{Charge Readout Performance}
\label{sec:charge}

\subsection{System Overview}
\mbox{Module-0} operation represents the first demonstration of the LArPix-v2 pixelated charge readout system in a tonne-scale LArTPC. Continuous acquisition and imaging of self-triggered cosmic ray data were successfully exercised, demonstrating the excellent performance of this technology. This section presents an array of studies of the charge readout system performance, including: pixel channel signal baselines and time stability, charge response as a function of track position and angle relative to the pixel plane, response uniformity across the instrumented area, ADC saturation, and overall calorimetric measurement performance.

\begin{figure}[htbp!]
 \includegraphics[width=0.95\linewidth]{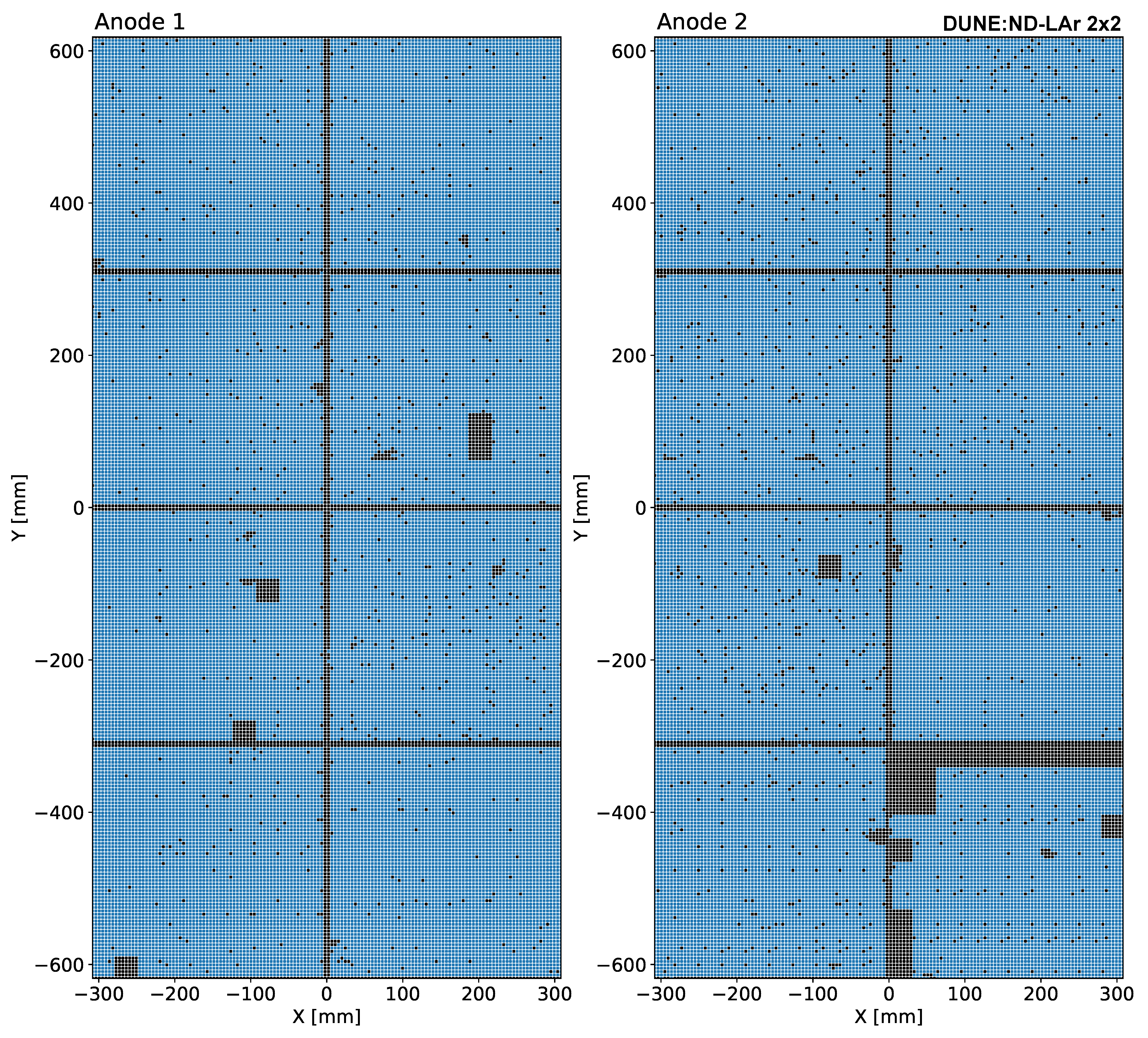}
 \caption{\label{fig:larpix-active-channel-map} Self-trigger active pixel channels (in blue) and inactive channels (in black). In these coordinates, $x$ is horizontal and $y$ is vertical, both parallel to the anode plane, and $z$ is the drift direction, perpendicular to the anode plane, completing a right-handed system. The origin is the center of the module.}
\end{figure}

In parallel to this successful series of technological achievements, this first large-scale integrated test highlighted areas for continued improvement in future iterations of the module design. This includes improved anode tile grounding and optimization of the pixel pad geometry. In the former case, enhancements to the grounding scheme will enable improved system-wide per-channel charge threshold sensitivity and system trigger stability, specifically allowing readout of the pixels on the edge of neighboring tiles, and mitigating the effects of triggering induced by system synchronization signals observed in the \mbox{Module-0} data. In the latter case, modifications to the pixel pad geometry will further minimize far-field current induction in the pixels, reducing the sensitivity of the readout system to drifting charge that is far from the anode plane. Additional improvements to the ASIC-related noise budget are planned for the next-generation LArPix design.
Of the total 78,400 instrumented pixel channels in \mbox{Module-0}, 92.2\% were enabled for LArTPC operation. The channels were disabled mainly due to limitations noted above --- grounding near tile edges (4.2\%), elevated noise levels due to signal pickup (3.1\%), high noise or leakage current (0.5\%) --- and their locations are illustrated in Fig.~\ref{fig:larpix-active-channel-map}. As noted above, no ASICs failed during \mbox{Module-0} operations.

\subsection{Noise and Stability}
Periodic diagnostics (pedestal) runs were taken to monitor the stability of the charge readout system. These diagnostic runs entailed issuing a periodic trigger on a per-channel basis in a round-robin fashion among channels on a single ASIC. In this way, sub-threshold charge was digitized to monitor
channel pedestal and the AC noise stability in time, with the ADC value returned by each digitization reflecting the sum of the quiescent pedestal voltage of the front-end amplifier and the integrated charge.
The distributions of ADC values collected during pedestal runs were in agreement with the design expectations, with a median value of $\sim78$ counts per channel, and pedestal
voltage varied by approximately 30~mV between channels. To determine the integrated charge, a correction for this pedestal value must be applied. We computed the channel-by-channel pedestal ADC value by using the truncated mean around the peak of the
ADC value distribution of each channel. The signal amplitude in mV was inferred based on the internal reference DAC values and the ASIC analog voltage, and a global gain value of 245 e$^-$/mV was then used to convert the signal amplitude to charge.

\begin{figure}[htbp!]
 \includegraphics[width=0.95\linewidth]{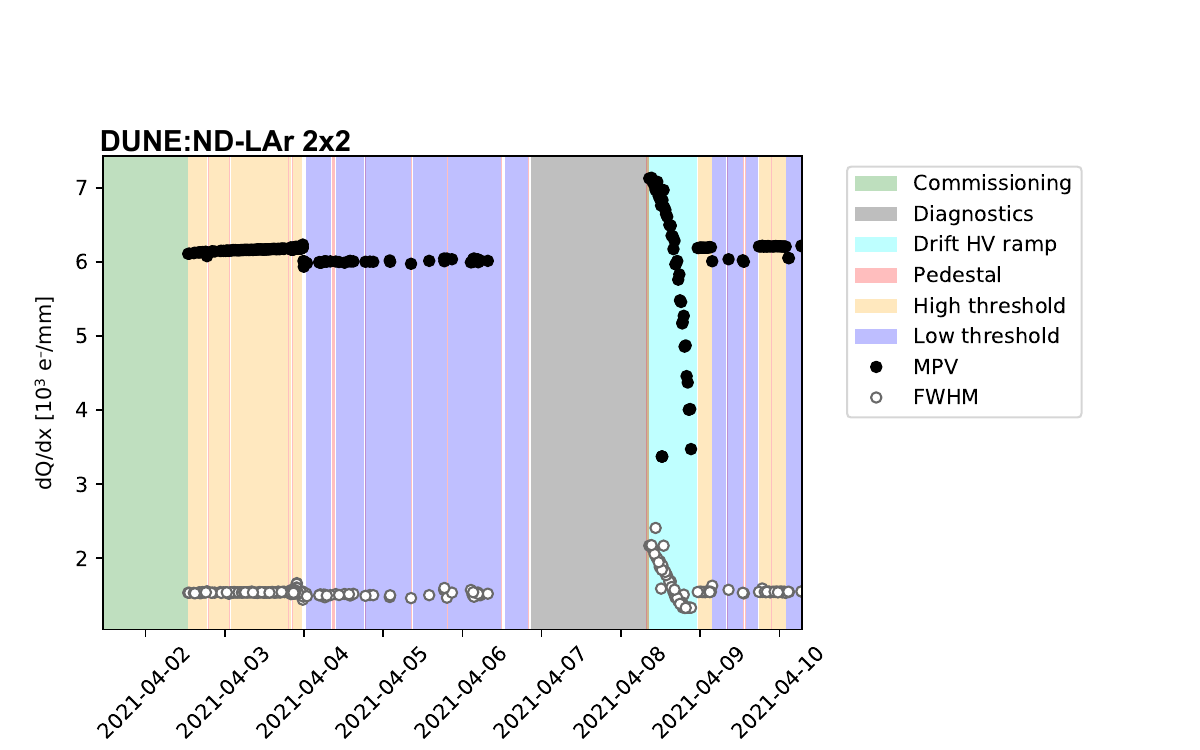}
 \caption{\label{fig:dqdx_stability} Most probable value (black circles) and full width at half maximum (white circles) of the $dQ/dx$ distribution for each data run. The system shows a good charge readout stability during data taking periods, both for {high threshold} (yellow bands) and {low threshold} (purple bands) runs.}
\end{figure}

\begin{figure}[htbp!]
 \includegraphics[width=0.7\linewidth]{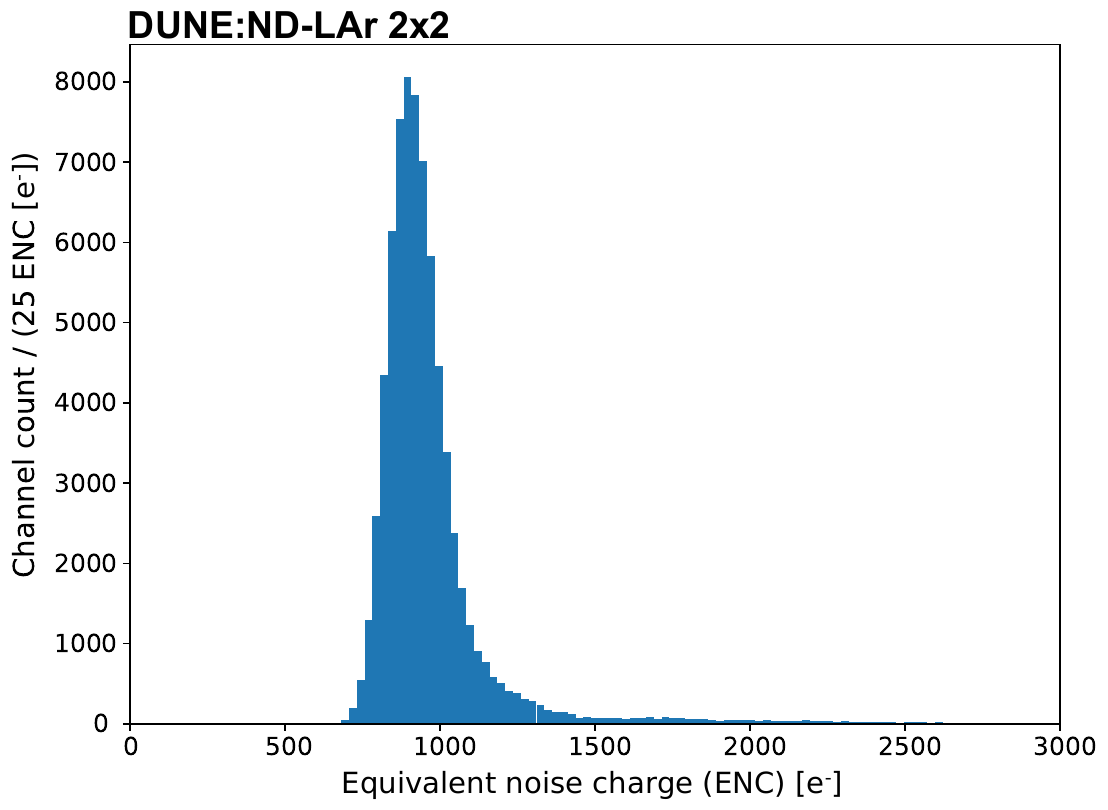}
 \caption{\label{fig:larpix-lar-ENC} LArPix channel noise in units of electron charge signal, as observed using periodic forced triggers. The total system noise is $\sim950~\mathrm{e}^-$, compared to a signal amplitude of $\sim1800~\mathrm{e}^-$ for a 4 GeV MIP track in ND-LAr's 3.7 mm pixel pitch.}
\end{figure}

Additionally, the stability of the charge readout over time was verified using cosmic ray data samples, by measuring the most probable value (MPV) and the full width at half maximum (FWHM) of the $dQ/dx$ distribution of minimum ionising particle (MIP) tracks for each data run, as shown in Fig.~\ref{fig:dqdx_stability}. To make these track-based measurements, 3D hits registered by the charge system are clustered together using the DBSCAN algorithm \cite{dbscan}. A principal component analysis of hits within each cluster then provides three-dimensional segments that we define as reconstructed tracks. The charge $dQ$ corresponds to the sum of the hits associated to the reconstructed track and the 3D reconstructed track length $dx$. The $dQ/dx$ distribution is then fitted with a Gaussian-convolved Moyal distribution~\cite{doi:10.1080/14786440308521076}, which is used to extract the MPV and the FWHM.
Total system noise contributes $\sim950$~e$^-$ equivalent noise charge (ENC) to each pixel hit, as assessed using periodic forced triggering of pixel channels in the absence of actual signals (Fig.~\ref{fig:larpix-lar-ENC}).
To put this metric in context, the intrinsic energy loss fluctuations associated with the charge from a 4~GeV MIP would be $\sim1800$~e$^-$ in ND-LAr's 3.7 mm pixel pitch. Therefore, the charge resolution is smaller than the intrinsic physical fluctuations for particle kinematics relevant to ND-LAr. 

\begin{figure}[htbp]
 \includegraphics[width=0.7\linewidth]{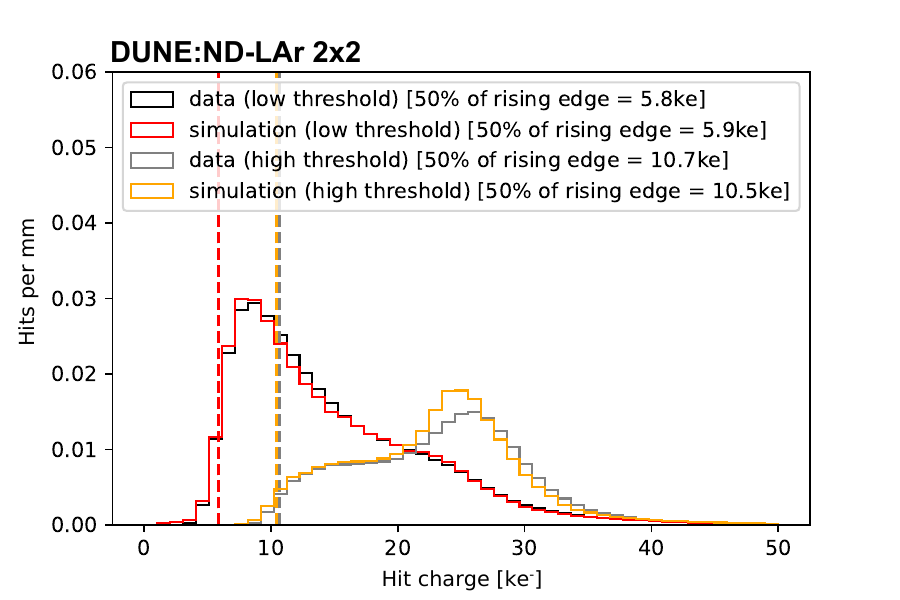}
 \caption{\label{fig:larpix-hit-charge-dist} Self-trigger charge distribution for MIP tracks measured in thousands of electrons (ke$^-$); 50\% of the rising edge are shown as indicators of the charge readout self-trigger thresholds. The low- and high-threshold curves are obtained from runs with the same 20 minute exposure. Each entry is normalized by hit charge over fitted track length. The MC simulation shown in comparison is described in Section~\ref{sec:cosmic-analysis}.}
\end{figure}

\begin{figure}[htbp]
 \includegraphics[width=0.7\linewidth]{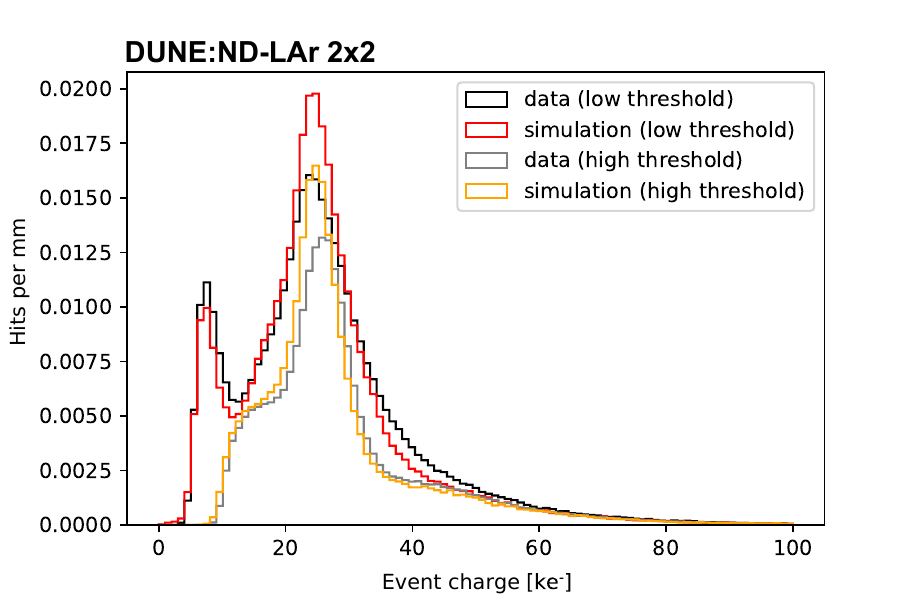}
 \caption{\label{fig:larpix-hit-charge-sum-dist} Total event charge per channel for MIP tracks measured in thousands of electrons (ke$^-$). The MC simulation shown in comparison is described in Section~\ref{sec:cosmic-analysis}.}
\end{figure}

Examining the corresponding charge in each pixel that has triggered (Fig.~\ref{fig:larpix-hit-charge-dist}), we identify a sharp rising edge corresponding to the self-trigger threshold at approximately $5.8\times10^3$ electrons (low threshold) and $11\times10^3$ electrons (high threshold). Above the self-trigger threshold, a peak at roughly $24\times10^3$ electrons corresponds to the typical charge deposited by a MIP crossing the full pixel pitch of 4.43~mm. Of note are the markedly different charge distributions of the high-- and low-threshold data. We find that for the low-threshold data, the average number of triggers per single channel for MIP energy deposition is substantially larger than for the high-threshold data, with mean values of 1.53 and 1.14 respectively. These numbers are well-reproduced by the Monte Carlo simulation (MC) described in Section~\ref{sec:cosmic-analysis}, with values of 1.52 and 1.12 respectively for a similar set of reconstructed MIP tracks. Summing the charge of all digitizations on each specific channel for a given event increases the similarity between the low-threshold data with the high-threshold data (Fig.~\ref{fig:larpix-hit-charge-sum-dist}). This is indicative of a ``pre-triggering'' effect, in which a channel is triggered by the induced signal generated by the drifting charge in advance of the charge signal arrival at the anode plane, thus motivating the reduction of far-field effects discussed above.

\begin{figure}[htbp]
 \includegraphics[width=0.65\textwidth]{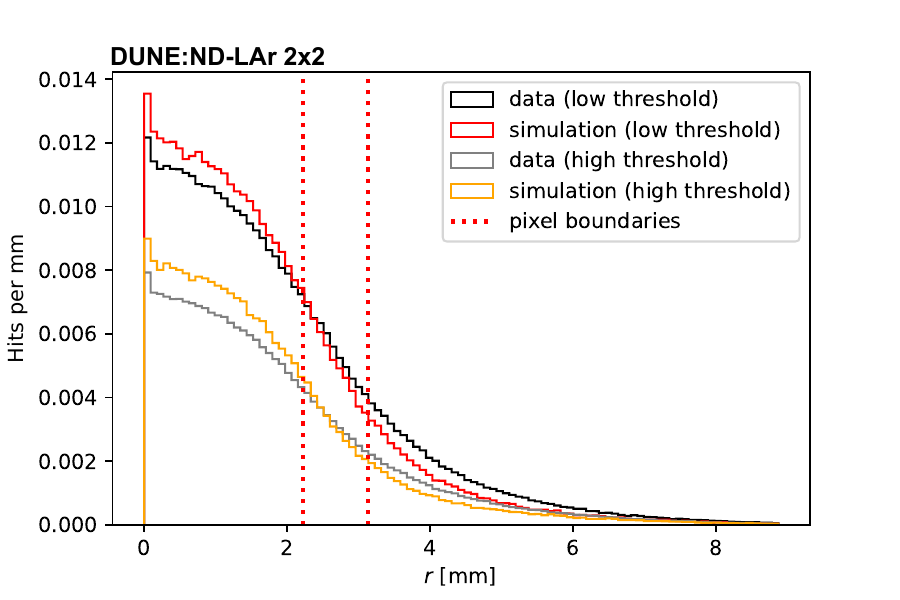}\\
 \includegraphics[width=0.65\textwidth]{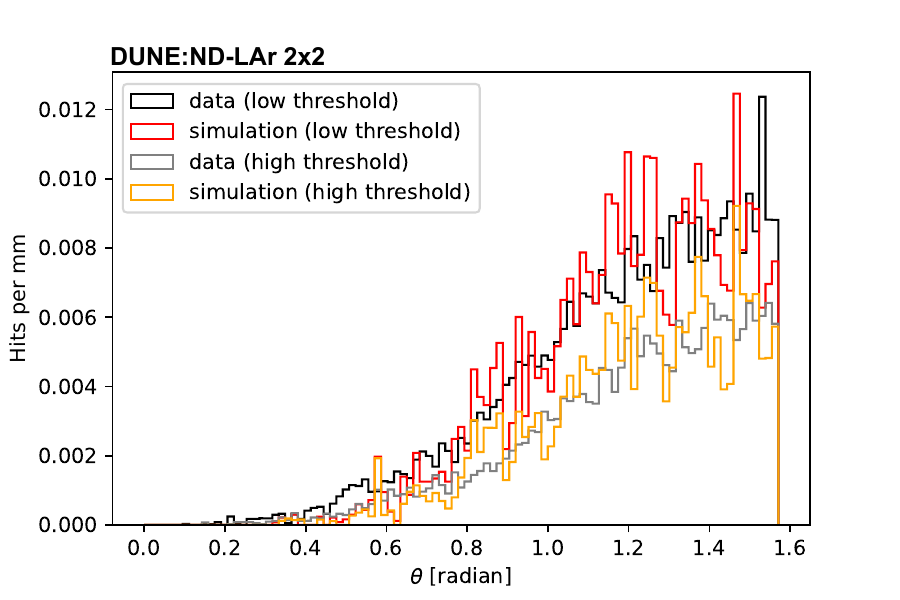}\\
 \includegraphics[width=0.65\textwidth]{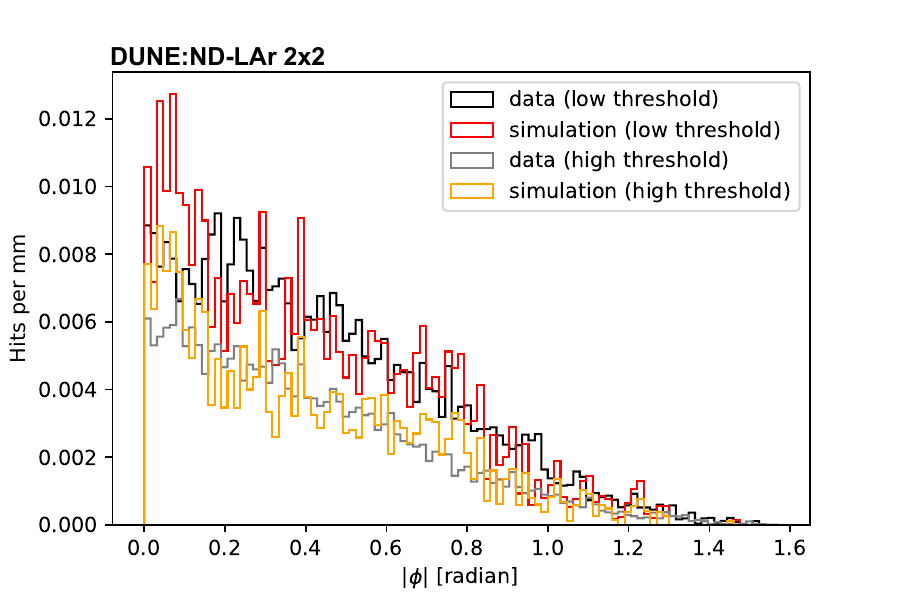}
 \caption{\label{fig:larpix-pixel-trigger-dist} Comparisons of response variation in the radial distance from the pixel center to the point of closest approach of the track projected onto the anode plane ($r$, top), the track inclination relative to the anode plane (polar angle $\theta$, middle), and the orientation angle of the track  projected onto the anode plane (azimuthal angle $\phi$, bottom). The MC shown in comparison is described in Section~\ref{sec:cosmic-analysis}.}
\end{figure}

\subsection{Pixel Charge Response}
To study the individual pixel charge response, we examine
the variation in response based on the track inclination relative to the anode plane (polar angle $\theta$), the orientation angle of the track 
projected onto the anode plane (azimuthal angle $\phi$), and the radial distance
from the pixel center to the point of closest approach of the track projected
onto the anode plane ($r$).
Fig.~\ref{fig:larpix-pixel-trigger-dist} shows the distribution of
these three quantities, normalized by the total track length. Generally, the
$\theta$ and $\phi$ distributions are comparable
between data and simulations. The $r$ distribution shows significantly more
triggers to peripheral tracks than simulated events. An overall normalization difference between high- and low-threshold data reflects the decreased sensitivity to tracks that clip the corners of the pixel. 

\begin{figure}[htbp!]
\includegraphics[width=0.495\linewidth]{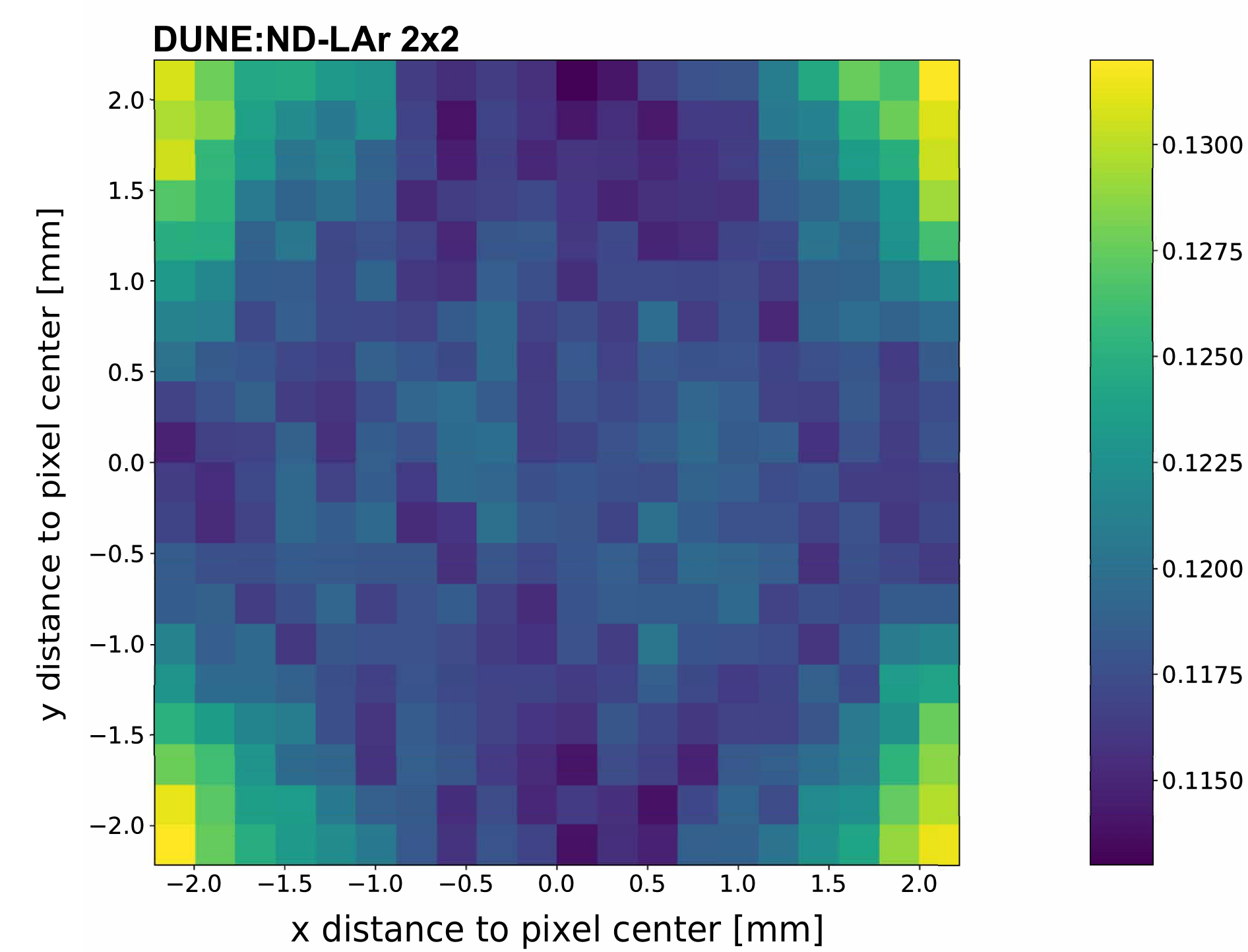}
\includegraphics[width=0.5\linewidth]{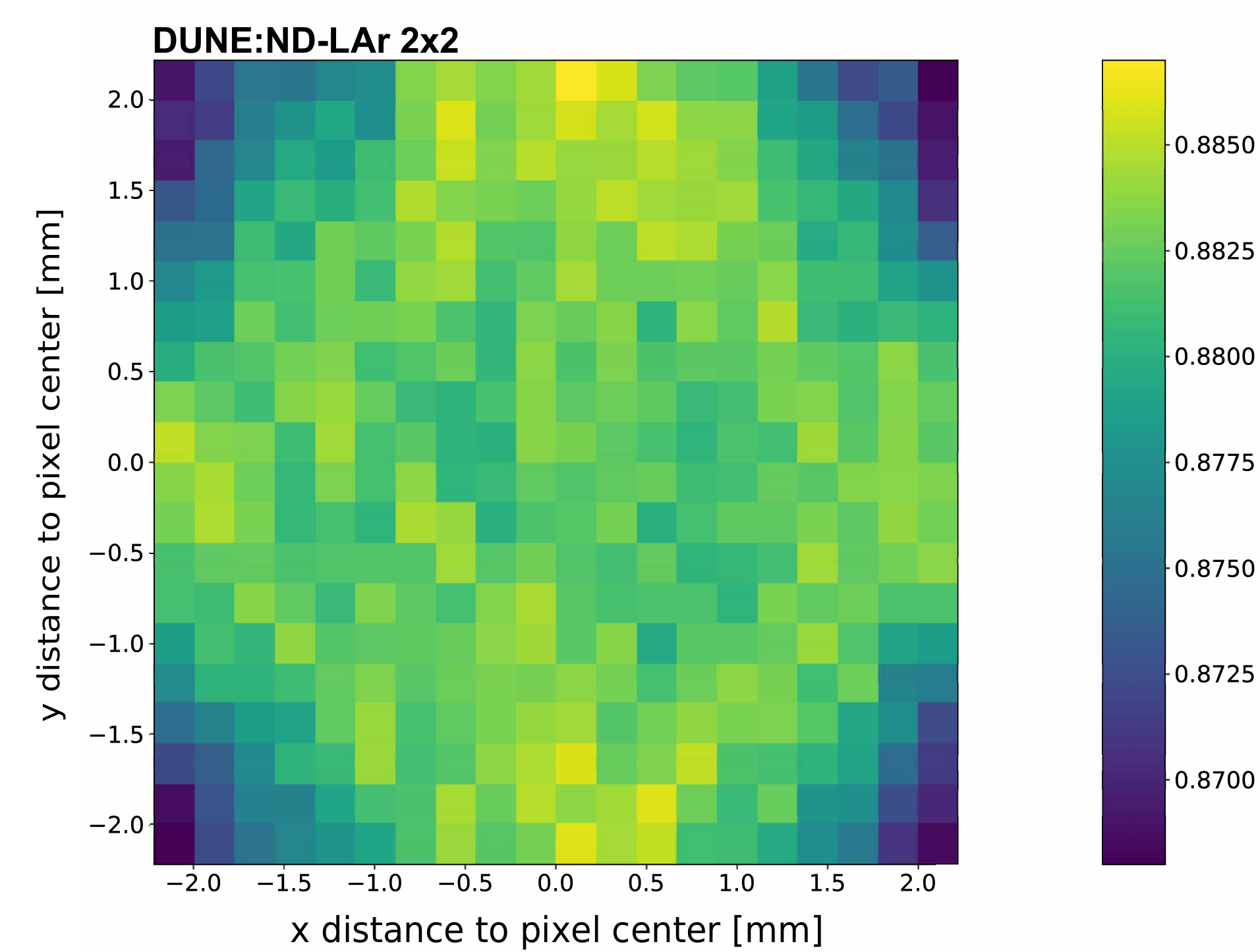}
\caption{\label{fig:pixel-to-hl-dist}Relative rate of pixel response as a function of the distance between Hough line segments and segment containing pixel's center for pixels on gaps, i.e. no charge response (left), and on tracks, i.e. with charge response (right) to the total.}
\end{figure}

A similar finding resulted from studying the distance between the MIP ionization axis and the center of the pixel.  This ionization axis can be inferred by performing a Hough transform algorithm (HTA) on the $x$, $y$, and estimated $z$ dimensions of the hit cloud.
A projection of the HTA line onto the pixel plane provides the minimum array of pixels along the axis that could have recorded some charge. This line is then divided into 0.1~mm segments longitudinally. Each individual segment's center then falls into a specific pixel, which is used to determine the distance between the segment center and the pixel center in $x$ and $y$.
The segments are split into three categories: (1) all segments as 
mentioned above independent of the recorded charge on that particular pixel, (2) those that fell into a pixel which did give a response, and (3) those in pixels that did not trigger. Prior to this categorization, all segments contained by pixels known to be inactive are excluded. In Fig.~\ref{fig:pixel-to-hl-dist}, 
the ratios of the number of segments in the latter two categories to the first one are shown. The four corners are over-represented for pixels that did not give a response but had the main ionization line crossing their pad. This quantifies the sensitivity of individual pixels to tracks clipping the corners. This difference in sensitivity is characterized by only a 3\% drop from pixel center to pixel edge where the minimum response is 85.5\%.

\begin{figure}[htbp!]
\includegraphics[width=0.99\linewidth]{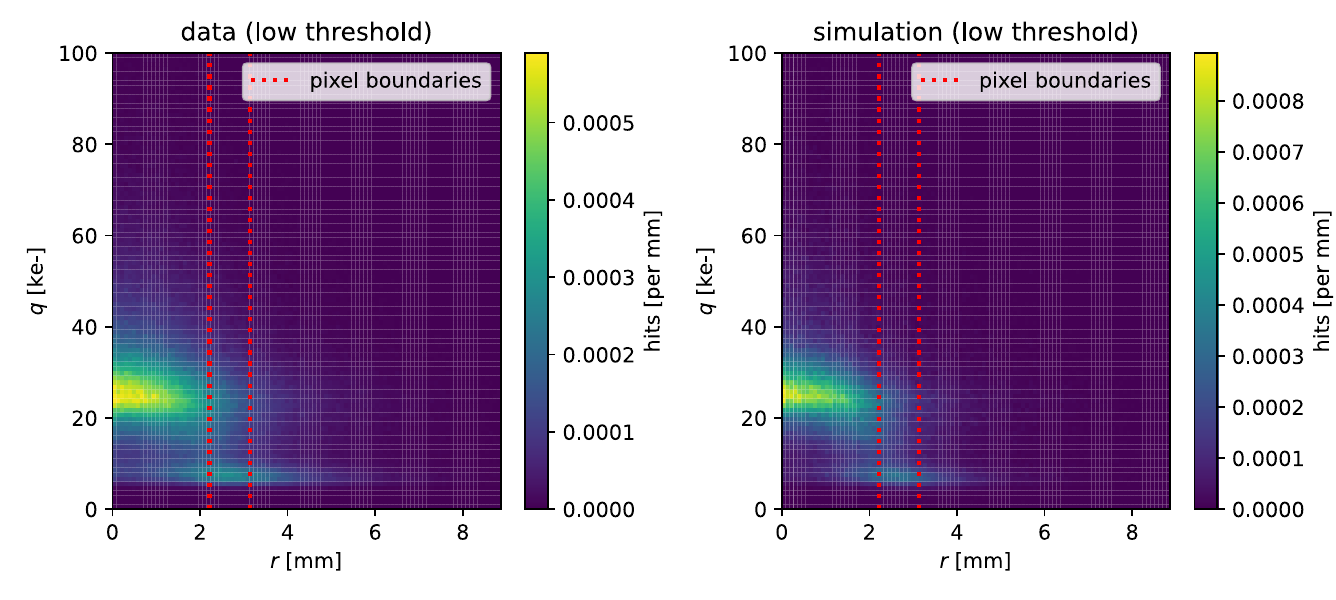}
\includegraphics[width=0.99\linewidth]{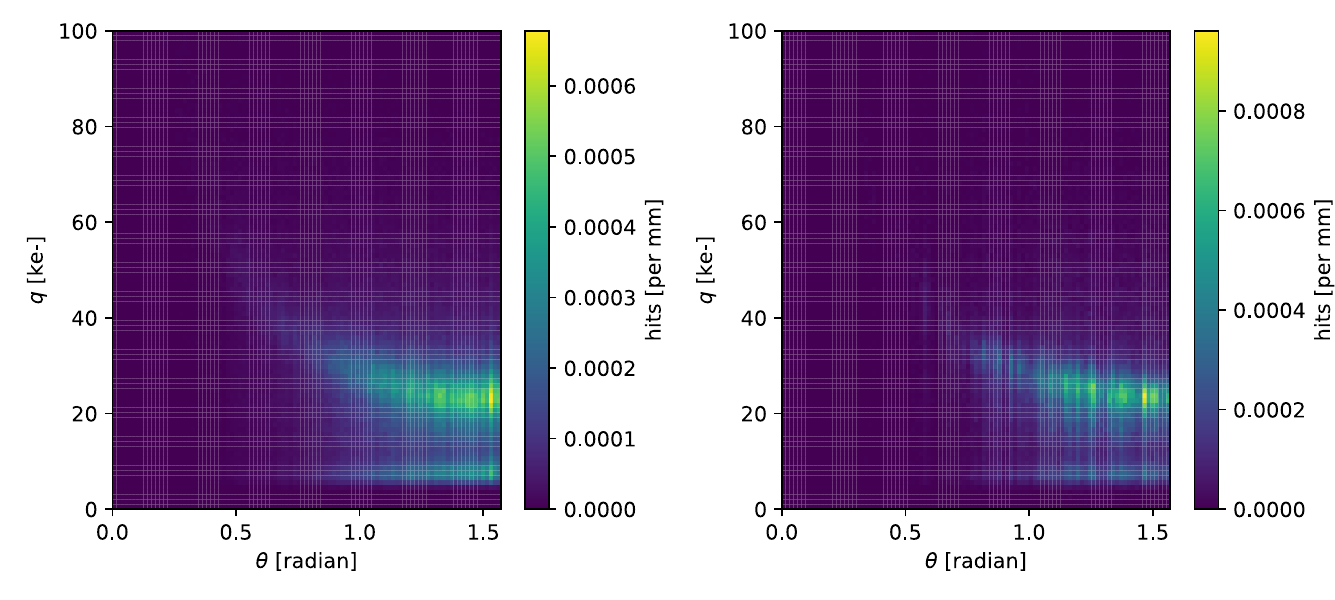}
\includegraphics[width=0.99\linewidth]{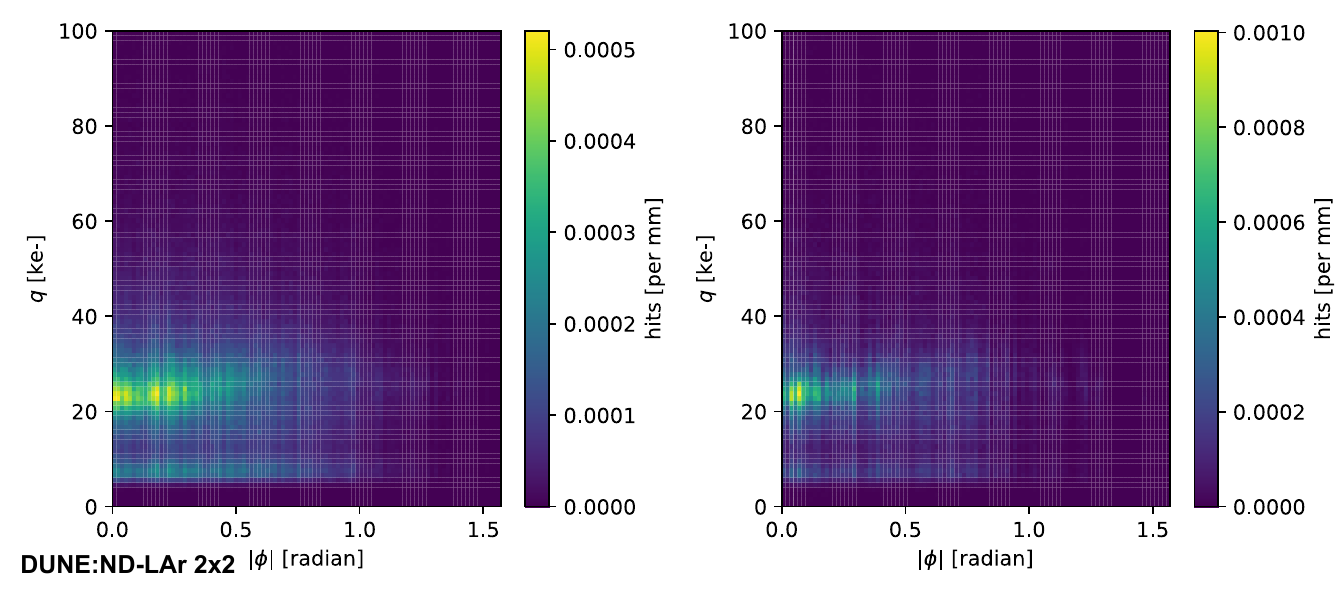}
\caption{\label{fig:larpix-pixel-q-dist-lt} Self-trigger charge distribution for MIP tracks with different track orientations with respect to the pixel, normalized to number of triggered channels per reconstructed track length. Low-threshold data are used. The MC simulation shown in comparison in the second column is described in Section~\ref{sec:cosmic-analysis}.}
\end{figure}

\begin{figure}[htbp!]
\includegraphics[width=0.99\linewidth]{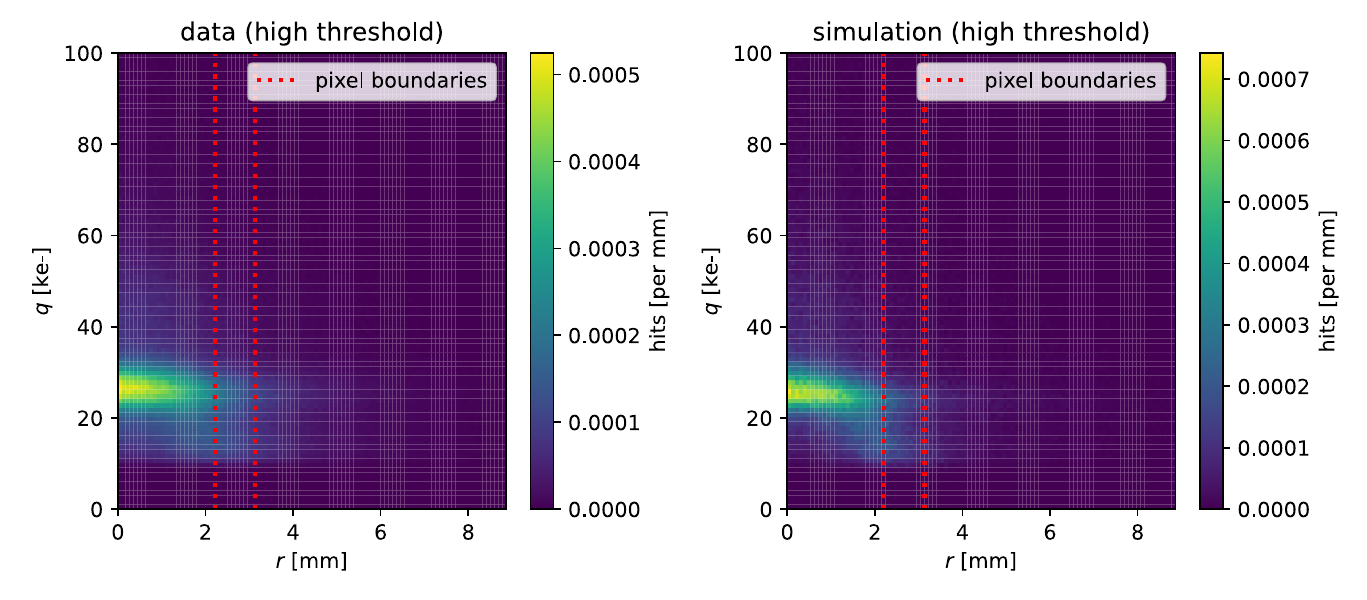}
\includegraphics[width=0.99\linewidth]{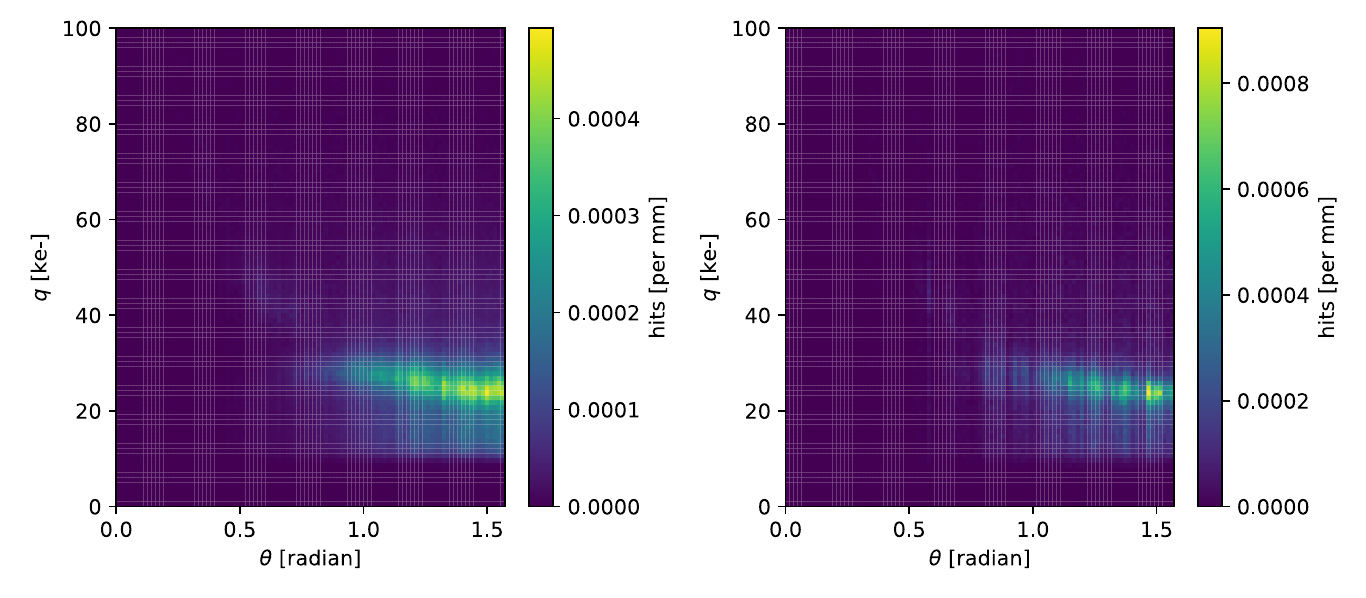}
\includegraphics[width=0.99\linewidth]{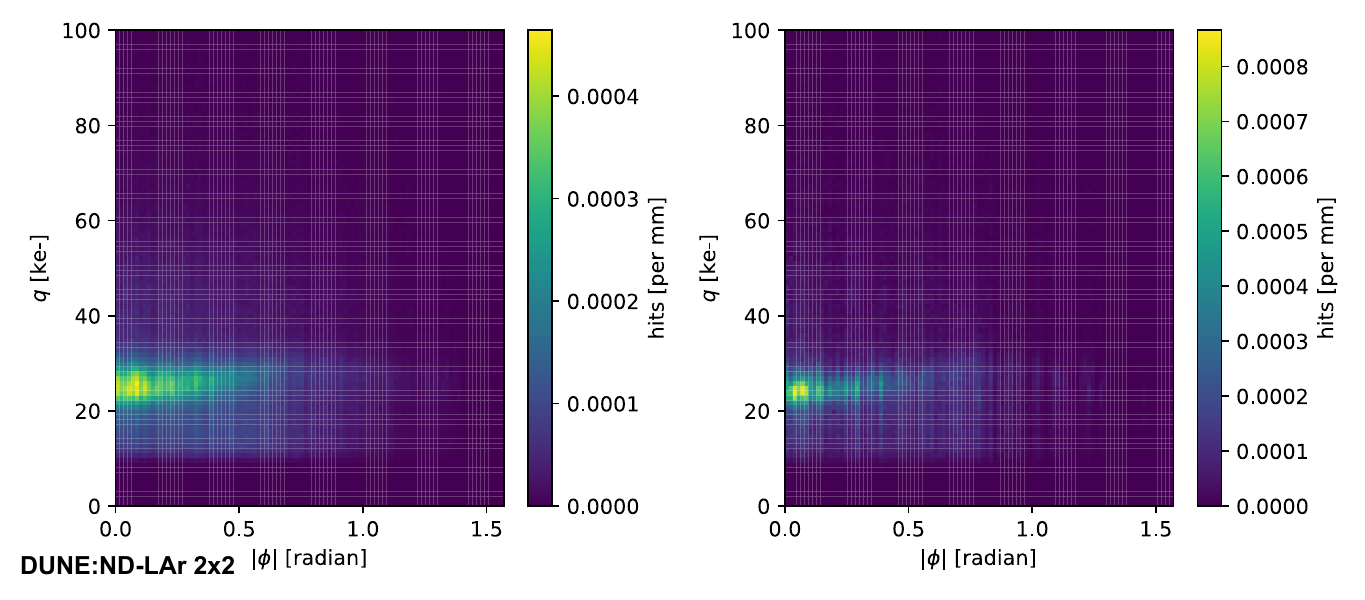}
\caption{\label{fig:larpix-pixel-q-dist-ht} Same as Fig.~\ref{fig:larpix-pixel-q-dist-lt} but for high threshold data.}
\end{figure}

\begin{figure}[htbp!]
\includegraphics[width=0.495\linewidth]{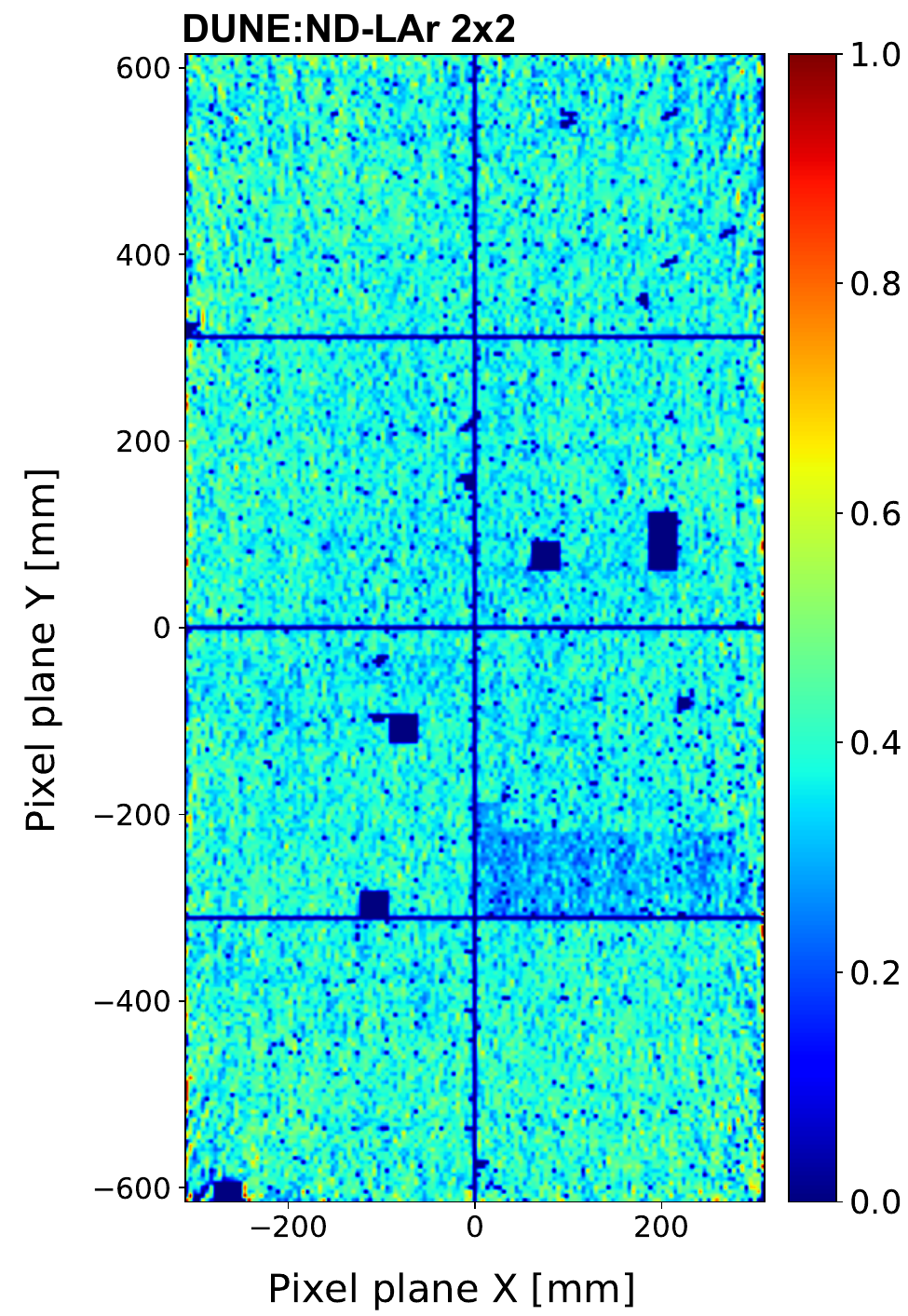}
\includegraphics[width=0.495\linewidth]{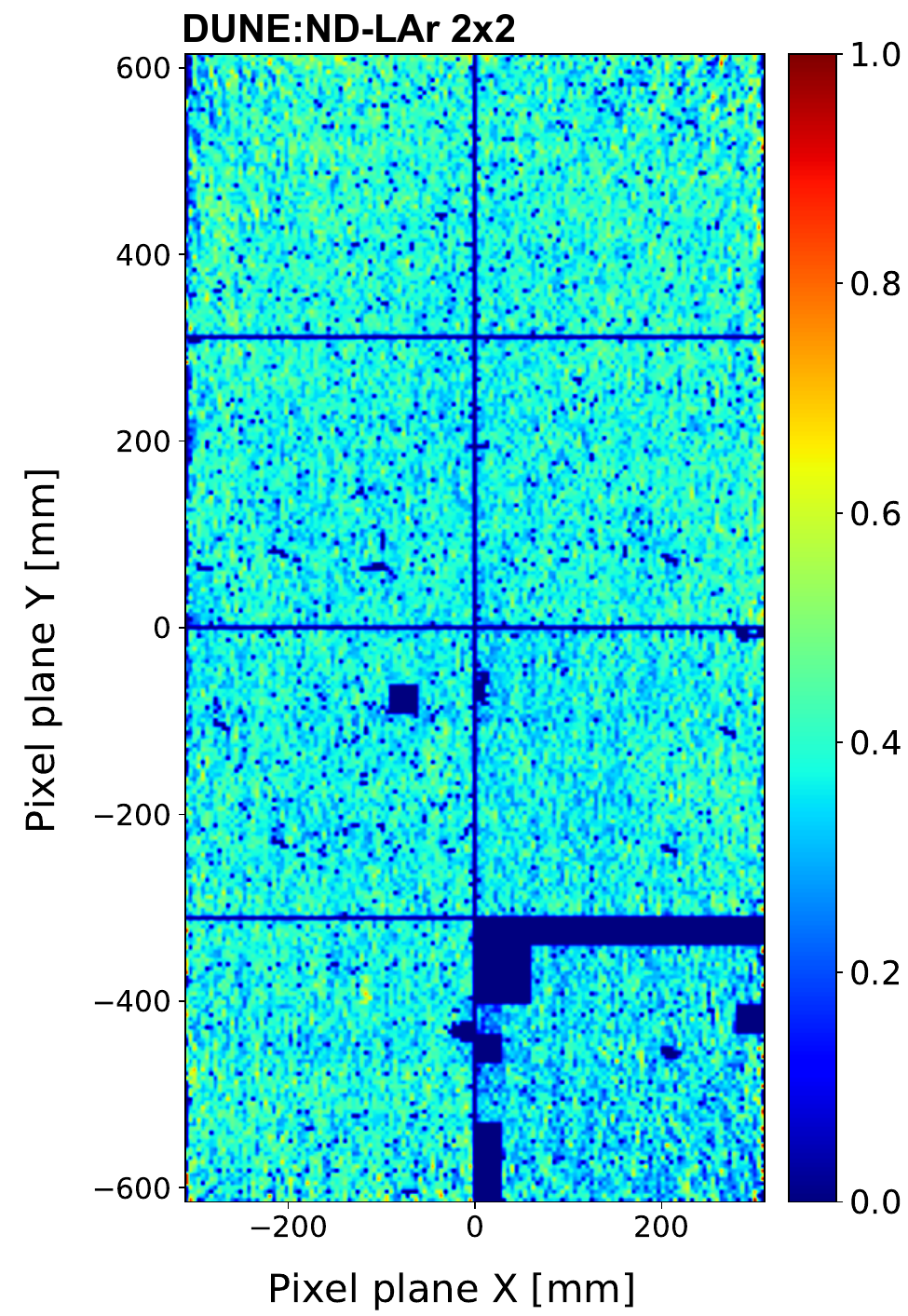}
\caption{\label{fig:mips-response-maps}MIP response maps for anode plane 1 (left) and anode plane 2 (right), showing the fraction of triggered hits on each pixel relative to the expected number based on reconstructed track trajectories.}
\end{figure}

Figs.~\ref{fig:larpix-pixel-q-dist-lt} and \ref{fig:larpix-pixel-q-dist-ht} show the charge distribution with respect to the track orientation for low- and high-threshold data, respectively.
Overall, similar features appear in each panel:
a prominent peak corresponding to the
charge deposited by a MIP across a single pixel width. In the $r$
distribution, a secondary distribution of low-charge hits is present, corresponding
to tracks that clip the corners of the pixel. This feature is also present
in the $\phi$ distribution as an increase in the spread of the 
charge as $\phi\to\pi/4$. The $\theta$ distribution
shows a characteristic increase in the charge as $\theta\to0$,
which corresponds to tracks perpendicular to the anode plane, where each pixel can see a contribution from a relatively long track length. A flattening of
the observed charge near $\theta=0.8$ is a threshold effect and
is not present in the low-threshold data.
To test the responsiveness of individual pixels and identify potentially malfunctioning
channels beyond those known to be inactive,
a MIP response map of the entire pixel plane was constructed. This map is the
ratio of recorded over expected
hits, and identifies regions on 
the pixel plane which are less responsive than others.
Both components start off with the same principle of performing an HTA on the $x$, $y$, and inferred $z$ dimensions of 
the hit cloud to obtain the MIP’s central ionization axis in 3D. This 
axis is then projected onto the pixel plane to result in a 2D line. Next, 
all hits within 8~mm of the line are selected and the maximum 
track width is set equal to the most distant point within this radius. 
To then obtain the first map, all pixels that recorded hits within a radius equal 
to the maximum track width of the projected line receive an entry. To construct 
the second map, all existing pixels within that same radius receive an entry. If a pixel is unresponsive, it will not show up in the first but will appear in the second, leading to a low ratio in that specific area.
Selection cuts place requirements on the straightness of tracks relative to the fit Hough lines
as well as the consistency with a roughly constant energy deposition profile, to ensure that the events analyzed consist primarily of MIP-like tracks.
Fig.~\ref{fig:mips-response-maps} shows the resulting MIP response maps for
both anode planes.

\subsection{Saturation}
\begin{figure}[htbp!]
\includegraphics[width=0.75\linewidth]{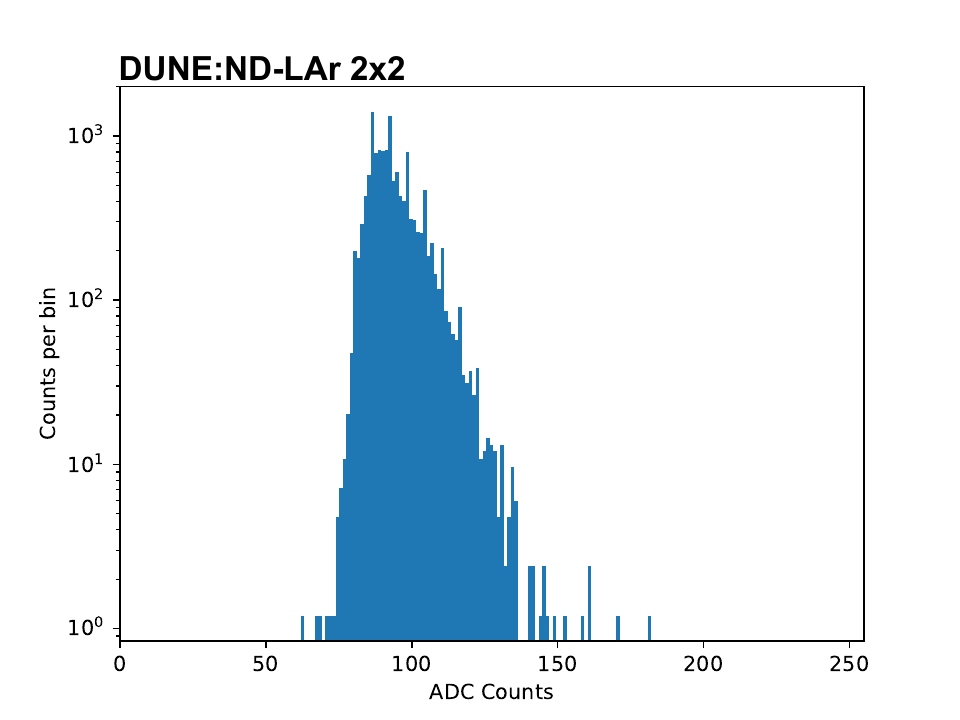}
\caption{Per-pixel ADC value distribution for cosmic ray events between 2 and 10 GeV. All signals are well within the ADC dynamic range of 0--256 counts.}
\label{fig:adc-cosmics}
\end{figure}

An additional consideration is saturation in the LArPix-v2 ASIC's 8-bit successive-approximation ADC, which is expected to occur when the charge on a given channel exceeds 200~ke$^-$ within a 2.6~$\mu$s time window.
A scan for events including saturated packets was performed over eight hours of cosmic ray data acquired at high gain and low threshold.
Packets within 1~s of a time synchronization pulse were found to include additional noise and saturation effects, and were excluded.
After accounting for this, a small fraction ($2.9\times10^{-6}$) of events with matching charge and light information contained a saturated ADC measurement. These events were manually inspected, and the saturation was clearly uncorrelated in space and time with the physical interactions, but rather they leaked into the event due to their proximity with a sync pulse.
With low thresholds, $<0.002$\% of triggers resulted in ADC saturation, again driven by the pulse-per-second sync signal;
channels 35-37 on all chips, which are located physically adjacent to the sync pulse pin, saturated most often and together accounted for 15\% of these saturated packets.
The ADC count distribution for events with deposited energy between 2 and 10~GeV is shown in
Fig.~\ref{fig:adc-cosmics}. These energies are of interest as they are representative of neutrino interactions at ND-LAr, and the distribution falls well within the dynamic range of the ADC.

\subsection{Calorimetric Response} 

\begin{figure}[htbp!]
 \includegraphics[width=0.95\textwidth]{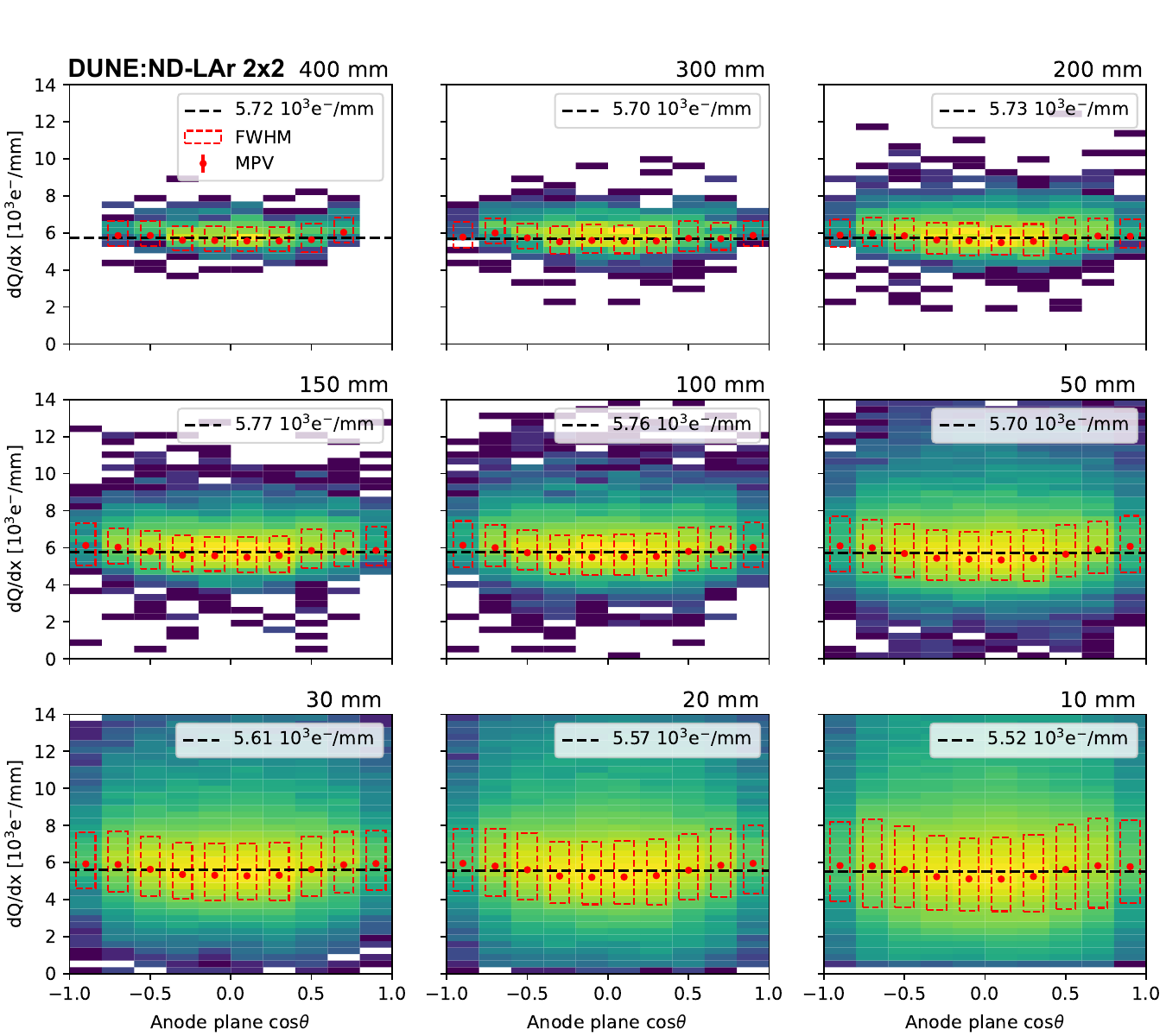}
 \caption{$dQ/dx$ measured for segments of different lengths as a function of the orientation relative to the anode planes. A value of $\cos\theta = 0$ corresponds to segments parallel to the anode plane. The distributions in each bin have been fitted with a Gaussian-convolved Moyal function. The red points correspond to the most probable value of the fitted distribution and the dashed rectangles correspond to the full width at half maximum. The dashed black line represents the average MPV.}
 \label{fig:dqdx_plane}
\end{figure}

\begin{figure}[htbp!] 
 \includegraphics[width=0.95\textwidth]{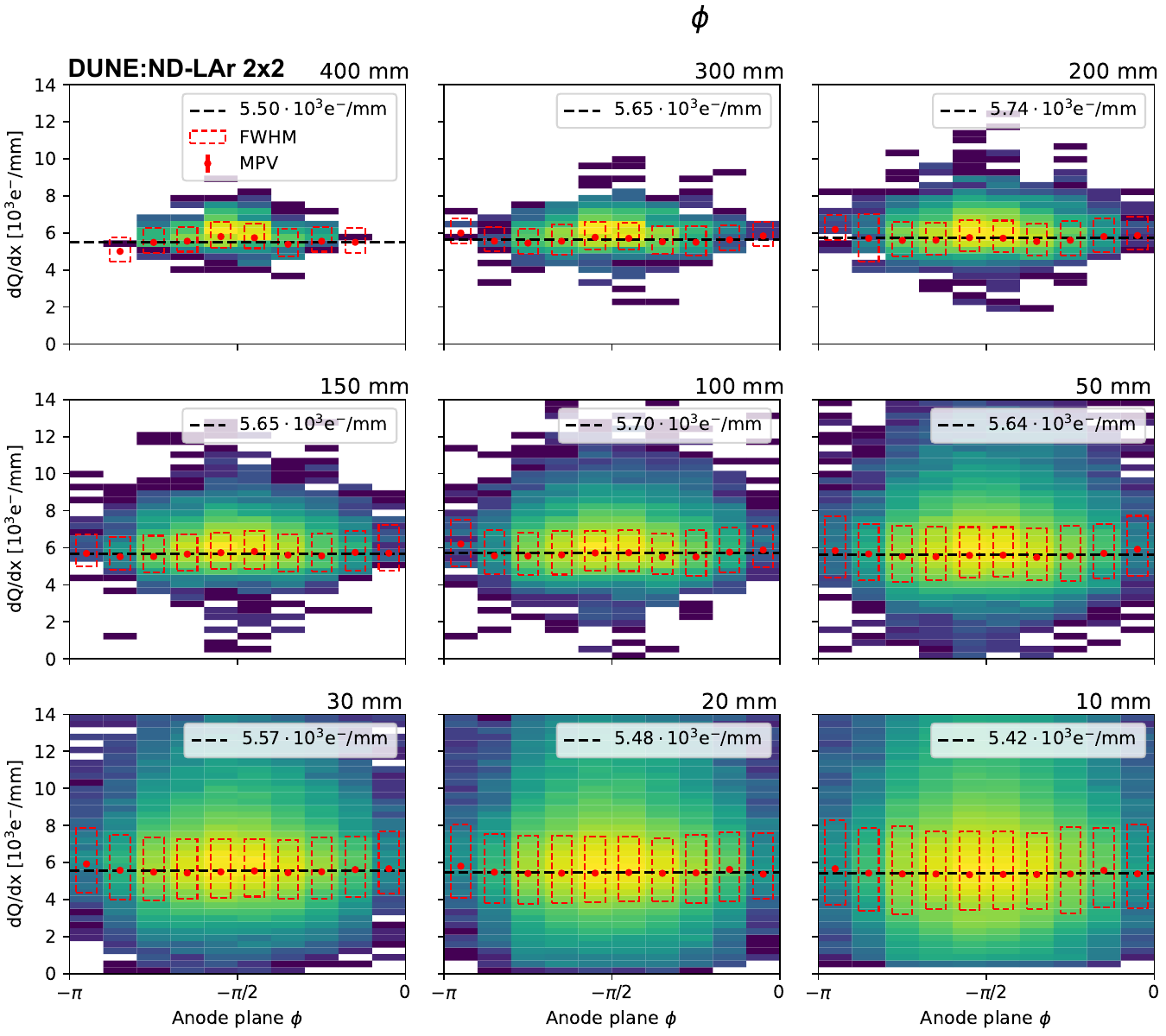}
 \caption{$dQ/dx$ measured for segments of different lengths as a function of the azimuthal angle $\phi=\mathrm{atan2}(y,x)$, where $y$ and $x$ are the components of the segment along the anode plane axes. The distributions in each bin are fitted with a Gaussian-convolved Moyal function. The red points correspond to the most probable value of the fitted distribution and the dashed rectangles correspond to the FWHM. The dashed black line represents the average MPV.}
 \label{fig:dqdx_phi}
\end{figure}

Finally, the calorimetric response of \mbox{Module-0} charge readout was also studied.
Figs.~\ref{fig:dqdx_plane} and \ref{fig:dqdx_phi} show the variation of the $dQ/dx$ for segments of different lengths relative to the track orientation, defined by the azimuth angle $\phi$ and the $\theta$ angle between the track and a vector normal to the anode plane. The reconstructed tracks used for this analysis come from the low threshold runs (see Section~\ref{sec:overview}). Events with more than 20 reconstructed tracks were excluded, since they often correspond to large showers or non-cosmic triggers. Tracks were required to be longer than 10~cm and to have at least 20 associated hits. They were then subdivided into segments of variable length from 10 to 400~mm and the distributions were fit with a Gaussian-convolved Moyal function. The MPV shows a slight dependence on $\cos\theta$, with tracks that impinge perpendicularly to the anode plane tending to have a larger amount of deposited charge per unit length. These data provide insight into subtle effects in the pixel charge response, such as those related to induction effects and electric field uniformity, and enable a data-driven calibration. 

\section{Light Readout Performance}
\label{sec:light}

\subsection{Overview}
The \mbox{Module-0} detector also provided a large-scale, fully integrated test of the light readout system, enabling a detailed performance characterization of the ArCLight and LCM modules, readout, DAQ, triggering, and timing with a large set of events. Using cosmic ray data and dedicated diagnostic runs under a variety of detector configurations, a suite of tests was performed to assess the charge spectrum, inter-- and intra-event timing accuracy, and photon detection efficiency. The subsequent matching of events between the charge and light system is considered in Section~\ref{sec:qlmatch}.

\subsection{Calibration}
\begin{figure}[htbp]
\includegraphics[width=0.48\linewidth]{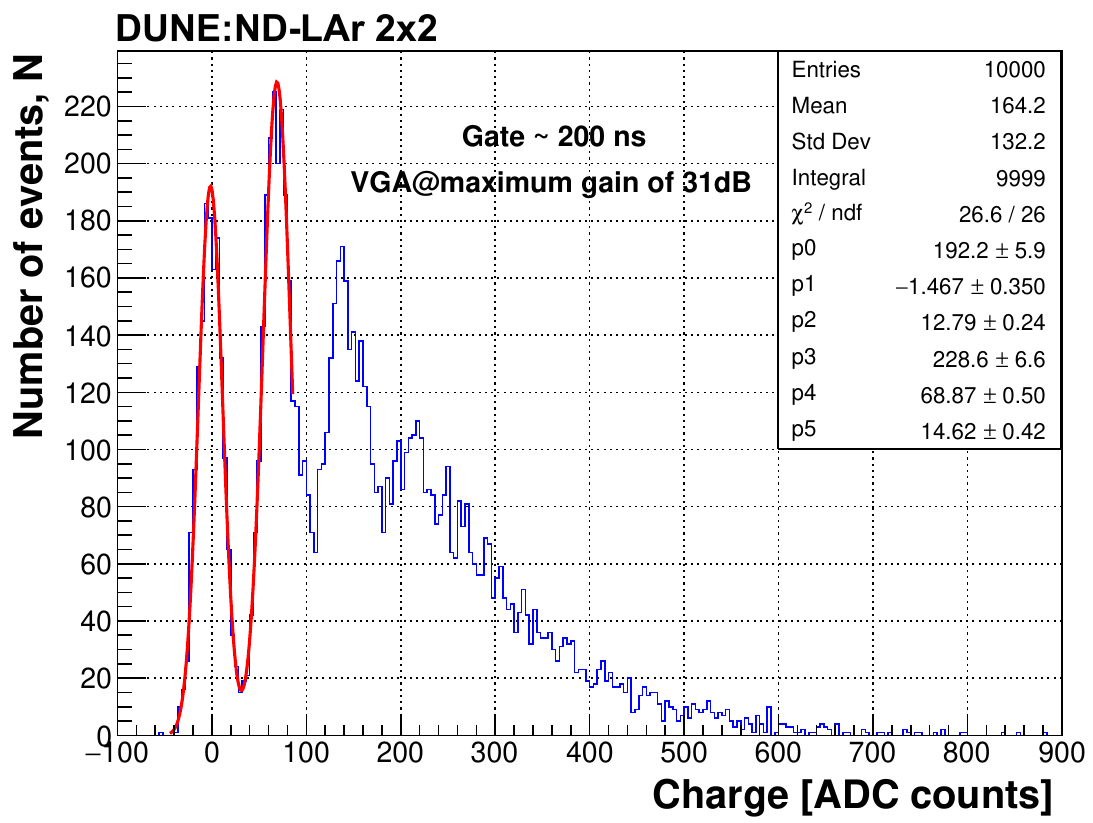}
\includegraphics[width=0.49\linewidth]{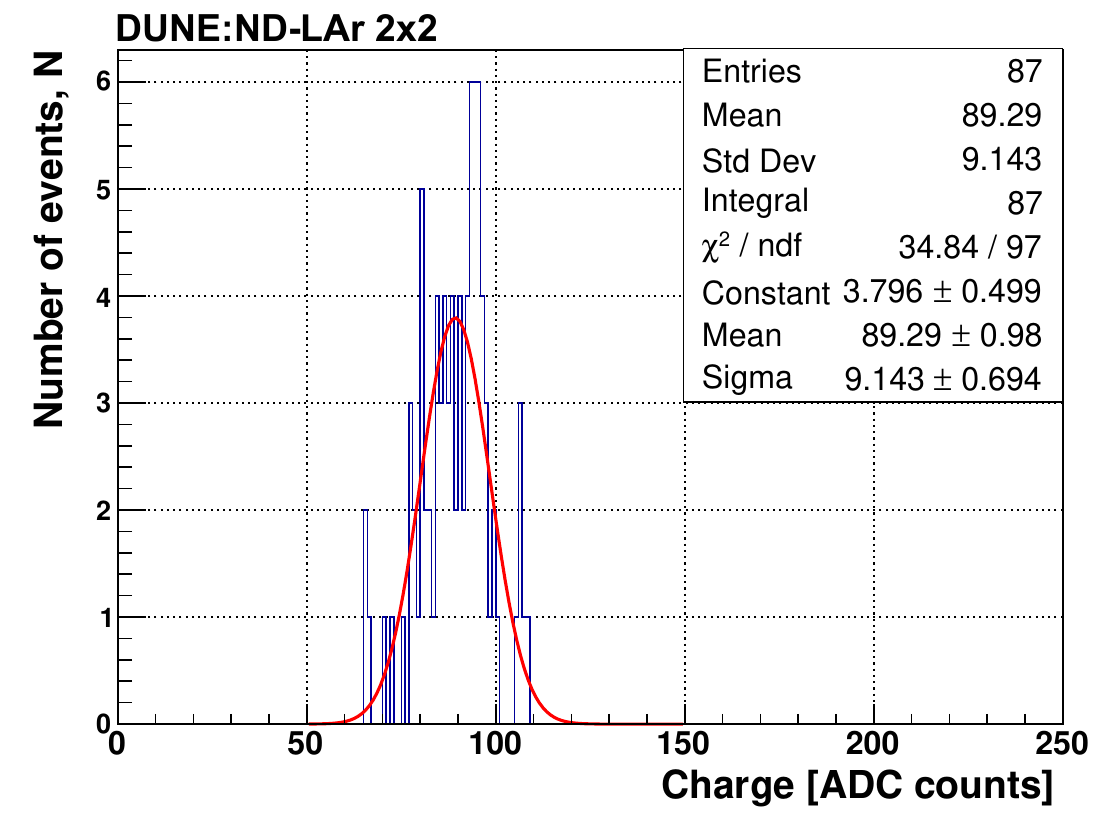}
\caption{\label{fig:gain_distr} Typical charge spectrum obtained during SiPM gain calibration (left); SiPM gain distribution
(right).
}
\end{figure}

Before collecting cosmic data, a SiPM gain calibration was performed using an LED source, where the bias voltage for each SiPM channel was adjusted to obtain a uniform gain distribution across the channels, as shown in Fig.~\ref{fig:gain_distr}. The amplification factors for the variable gain amplifiers used in the SiPM readout chain were also tuned, and set to maximum (31~dB) except for LCM channels (21~dB) during cosmic ray data taking, to adjust signals to the input dynamic range of the ADC. LCMs were used to provide an external trigger to the charge readout system, with an effective threshold of about 30 photoelectrons (p.e.). The trigger message, written into the continuous self-triggered data stream of the charge readout system, provides a precise timestamped flag for identifying coincidences between charge and light readout.

\subsection{Time Resolution}

\begin{figure}[htbp]
\includegraphics[width=0.47\linewidth]{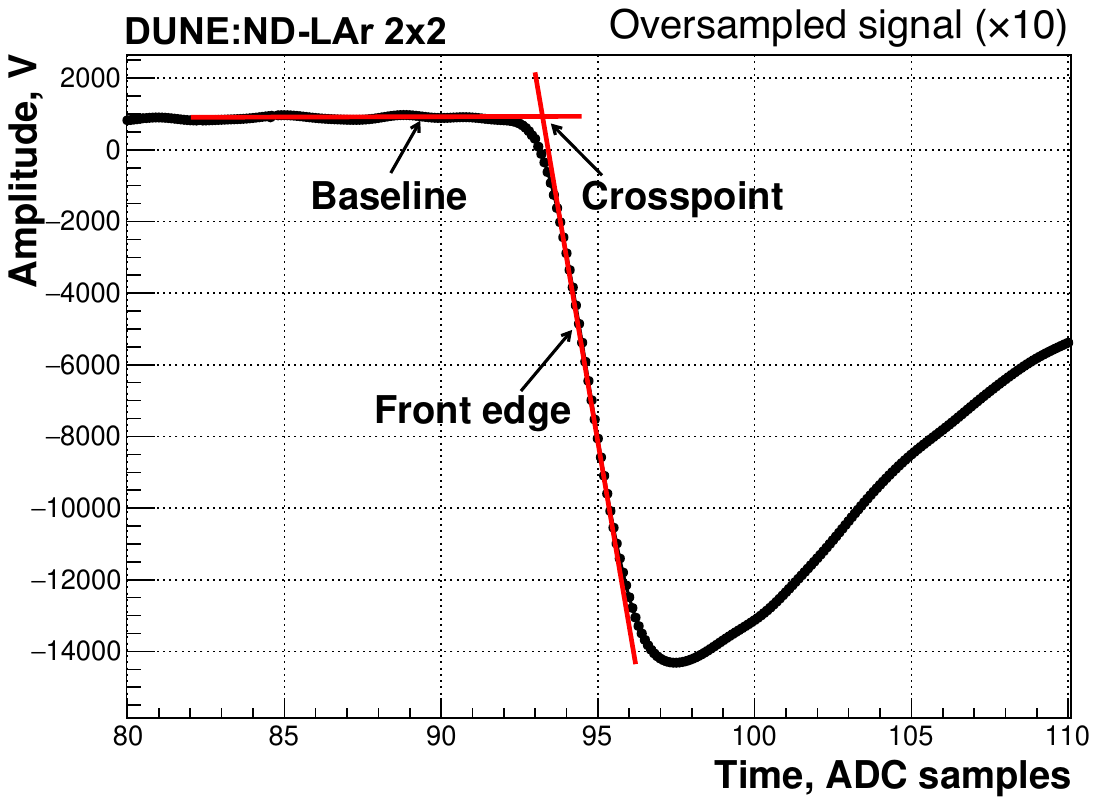}
\includegraphics[width=0.49\linewidth]{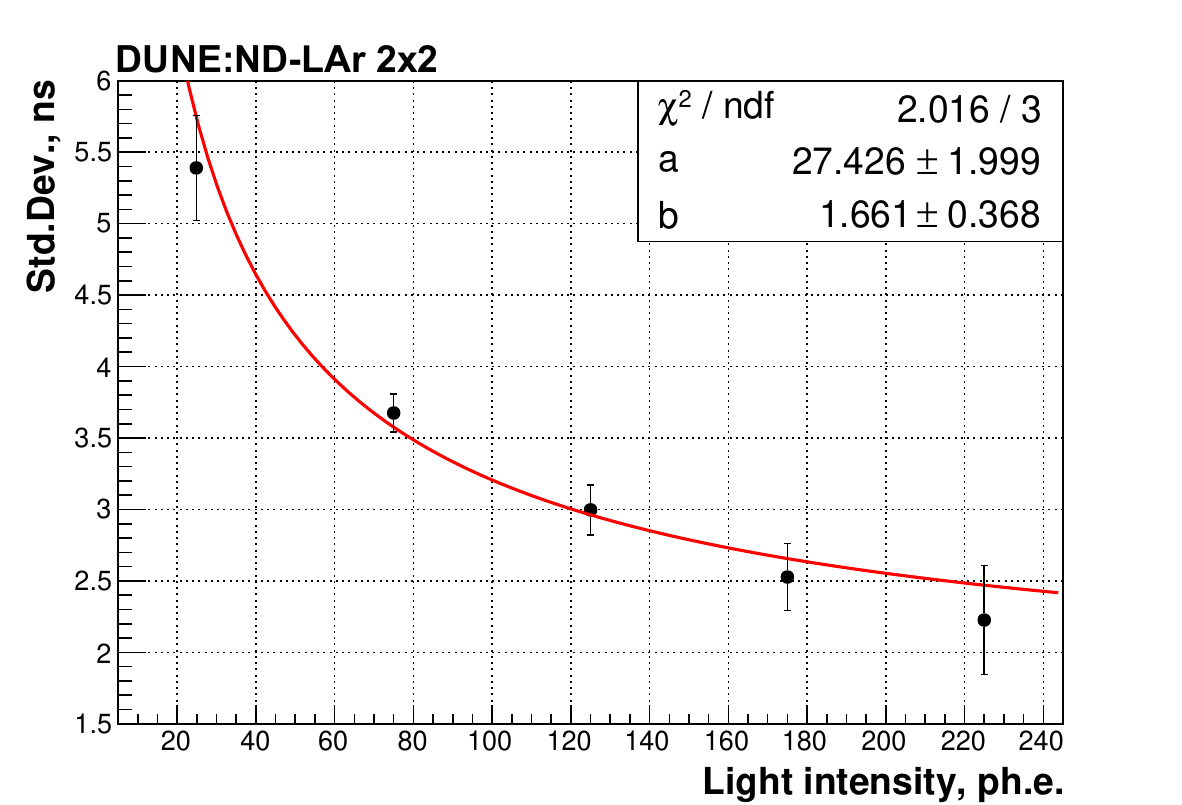}
\caption{\label{fig:time_resolution} Oversampled signal using Fourier transformation. Red lines show the linear approximations of the rising edge and the baseline (left). The time resolution between two LCMs (LCM-011, LCM-017) as a function of the signal response (right).}
\end{figure}

\begin{figure}[htbp]
\includegraphics[width=0.485\linewidth]{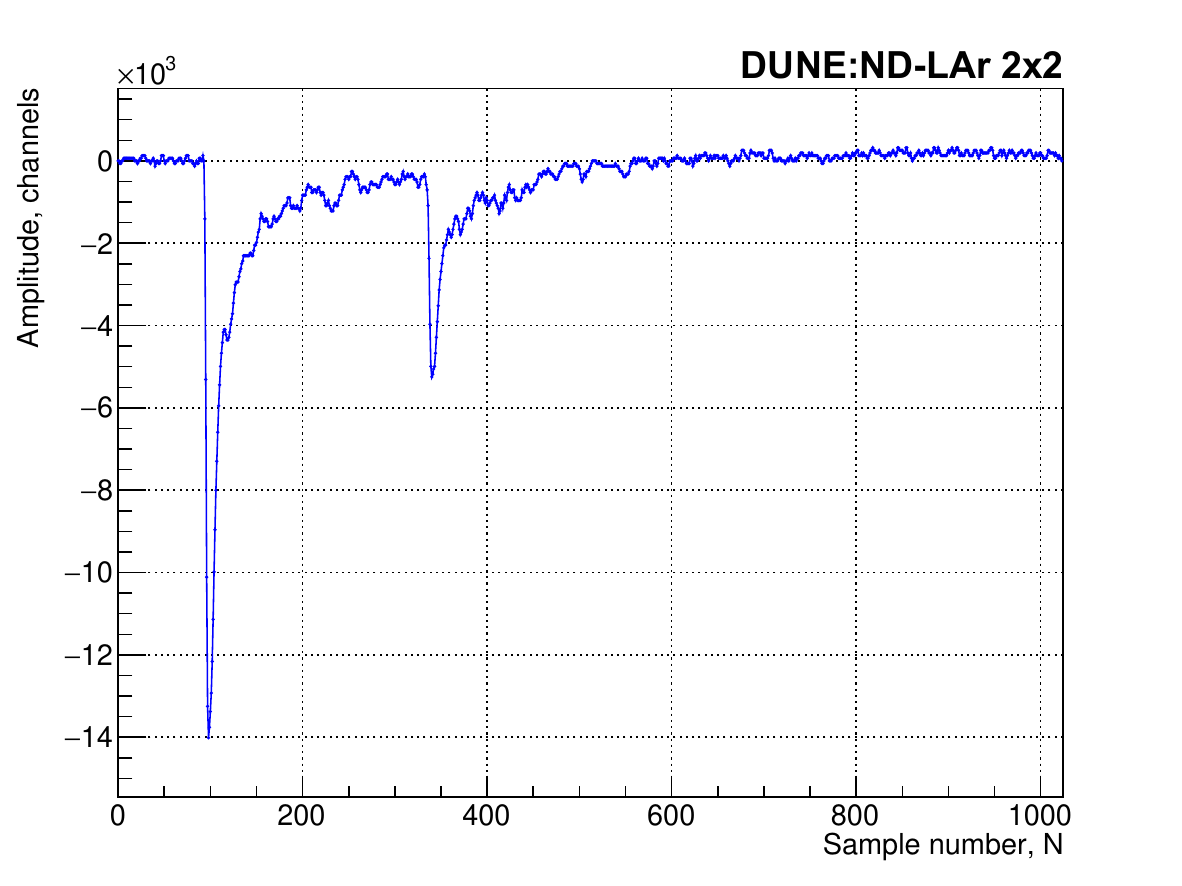}
\includegraphics[width=0.485\linewidth]{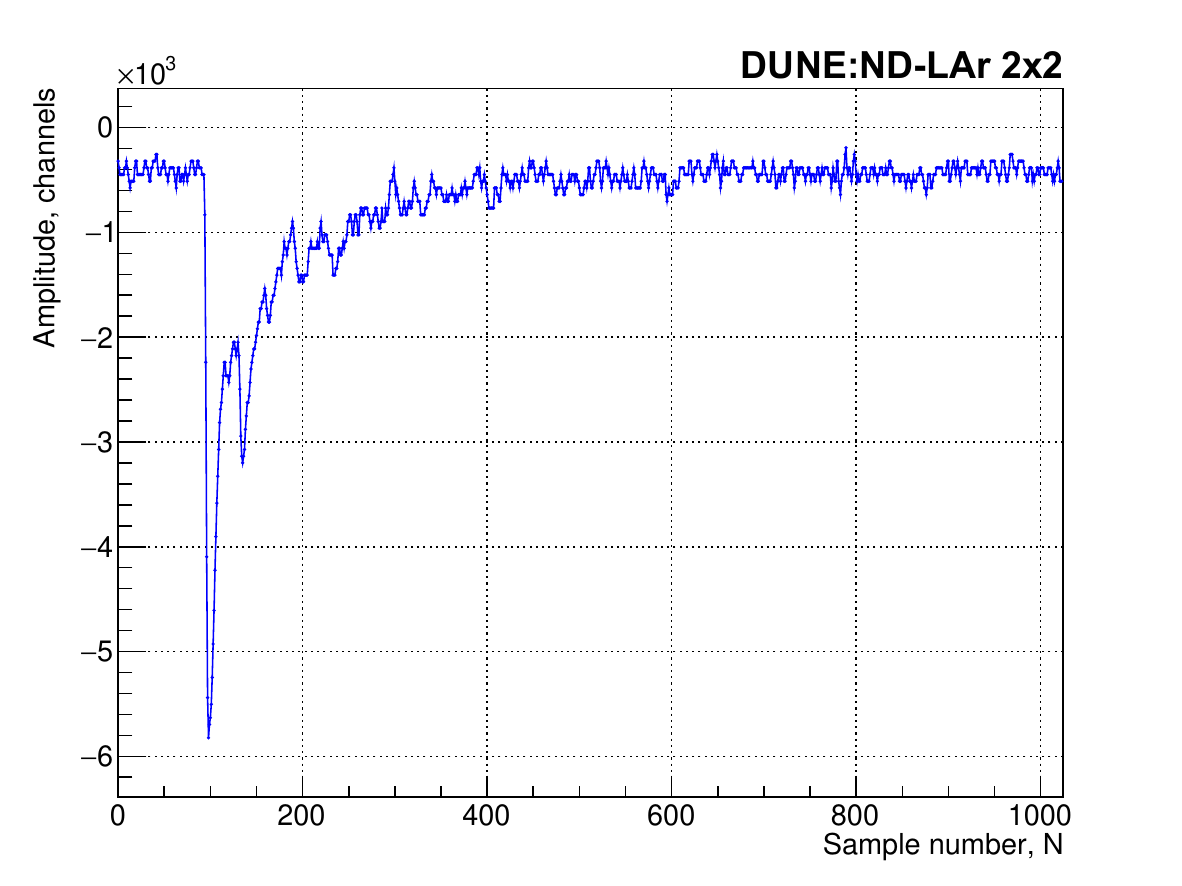}
\caption{\label{fig:muon_and_michel} Two examples showing signals of the stopping muon and delayed Michel electron detected by the LCM. The waveforms were digitized at 10~ns intervals.}
\end{figure}

Events induced by cosmic muons traversing the TPC volume were used to extract the time resolution of the light detectors. The time measurement proceeds as follows: each waveform is oversampled through a Fourier transform to increase the number of points on the rising edge, enabling a good linear fit of it. Then, a linear fit to the baseline is performed, and the crossing point of the rising edge of the signal with the baseline is calculated, providing a robust single-channel event time. This process is illustrated in Fig.~\ref{fig:time_resolution} (left). The extracted time resolution for a pair of neighboring LCM channels is shown in Fig.~\ref{fig:time_resolution} (right) as a function of the signal amplitude. This quantity is obtained by taking the standard deviation of the time difference recorded between the two channels over multiple events without any time-of-flight corrections. For large signals, this resolution approaches $\sim2$~ns.
An example application of the excellent timing resolution for the LCMs is the identification of Michel electrons from stopping muon decays, where the relative timing between the muon and electron signals is dominated by the mean lifetime of the muon, $\tau\sim2.2~\mu$s.
Two  examples of signals from a stopping muon and a delayed Michel electron detected by the LCM are shown in Fig.~\ref{fig:muon_and_michel}. Since the muon decay time is variable but follows a well-understood exponential distribution, such events may be used, for example, to study event pile-up in neutrino interactions.

\subsection{Efficiency}

To assess the efficiency of the LRS, the scintillation light induced by tracks reconstructed from the TPC charge readout data is used. In particular, cosmic muon tracks crossing the entire detector vertically are considered.
In a 3D simulation, the charge of a track is discretized to single points with a \SI{1}{\milli\metre} resolution along the track, assuming an infinitely thin true trajectory.
For each point in this voxelized event, the solid angle to the light detector in the detector module is then calculated. Next, assuming isotropic scintillation light emission, the solid angle can  be used to compute the geometrical acceptance of the light for each detector tile.
The number of photons hitting the detector surface is estimated by multiplying the geometrical acceptance by the number of emitted photons per unit track length and integrating over the full track length. Here, the number of emitted photons per unit track length has been calculated for the nominal electric field intensity of \SI{0.5}{\kilo\volt\per\centi\metre}~\cite{bnl_lar_table}.
Rayleigh scattering, a small effect over the relevant distance scales, is neglected in this calculation.

\begin{figure}[htbp]
\vspace{5mm}
\includegraphics[width=0.495\linewidth]{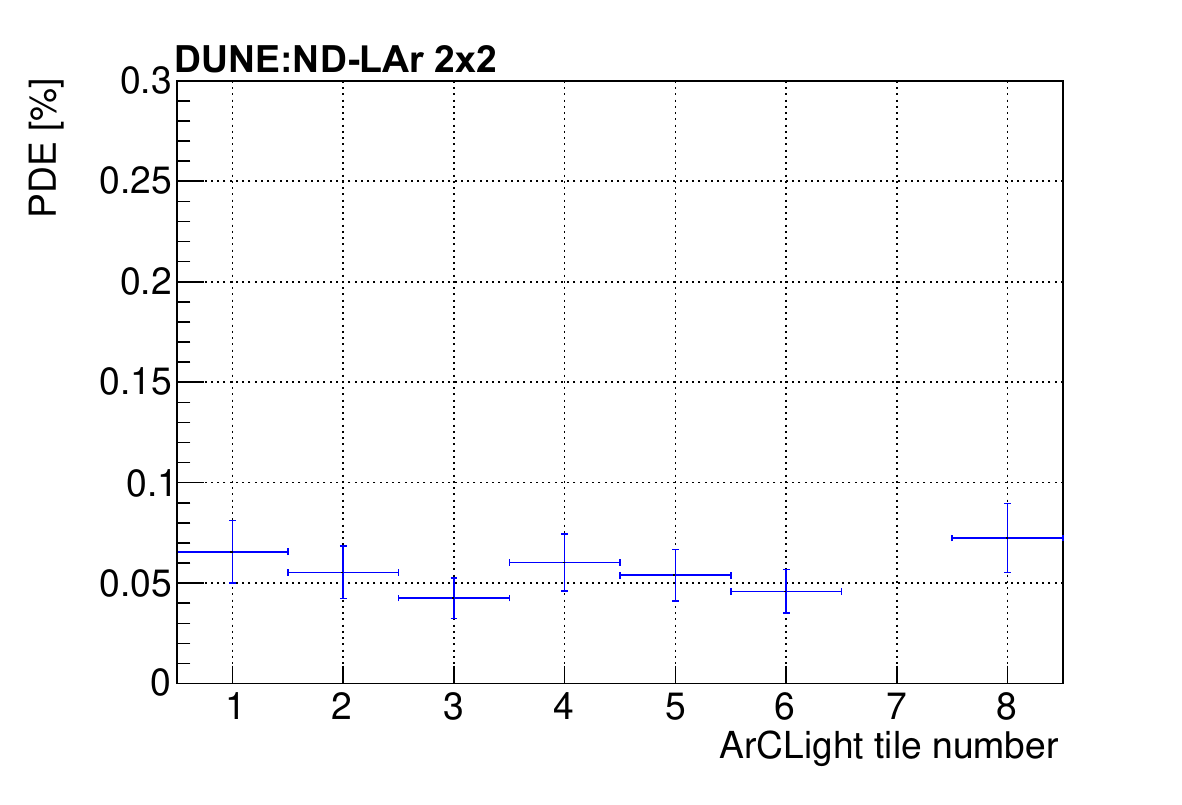}
\includegraphics[width=0.495\linewidth]{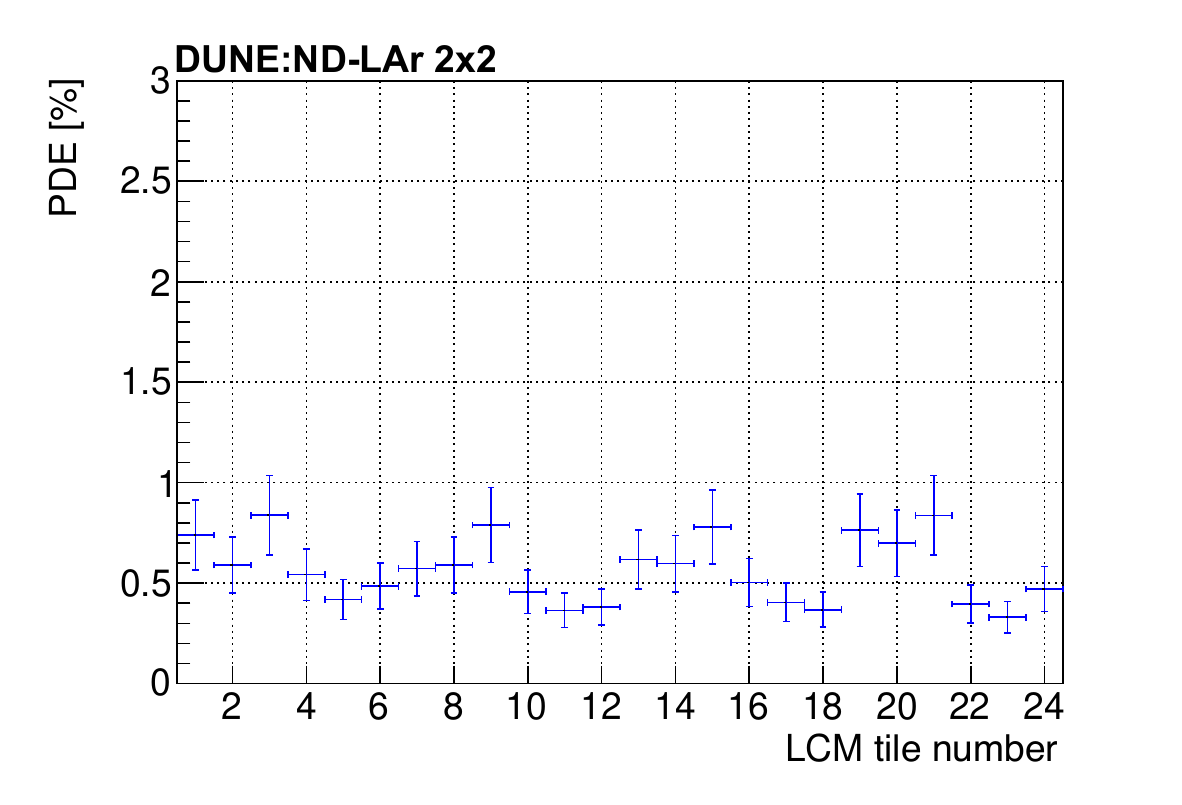}
\caption{\label{fig: PDE} Absolute PDE for each ArCLight (left) and LCM (right) tile (arbitrary numbering). ArCLight tile 7 was disabled during Module-0 data taking. The LCM tiles are placed in sets of 3 to cover the same area as one ArCLight tile.}
\end{figure}

The photon detection efficiency (PDE) of the light detection system can be estimated by comparing the measured number of p.e. and the estimated number of photons hitting the detector surface, as obtained from the simulation described above.
Since the waveforms obtained with the light detectors have been integrated using a limited gate length, the actual scintillation light might be underestimated.
This was corrected by multiplying the number of reconstructed photons by an integration gate acceptance factor, which is calculated based on the detector response and the scintillation timing characteristics.
Fig.~\ref{fig: PDE} shows the measured PDE for all ArCLight and LCM modules used in the \mbox{Module-0} detector.
The LCM shows an average PDE of 0.6\%, which enables a light trigger for events depositing MeV-scale energies, with an accurate scintillation amplitude and energy reconstruction.
The PDE of the ArCLight modules is about a factor of \SI{10}{} lower than the corresponding value obtained with the LCMs, which allows for a larger dynamic range.
The ArCLight technology additionally enables a high position sensitivity, which can be used to accurately triangulate the origin of the scintillation light emission point~\cite{arclight}. For the LCM it can be observed that tiles placed at the top (see Fig.~\ref{fig: PDE} (right), LCM groups 4--6, 10--12, 16--18, and 22--24) of the TPC show a systematically lower PDE with respect to tiles placed in the middle of the TPC. This can be explained by an anisotropy of light collection of LCM with respect to the angle of incoming photons, driven by structural non-uniformity of fibers and spaces. The absence of non-uniform effects in the ArCLight tiles due to reflections on the TPC structure or Rayleigh scattering, meanwhile, further indicates that these effects are negligible within the experimental uncertainties.
In \mbox{Module-0}, a Hamamatsu MPPC S13360-6025 \cite{MPPC} is used. By replacing the SiPM for future modules with the MPPC S13360-6050 with higher efficiency, the overall PDE would improve by a factor of 1.6 to yield a LCM efficiency of about 1\%.

\section{Measurements with Cosmic Ray Data Samples}
\label{sec:cosmic-analysis}

The following sections discuss the analyses performed using reconstructed tracks from the large cosmic ray data set collected during the \mbox{Module-0} run. As discussed in
Section~\ref{sec:overview}, the \mbox{Module-0} detector incorporates several novel
technologies for the first time in a LArTPC of this scale. These studies assess the
performance of the fully-integrated system, including the LArPix charge readout
with a very large channel count, the high-coverage hybrid LCM and ArCLight photon
detection systems, and their matching; the capability to achieve
the necessary levels of LAr purity for physics measurements without prior evacuation of the
cryostat; and the degree of drift field uniformity achievable with the low-profile
resistive shell field cage. Detailed studies of each of these key detector
parameters demonstrate excellent performance of the integrated system relative to
the requirements in view of the operation for the DUNE ND-LAr.

In support of these studies, a sample of cosmic rays has been simulated using CORSIKA \cite{Heck:1998vt}, a program for detailed simulation of extended air showers. The passage of the particles through matter has been simulated using a Geant4-based Monte Carlo~\cite{AGOSTINELLI2003250}. The detector simulation has been performed with \texttt{larnd-sim} \cite{larndsim,larndsim2}, a set of highly-parallelized GPU algorithms for the simulation of pixelated LArTPCs.
A track-fitting algorithm is applied to provide an estimate of the particle track angle and location. First, a 3D point cloud is reconstructed using the unique 
channel index to determine the position transverse to the anode and the drift
time. DBSCAN ($k=5$, $\epsilon=2.5$~cm) \cite{dbscan} is used to find the hit
clusters. The cluster radius ($\epsilon$) was tuned using the $k=5$th-neighbor distance of 3D points from a typical run. Each cluster is then passed through a RANSAC line fit \cite{Fischler:1981} with an outlier radius of $\rho=8$~mm and 100 random samples. This provides a set of highly-collinear points which constitute the reconstructed track.

\subsection{Electron lifetime}
\label{sec:electron-lifetime}

The amount of charge collected by the readout system depends heavily on the electron lifetime, $\tau$, in the argon of the TPC volume. The electron lifetime parameterizes (in units of time) how much charge is lost due to attachment to electronegative impurities in the argon, such as oxygen or water, during the drift of the deposited ionization charge toward the anode. The charge measured at the anode, $Q$, is given by
\begin{equation}
\label{eq:lifetime}
Q = e^{-t/{\tau}} \cdot R \cdot Q_0,
\end{equation}
where $Q_0$ is the  amount of the primary ionization charge deposited by a particle in the liquid argon, $R$ is the recombination factor that describes the fraction of charge that survives prompt recombination of the ionization with argon ions prior to drift, and $t$ is the drift time from the point of original charge deposition to detection in the anode plane.
Measuring signals originating across the entire TPC via the charge readout system requires a sufficient electron lifetime in the detector. For the DUNE ND-LAr detector this requirement is $>\SI{0.5}{ms}$ at a drift electric field of $\SI{500}{V/cm}$; this relatively low value compared to other large LArTPC detectors~\cite{DUNE:2020cqd,Adams_2020,BETTINI1991177} is due to the relatively short maximum drift length of DUNE ND-LAr ($\sim50$~cm) and allows ND-LAr to meet the charge attenuation performance of the far detector, which specifies a 3~ms lifetime in a detector with a 3.5~m drift length at a 500~V/cm drift field \cite{DUNE:2020txw}. 
A measurement of the electron lifetime with \mbox{Module-0} has been carried out to confirm that the materials used in the detector, which will be similar to those of DUNE ND-LAr, are compatible with the argon purity requirement. Additionally, tracking this parameter as a function of time is necessary to provide a calibration of charge scale for other measurements carried out using the \mbox{Module-0} charge data.

\begin{figure}
 \includegraphics[width=0.99\textwidth]{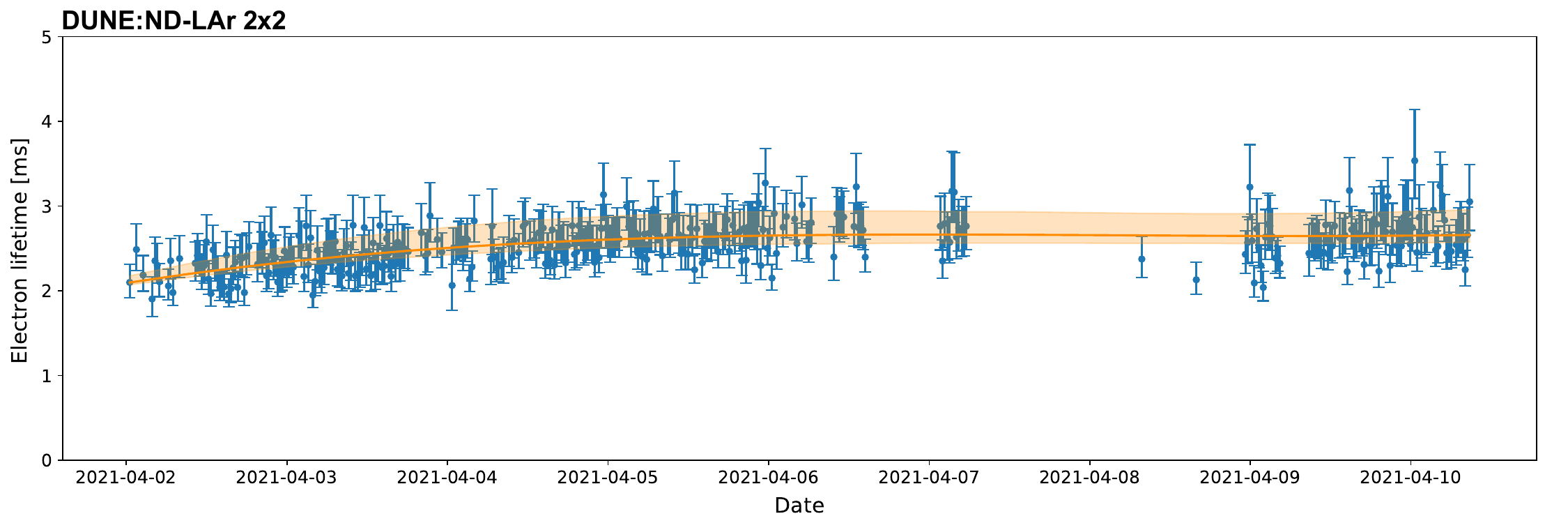}
 \includegraphics[width=0.99\textwidth]{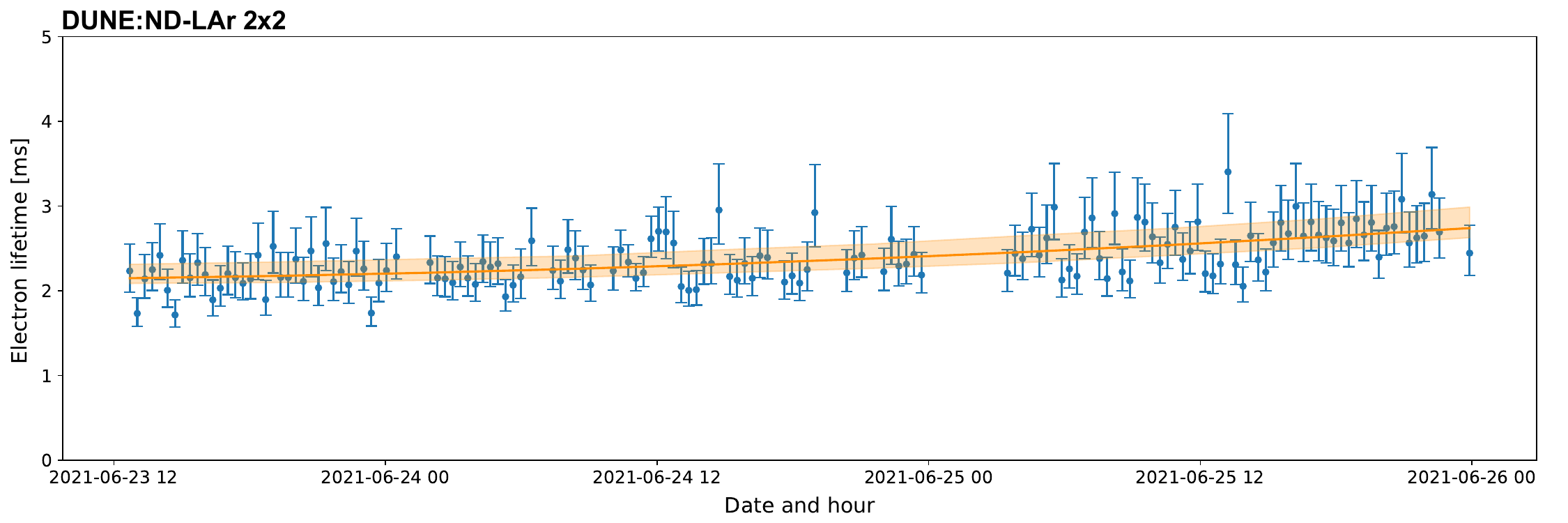}
 \caption{Extracted electron lifetime as a function of time during Module-0 Run 1 (top) and Run 2 (bottom), with the average uniformly exceeding 2~ms in both cases.}
 \label{fig:electron-lifetime}
\end{figure}

As seen in Eq.~\ref{eq:lifetime}, charge measurements at the anode depend both on the electron lifetime and the recombination factor.  However, by measuring $Q$ as a function of the drift time for a collection of cosmic muon tracks that span the entire drift distance, the dependence on $R$, which is independent of drift time, can be ignored as an overall normalization factor. Additionally, a more fitting quantity to use in this study is $dQ/dx$, the measured charge per unit length along the cosmic muon track, given the dependence of the amount of charge seen by a single pixel channel on the orientation of each track.
The electron lifetime for each \mbox{Module-0} data run at a drift electric field of $\SI{500}{V/cm}$ is measured by applying an exponential fit to the mean $dQ/dx$ of muon track segments as a function of drift time to the anode, assuming a uniform $dQ/dx$.  A sample of anode-cathode-crossing tracks
is used for this measurement; these tracks span the entire drift distance and the absolute drift time associated with each part of the track is known for this track sample.
The electron lifetime values measured in \mbox{Module-0} were consistently above 2~ms for the duration of the run, thus satisfying the $\tau>0.5$~ms requirement. This trend continued in the second run (Run 2) of \mbox{Module-0}, where cryogenic operations differed from those in Run 1. Run 1 achieved LAr purity through cryostat evacuation before cooldown and LAr filling, while Run 2 made use of a piston purge procedure (repeatedly purging the volume with clean gas), as this is the anticipated approach for the full-scale cryostat of ND-LAr. A recirculation system with filtration was operational during both runs. Results are shown in Fig.~\ref{fig:electron-lifetime}.

\begin{figure}
 \includegraphics[width=0.49\textwidth]{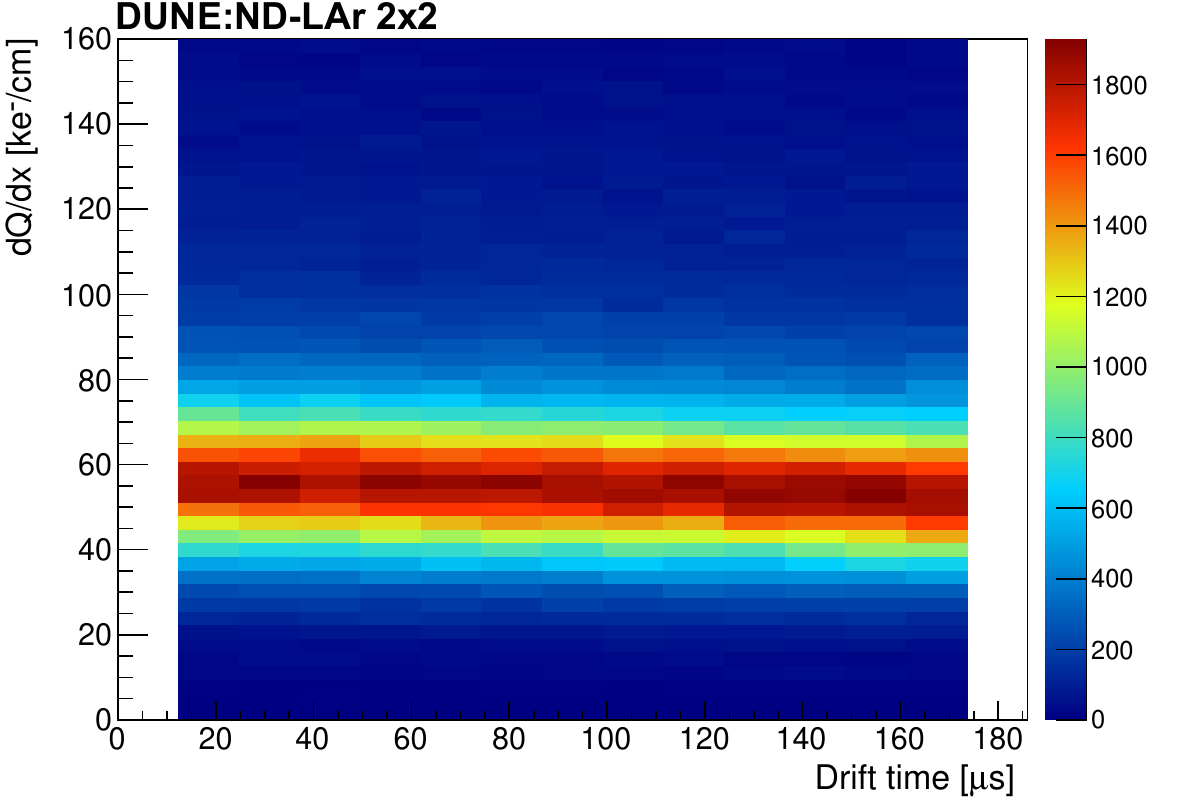}
 \includegraphics[width=0.49\textwidth]{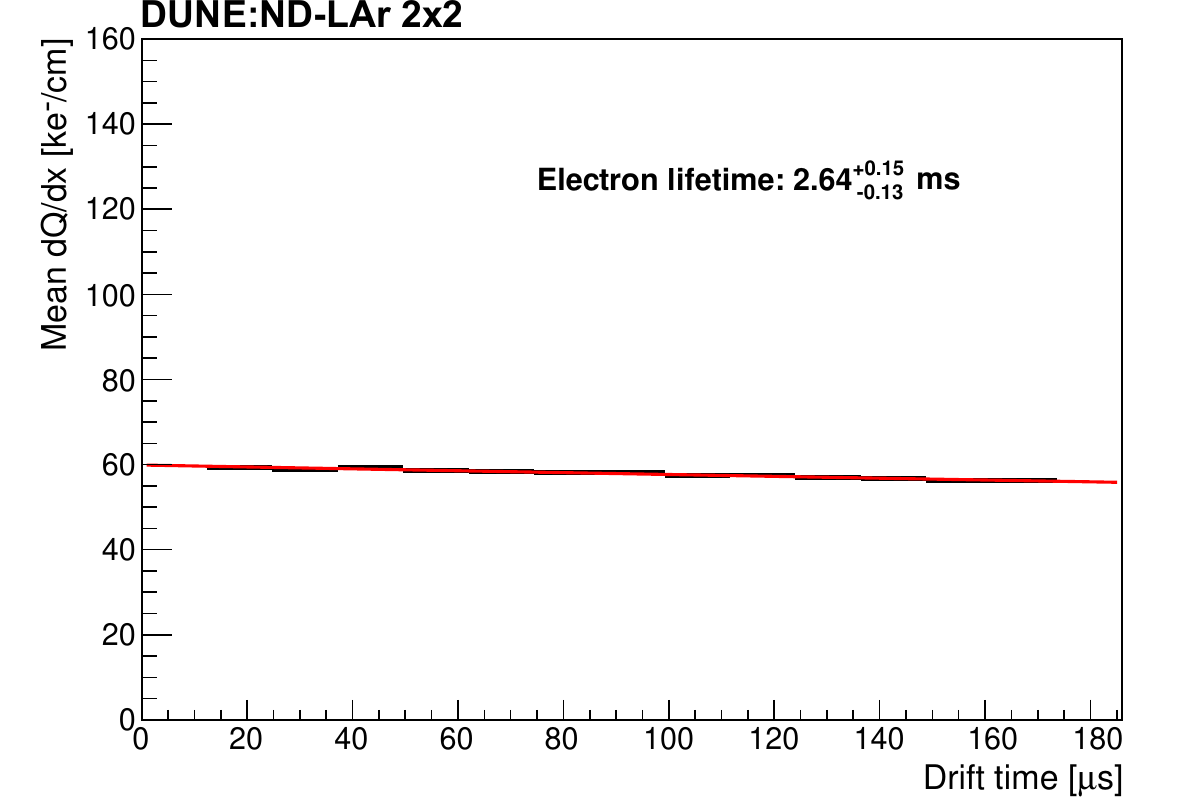}
 \caption{Measured $dQ/dx$ versus drift time for ionization associated with anode-cathode-crossing muon tracks (left); mean $dQ/dx$ versus drift time, along with exponential fit, for the same track sample (right).}
 \label{fig:electron-lifetime-dqdx}
\end{figure}

\subsection{Electric field uniformity}
\label{sec:efield}

The magnitude of electric field distortions due to space charge effects for \mbox{Module-0} are expected to be much smaller than other, larger LArTPC detectors running near the surface, such as MicroBooNE~\cite{Abratenko_2020} and ProtoDUNE-SP~\cite{Abi_2020}. This is due to the relatively small maximum drift length of \SI{\sim30}{cm} of \mbox{Module-0}, compared to \SI{\sim2.5}{m} for MicroBooNE and \SI{\sim3.6}{m} for ProtoDUNE-SP. Even for a maximum drift length of \SI{\sim50}{cm} that is anticipated for DUNE ND-LAr, the impact from space charge effects is expected to be negligible; the fact that ND-LAr will operate \SI{65}{m} underground will reduce this effect further due to the smaller flux of cosmic muons. 
However, it is possible that electric field inhomogeneities arise in the \mbox{Module-0} detector from other sources.  In particular, it is important to determine whether or not the field cage design causes significant distortions of the electric field, which can alter the trajectories ionization electrons take while drifting to the anode plane. Such distortions could lead to incorrect reconstruction of the true position of original energy depositions in the detector due to primary particles ionizing the argon, consequently impacting their trajectory and energy reconstruction.  Furthermore, associated modification to the electric field intensity throughout the detector can lead to significant impact on the amount of electron-ion recombination experienced by ionization electrons, leading to bias in reconstructed particle energy scale or degradation of reconstructed particle energy resolution. The use of the novel resistive field cage technology in \mbox{Module-0}, as is anticipated for DUNE ND-LAr, provides an important opportunity to study the impact on electric field homogeneity.

\begin{figure}
 \includegraphics[width=0.49\textwidth]{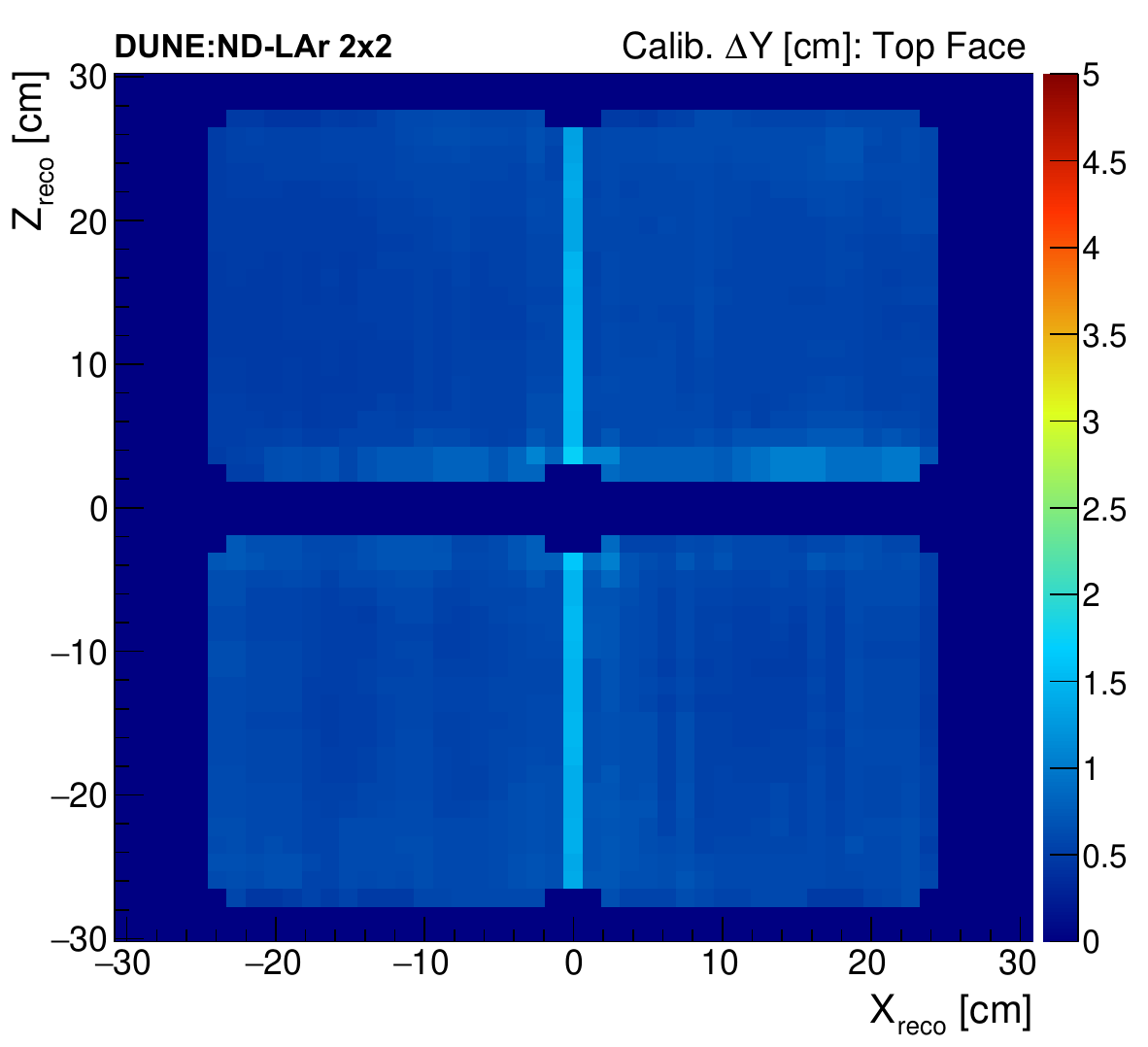}
 \includegraphics[width=0.49\textwidth]{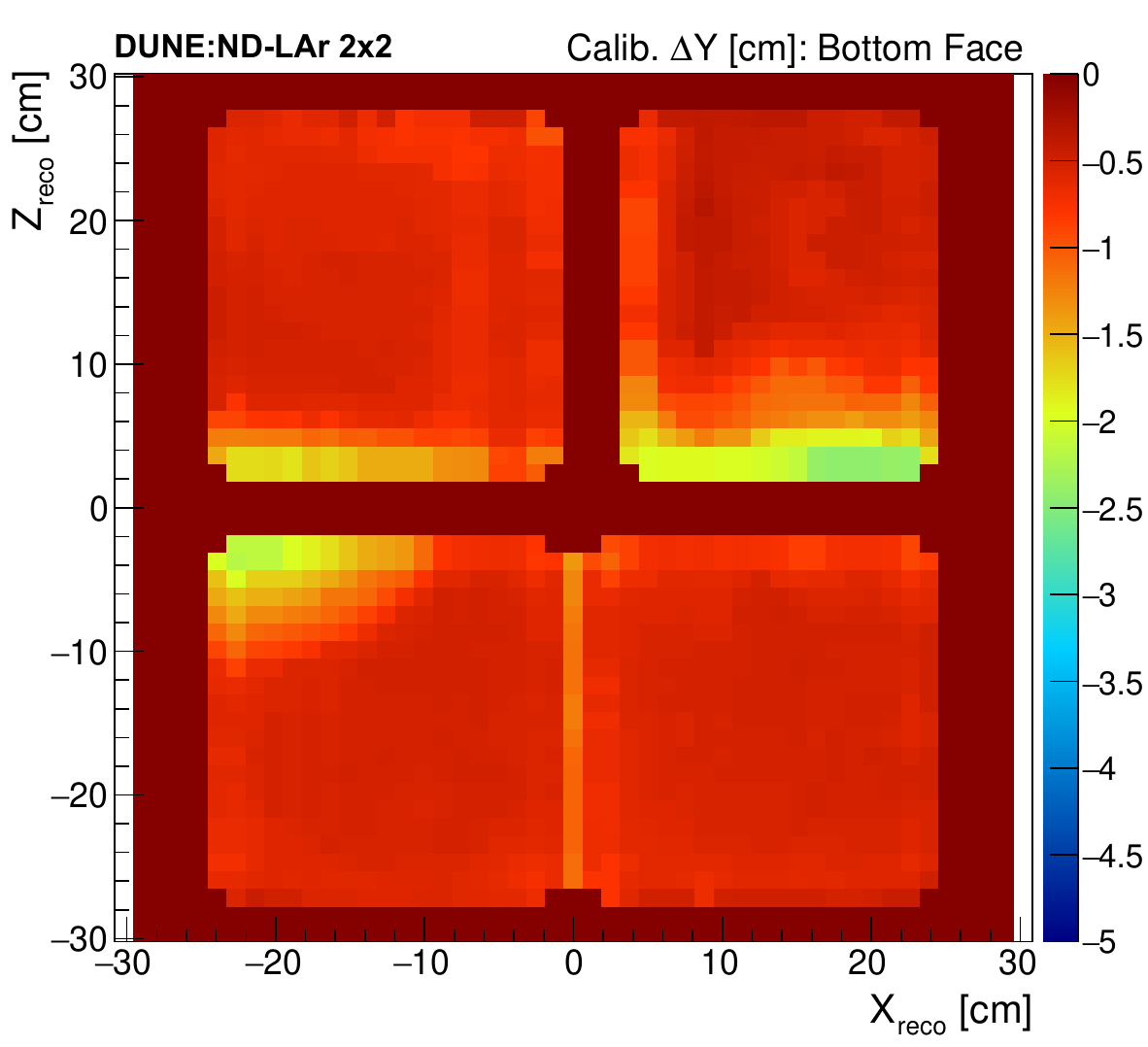}
 \caption{Average spatial offsets measured at the top (left) and bottom (right) of the Module-0 detector. These offsets in cm are measured with respect to the location of the pixel channels at the edge of the detector. }
 \label{fig:efield-map-tb}
\end{figure}

\begin{figure}
 \includegraphics[width=0.49\textwidth]{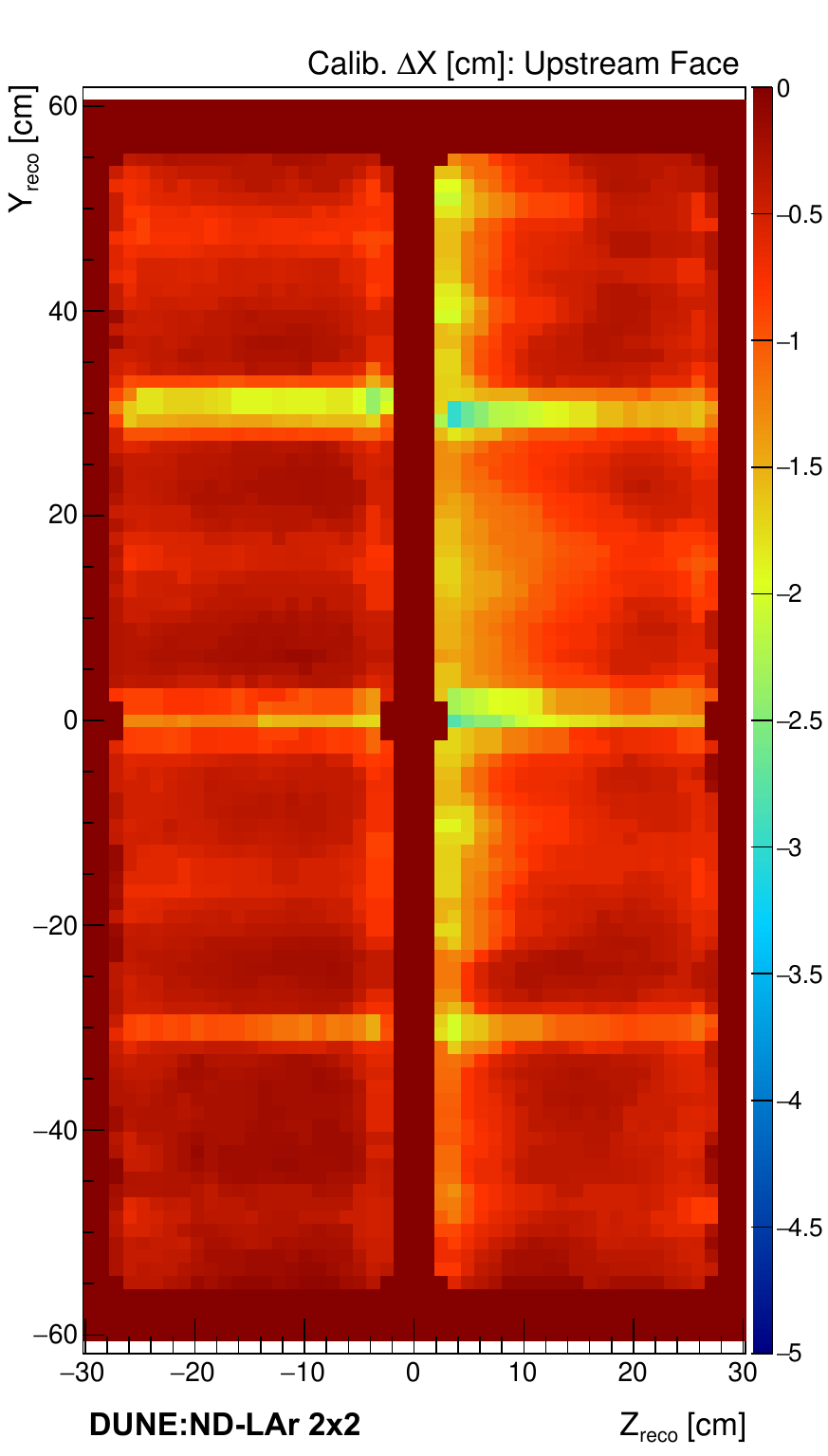}
 \includegraphics[width=0.49\textwidth]{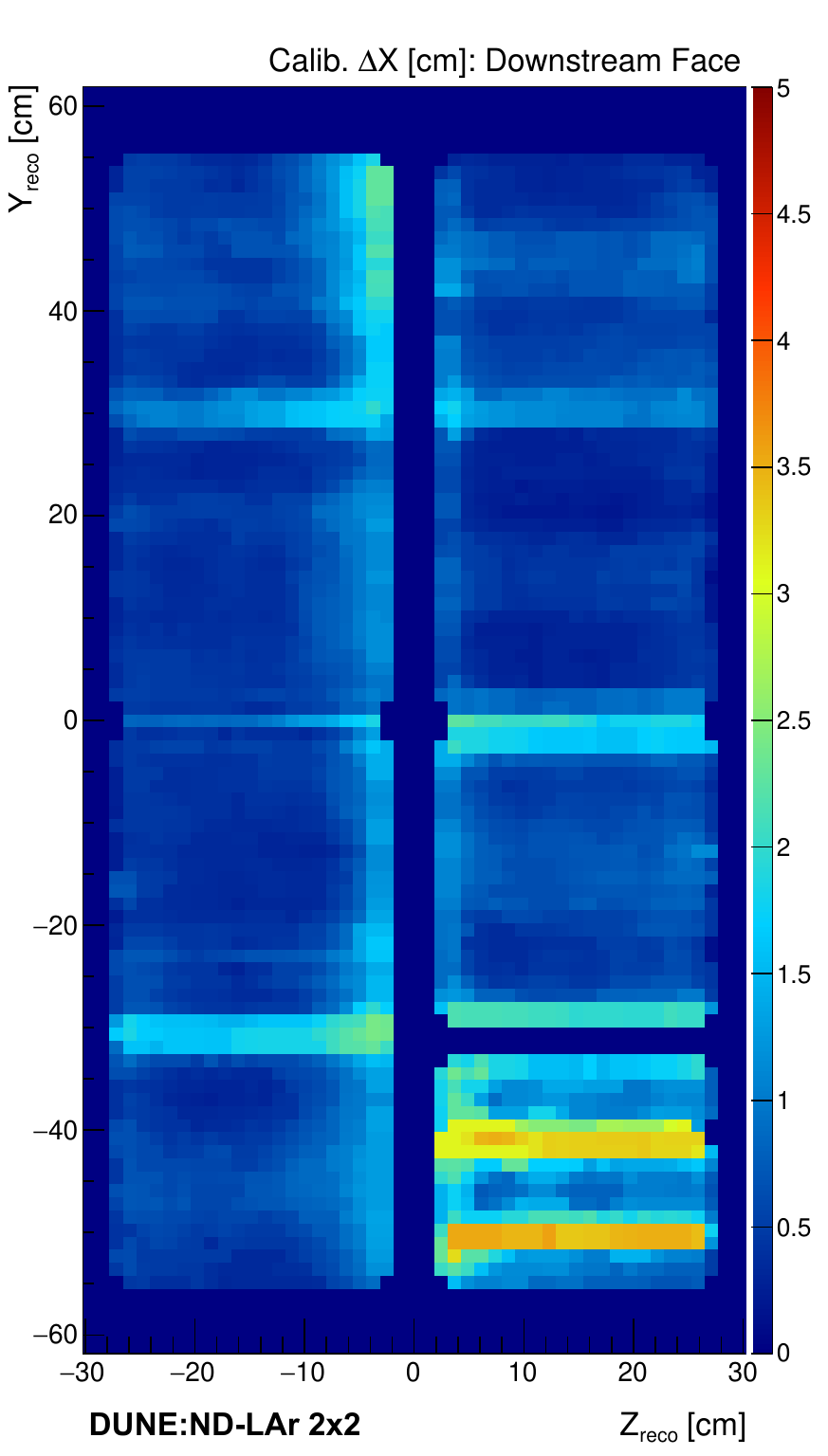}
 \caption{Average spatial offsets measured at the front (left) and back (right) of the Module-0 detector. These offsets in cm are measured with respect to the location of the pixel channels at the edge of the detector.}
 \label{fig:efield-map-sides}
\end{figure}

Following the methodology developed by the MicroBooNE experiment for analysis of space charge effects~\cite{Abratenko_2020}, electric field distortions are probed using end points of through-going cosmic muon tracks in \mbox{Module-0} data. Tracks passing through an anode plane and another face of the detector that is not the other anode plane are selected for this study, providing a known absolute drift time associated with each part of the track via subtracting the time associated with the anode side of the track. The track end point associated with the non-anode side of the anode-crossing track is then probed by measuring the transverse (i.e., perpendicular to the drift direction) displacement from the edge of the TPC active volume, as measured from the $y$ value (TPC top and bottom) or $x$ value (TPC front and back sides, perpendicular to the drift direction) of the pixel channels at the edge of the detector. The average transverse displacement is recorded as a function of the two directions within the TPC face for all four non-anode faces of the \mbox{Module-0} TPC.  If there are no electric field distortions in the detector, there would be no inward migration of ionization electrons during drift, leading to zero transverse displacement of ionization charge with respect to the TPC face for this sample of through-going muon tracks (contamination from stopping muons is expected to be less than 1\%).
The result of the average transverse displacement measurement is shown for the TPC top and bottom in Fig.~\ref{fig:efield-map-tb} and for the TPC front and back in Fig.~\ref{fig:efield-map-sides}.  A few features not associated with electric field distortions in the detector should be pointed out. First, there are gaps in coverage near the anode planes ($z$ values of roughly $\pm \SI{30}{cm}$) due to a requirement in the track selection that the non-anode side of the track is at least \SI{5}{cm} away from both anode planes, and near the pixel plane edges (edges of the TPC face) due to a requirement that the non-anode side of the track is not located within \SI{1}{cm} (\SI{2}{cm}) of these features. These selection criteria were introduced to minimize contamination of the sample from poorly-reconstructed muon tracks.  
Some residual contamination is seen near the edges of the pixel planes, where the measured average transverse spatial offset is artificially large due to edge channels of the pixel planes being turned off for data-taking, leading to the ends of tracks being clipped off near the edges of pixel planes.  Second, the two horizontal bands in the bottom right corner of the right side of Fig.~\ref{fig:efield-map-sides} are associated with a known grounding issue of an ArCLight unit in this part of the detector. The vertical gap in the right panel of Fig.~\ref{fig:efield-map-tb} is due to inactive channels in this region of the anode plane (see Fig.~\ref{fig:larpix-active-channel-map}).

\begin{figure}
 \includegraphics[width=0.99\textwidth]{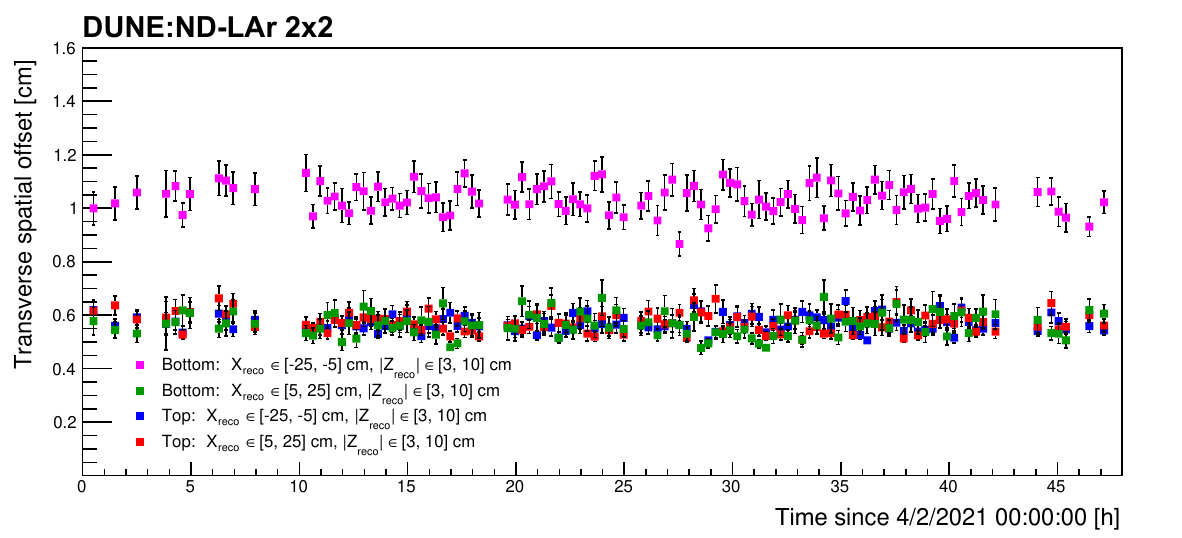} \\
 \includegraphics[width=0.99\textwidth]{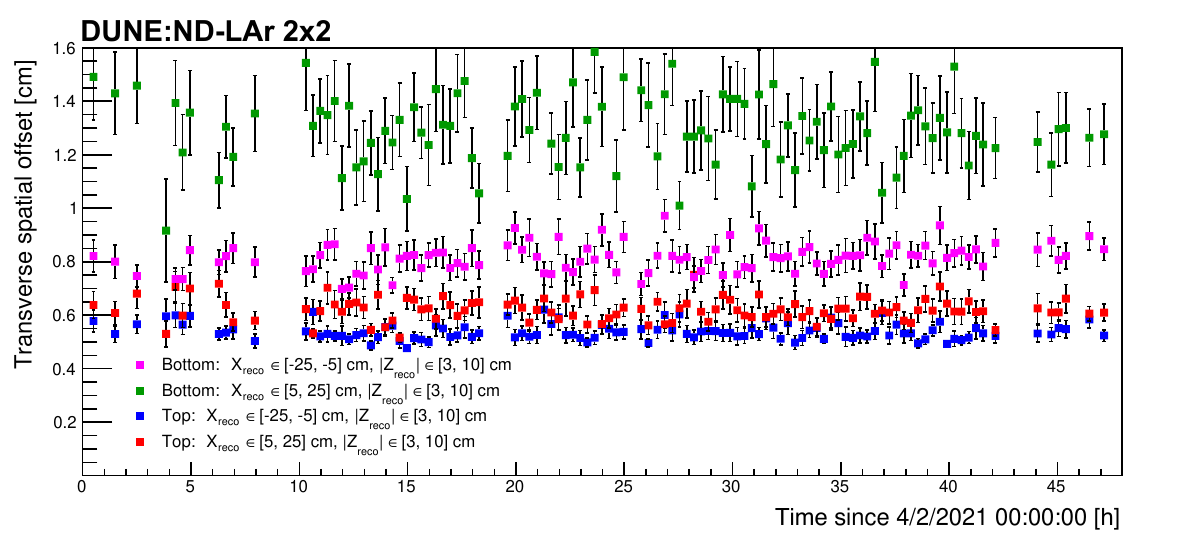}
 \caption{Time dependence of spatial offsets in the $-z$ (top) and $+z$ (bottom) drift volumes.  These offsets are measured with respect to the location of the pixel channels at the edge of the detector.}
 \label{fig:efield-timedep}
\end{figure}

After accounting for these two artifacts, non-negligible transverse spatial offsets are observed near the cathode (central horizontal lines in Fig.~\ref{fig:efield-map-tb}, central vertical lines in Fig.~\ref{fig:efield-map-sides}), roughly \SI{1}{cm} on average but as large as \SI{2.5}{cm} in some places in the TPC.  After adding an additional \SI{\sim1}{cm} to these measurements to account for the separation between the edge pixel channels and the field cage (or light detectors in the case of the front and back of the TPC), the average (maximum) transverse spatial offset experienced by drifting ionization charge originating near the cathode is roughly \SI{2}{cm} (\SI{3.5}{cm}). Ascribing this transverse drift to an additional electric field component strictly in the direction transverse to the TPC faces, the average (maximum) transverse electric field magnitude leading to this amount of inward drift of ionization charge is roughly \SI{30}{V/cm} (\SI{60}{V/cm}).  The associated average (maximum) impact to the electric field magnitude in the detector is 0.2\% (0.7\%).  This is below the conservative physics requirement of $1$\% maximum allowed deviation of the electric field magnitude within $95$\% of the detector volume, indicating that the design of the field cage is adequate for the physics goals of DUNE ND-LAr. It is worth pointing out that this physics requirement for electric field distortions corresponds to after detector calibrations have been carried out, while the measurements presented here have no calibration applied. It is thus expected that the calibrated electric field map would be even more homogeneous at DUNE ND-LAr.
An additional study is carried out to determine if the small electric field distortions in the \mbox{Module-0} detector vary substantially over time.  A substantial time dependence of the electric field distortions may complicate efforts to obtain a calibrated electric field map in the DUNE ND-LAr detector using cosmic muons, neutrino-induced muons, or dedicated calibration hardware.  Average transverse spatial offsets were measured at four different places on each side of the \mbox{Module-0} cathode as a function of time, spanning two full days of data-taking.  The results of the study are shown in Fig.~\ref{fig:efield-timedep}.  No substantial time dependence of transverse spatial offsets is observed ($< \SI{0.2}{cm}$), indicating that calibration of the underlying electric field distortions is achievable by averaging measured spatial offsets over at least a few days of data-taking. A study of electric field stability over longer periods of time is planned in future prototyping of the DUNE ND-LAr detector concept.

\subsection{Charge-light matching}
\label{sec:qlmatch}
\begin{figure}
 \includegraphics[width=0.49\textwidth]{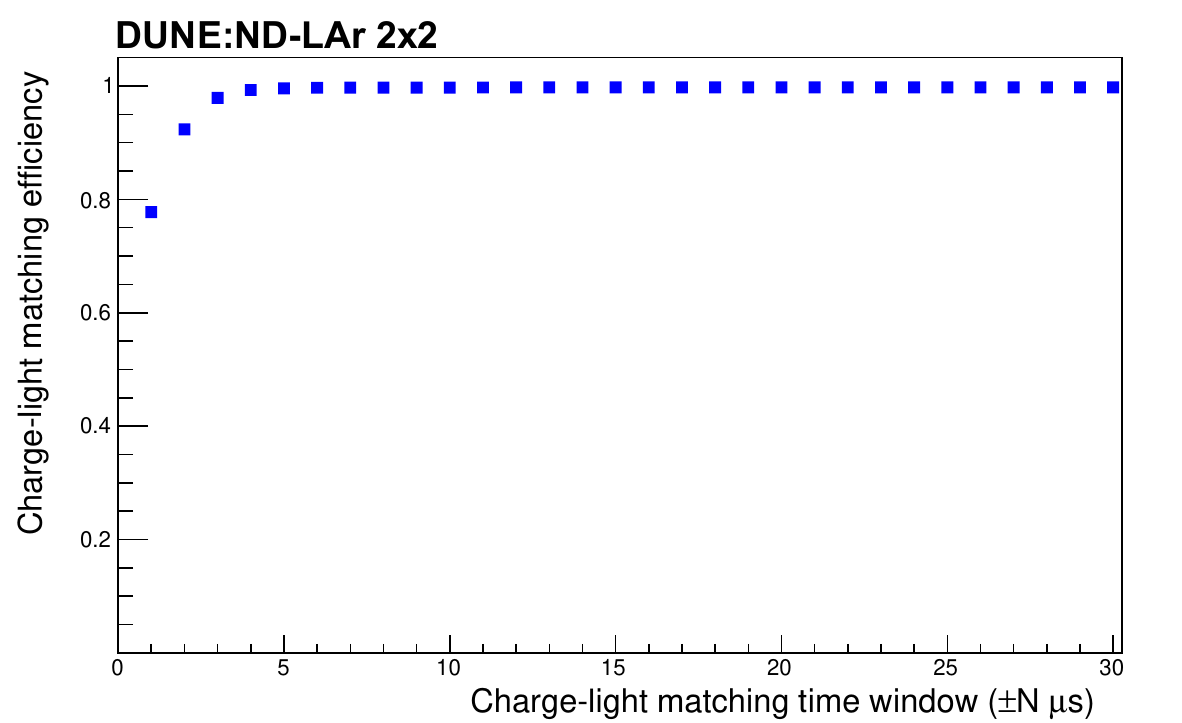}
 \includegraphics[width=0.49\textwidth]{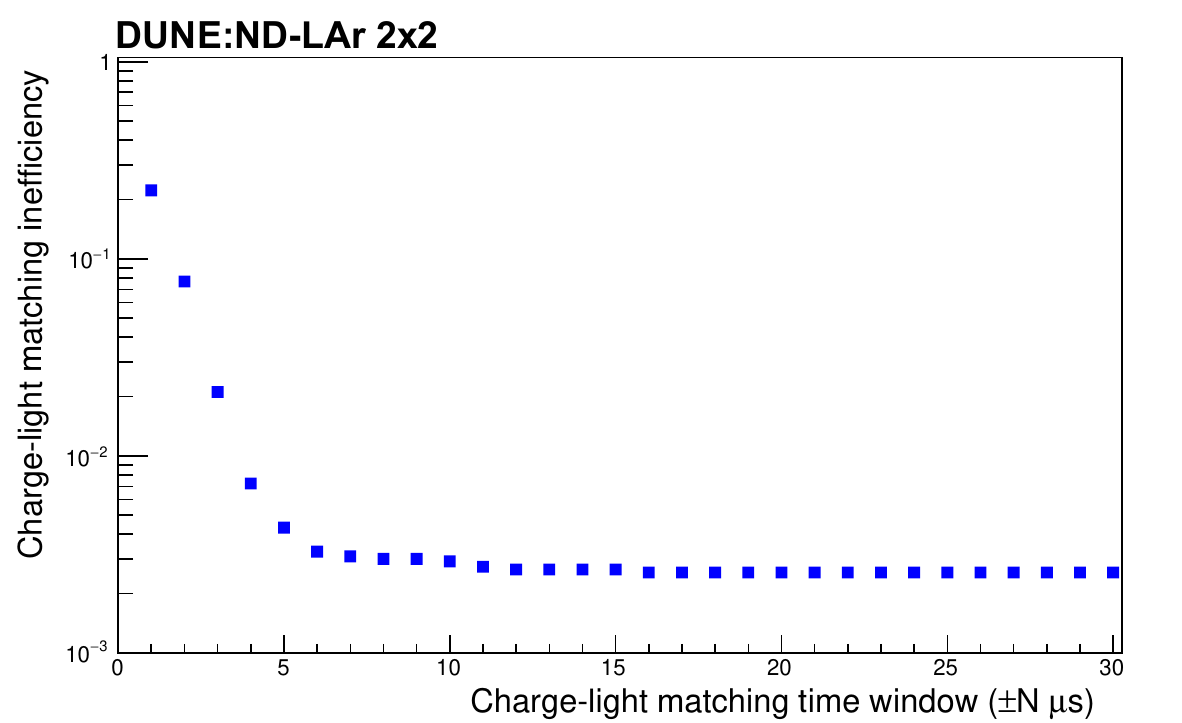}
 \caption{Charge-light matching efficiency in linear scale (left) and inefficiency in logarithmic scale (right) for light detector triggers matched to the arrival time of charge at the anode side of anode-cathode-crossing tracks.}
 \label{fig:ql-match-eff}
\end{figure}

\begin{figure}
 \includegraphics[width=0.5\textwidth]{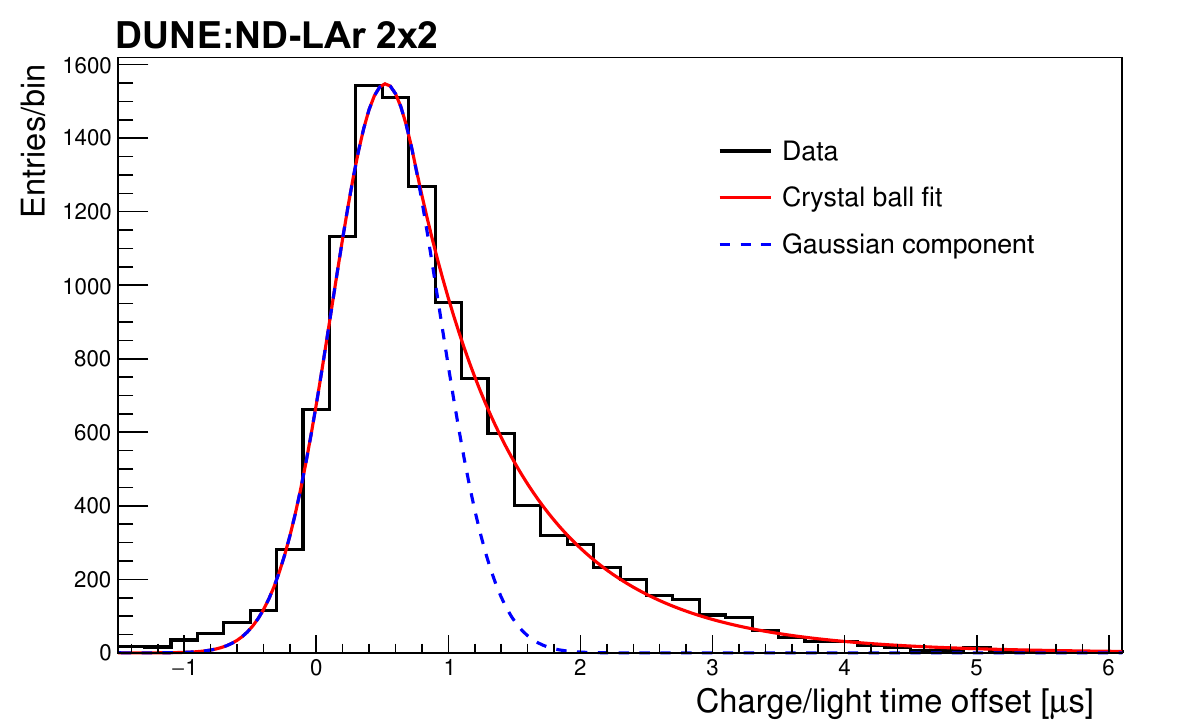}
 \caption{Time offset distribution for light detector triggers matched to the arrival time of charge at the anode side of anode-cathode-crossing tracks (charge minus light).}
 \label{fig:ql-timing-res}
\end{figure}

Efficient matching between signals in the charge and light readout systems is essential, as this enables the use of light to disambiguate pile-up of separate neutrino interactions within a single beam spill. The unique association between charge and light signals is a nontrivial problem in a large-volume LArTPC, especially in an environment with a high rate of neutrino event pile-up, such as DUNE ND-LAr. This motivates the modular design, where the full active volume is composed of an array of optically-isolated TPC volumes, each with high coverage of optical detectors with fast timing and good spatial resolution.
Charge-light matching in \mbox{Module-0} has been accomplished via association of precision GPS-synchronized timestamps in the two systems. 
Here, two performance metrics are considered: the efficiency of matching for a selection of tracks as a function of
the allowed coincidence time window and the resolution in terms of the offset between
the two systems' timestamps. Fig.~\ref{fig:ql-match-eff} shows the matching efficiency for varying definitions of
the allowed time window for coincidence formation, for a selection of
anode-cathode-crossing muon tracks. The overwhelming majority of these are single tracks, as the probability of having another event in the same $\sim$200~$\mu s$ window is very small. For conservative matching parameters,
an efficiency of $\geq99.7$\% is found. In this study, the timing resolution is
limited by the spatial resolution of the tracking from the charge readout,
not by the intrinsic light detector timing resolution, which is discussed in
Section~\ref{sec:light}.
Next, Fig.~\ref{fig:ql-timing-res} illustrates the relative time offset between the
two systems for the \mbox{Module-0} prototype, again for a selection of anode-cathode-crossing
tracks. The distribution exhibits a Gaussian core and a tail.
The asymmetric tail of the distribution, captured by a Crystal Ball fit~\cite{crystalballfunc}, is due to
track truncation near the boundaries of the pixel planes. The Gaussian component of
the Crystal Ball fit is also shown; the standard deviation of the Gaussian, \SI{0.4}~$\mu$s, is identified as the charge readout timing resolution.
The physics requirements for ND-LAr require that the resolution in the drift dimension
be at least as precise as that across the anode plane, i.e. the pixel pitch divided
by $\sqrt{12}$, or \SI{1.3}{mm}.
The resolution extracted in \mbox{Module-0} corresponds to \SI{0.6}{mm} at a drift electric
field of \SI{500}{V/cm}, thus meeting the requirement.


\subsection{Correlation of the charge and light yield}
\label{sec:qlcorr}

\begin{figure}
    \includegraphics[width=0.95\textwidth]{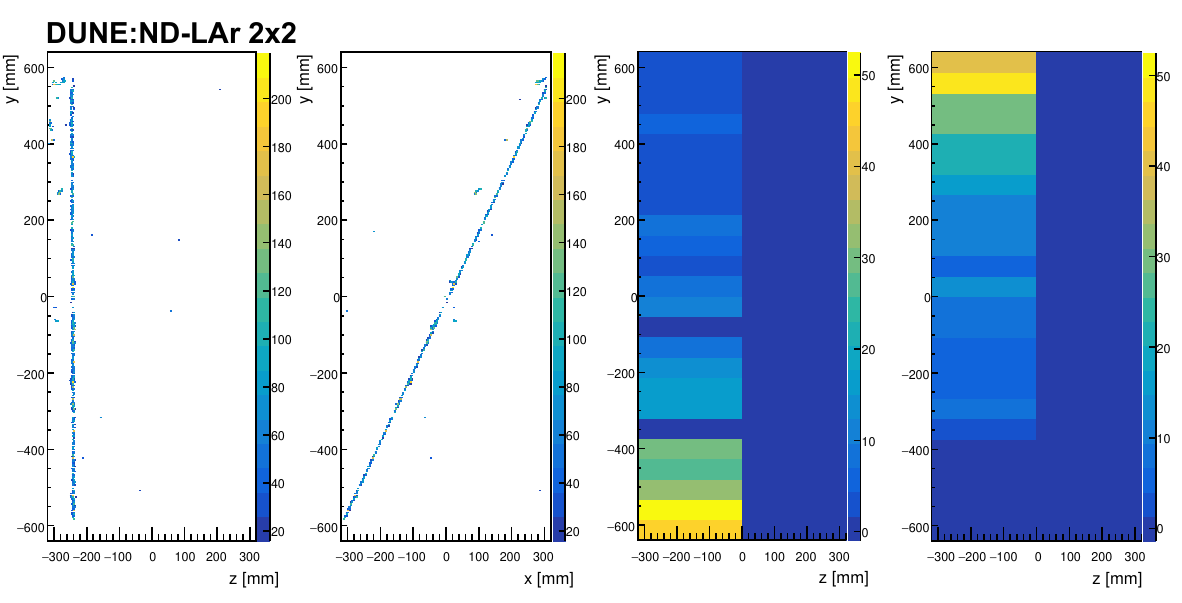}
    \caption{Charge-light matched event display of a cosmic muon track. The left two panels show the TPC charge readout, in a $z-y$ project (left) and $x-y$ projection (center left). The right two panels show the light detector responses for the arrays at $-x$ (center right) and $+x$ (right), with each bin along the vertical axis representing the strength of signal read by individual SiPMs.}
    \label{fig:ql_matched_display}
\end{figure}

Matched charge and light events as shown in Figure~\ref{fig:ql_matched_display} provide another data sample which may be used to study the correlation in the relative charge and light yields in the detector. These yields are related to electric-field dependent recombination effects. 

To describe the recombination mechanism in LAr we formalize the ionization and excitation states generated by the deposited energy of a traversing particle as follows:
\begin{equation}
    \label{eq:QL_sum}
    N_i + N_{ex} = QY + LY,
\end{equation}
where the sum of available ionization ($N_i$) and excitation ($N_{ex}$) states determines the total number of electrons ($QY$) and photons ($LY$) generated in LAr. 
The number of ionization states $N_i$ is given by
\begin{equation}
    N_i = \frac{E_{dep}}{W_i} \, , \,  W_i = 23.6\, \mathrm{eV},
\end{equation}
where $W_i$ is the ionization work function \cite{Shibamura:1975zz} and $E_{dep}$ is the deposited energy.
In the absence of charge attenuation and impurities, the total charge $Q$ arriving at the anode depends only on the initially-produced ionization charge $Q_0=N_ie$ as
\begin{align}
 \label{eq:LY_eq0}
    QY &= N_i \cdot R_c,  \\
    LY &= N_i \left( 1 + \frac{ N_{ex} }{N_i} - R_c \right),
\end{align}
where the charge recombination factor $R_c$ is dependent on the electric field $\epsilon$, and $e$ is the electron charge. In the presence of impurities, the electron lifetime correction is applied first; see Eq.~\ref{eq:lifetime}.
Increasing $\epsilon$ leads to less recombination between argon ions and ionization electrons, and thus more free charge carriers are present in the TPC drift field, increasing the total detected charge at the anode plane.
At the same time, a reduced charge recombination factor corresponds to less scintillation light produced within the TPC, leading to a decrease of the light yield at higher electric fields, as expressed by Eq.~\ref{eq:LY_eq0}.
Hence, the amount of charge yield and the amount of light yield observed in the detector are expected to be anti-correlated. To describe the recombination of electron-ion pairs, we focus on the most commonly used models, namely the Box \cite{Thomas:1987zz} and the Birks' models \cite{Birks:1951boa}, and compare the results of \mbox{Module-0} measurements with those of the ICARUS \cite{Amoruso:2004dy} and ArgoNeuT \cite{ArgoNeut_2013} experiments.
The Box model assumes zero electron diffusion, zero ion mobility, and a distribution of ionization electrons that are uniformly produced within a 3D box along the path of the ionizing particle.
The collected charge $Q$ is given by
\begin{equation}
Q = Q_{0} \cdot \dfrac{A_{\rm{Box}}}{\xi} \cdot \ln \left( \xi \right),
\label{eq:box_model}
\end{equation}
where $Q_{0}$ denotes the primary ionization charge and $\xi$ is
\begin{equation}
\xi = \dfrac{N_{0} K_{r}}{4 a^2 \mu \epsilon},
\end{equation}
where $a$ is the linear size of the charge `box', $N_{0}$ denotes the number of electrons in the box and $K_r$ is the recombination rate constant.
$\mu$ and $\epsilon$ define the electron mobility and the electric field, respectively.
Note that in the limit of an infinite electric field intensity $\epsilon$, the collected charge at the anode plane corresponds to the initially produced charge, $Q_{0}$. Birks' model describes the collected charge $QY$ as
\begin{equation}
\label{eq:Birk_model}
QY = N_{i} \cdot \frac{A_{\rm{Birks}}}{1 + \frac{k_{B}}{\epsilon} \cdot \frac{dE}{dx}} = \frac{Q_{0}}{e}R_{c},
\end{equation}
where $A_{\rm{Birks}}$ and $k_{B}$ are fitting constants. In this formulation of the Birks' model, for infinite electric field intensities $\epsilon \rightarrow \infty$, the recombination factor does not go to $1$ and is limited to $R_c \rightarrow A$. 
We can now express the light yield as
\begin{equation}
 \label{eq:LY_eq1} 
    LY = N_i \left( 1 + \frac{N_{ex}}{N_i} - \frac{A_{\rm{Birks}}}{1 + \frac{k_{B}}{\epsilon} \cdot \frac{dE}{dx}} \right).
\end{equation}
However, since the fraction of excited states $\frac{N_{ex}}{N_i}$ is not precisely known, the commonly used model for description of the light yield in scintillating materials uses the following formulation: 
\begin{align}
   LY &= L_0 \left( 1- \alpha R_{c} \right) = L_0 R_{L}, \label{eq:LY_eq2} \\
   L_0 &= \frac{E_{dep}}{W_L} \, , W_L = 19.5 \, \rm{eV},
\end{align}
where $L_0$ denotes the number of scintillation photons at zero electric field intensity, $\alpha$ is a constant fitted to the data and $W_L$ is the scintillation work function \cite{Doke_2002}. This formulation is used in this analysis to evaluate the parameters in the Birks' model for the light yield.

To study the charge and light correlation in Module-0, data samples at different electric field intensities ranging from \SI{0.05}{\kilo\volt\per\centi\metre} to \SI{1.00}{\kilo\volt\per\centi\metre} were acquired and analysed.
These events contain information about the collected charge and scintillation light.
A selection of vertical through-going tracks, as expected from MIP muons, was used to extract the collected charge and light per unit length of the track. 
For the measurement of the collected charge per unit track length, the track was divided into 2 cm segments and the total charge collected each the segment was divided by the segment length. Then, the light yield per unit track length is extracted as:
\begin{equation}
    \label{eq:LightYield}
    \frac{dL}{dx} = 
    \frac{L_{\mathrm{detected}}}{\int \Omega dl \times \mathrm{PDE} \times G}.
\end{equation}
The factors in this expression include the geometrical acceptance $\int \Omega dl$, the readout gate acceptance $G$, and the overall PDE of each tile reported in Section~\ref{sec:light}.
The geometrical acceptance was computed based on the charge data and the track segment position with respect to a light detection tile, integrated over the track length. The readout gate acceptance is an estimation of the fraction of photons which reach the SiPM within the readout integration gate of $500$~ns. The gate acceptance was measured using the average waveform of the light signals in \mbox{Module-0} data to be $\sim 64\%$ for both the LCM and ArCLight modules.

\begin{figure}
    \includegraphics[width=0.49\textwidth]{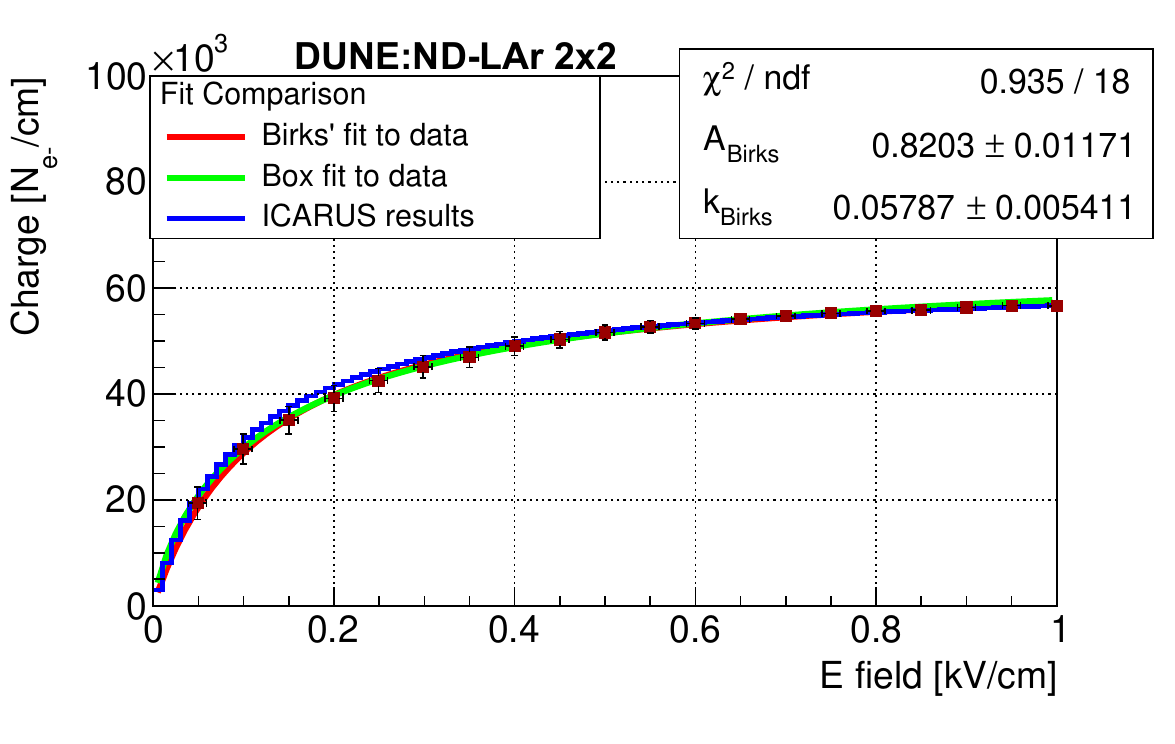}
    \includegraphics[width=0.49\textwidth]{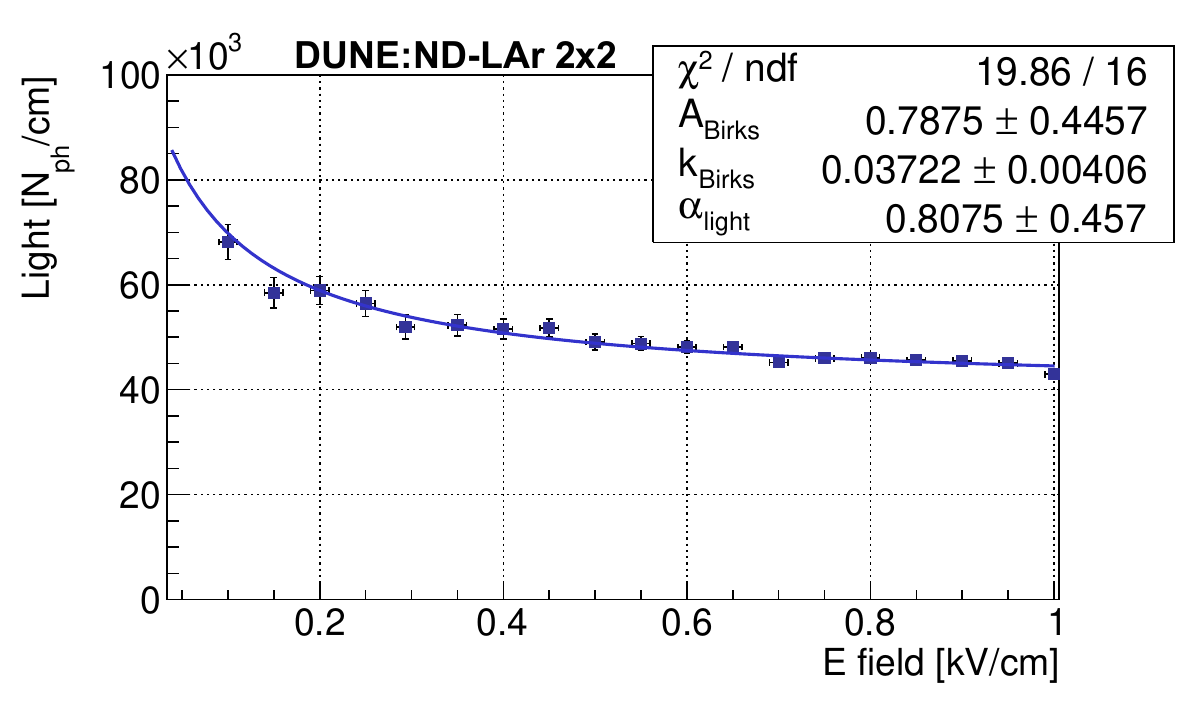}
    \caption{Charge yield as a function of the electric field intensity fitted with the Box and Birks' models, and compared to ICARUS results (left); Light yield as a function of the electric field intensity fitted separately with the Birks' model (right).}
    \label{fig:charge_light_separate_fits}
\end{figure}

The $dQ/dx$ and $dL/dx$ distributions are well-described by a Landau-convolved Gaussian function, which is used to extract the most probable value (MPV). 
We note that the fits are performed on raw data, i.e. without additional calibration of the track $dE/dx$. Due to uncorrected charge losses, the extracted MPV values for charge measurements should be compared with an effective value of $\sim1.8$~MeV/cm, while MPVs corresponding to light measurements correspond to an effective $dE/dx\sim2.1$~MeV/cm.
The dependence of the charge yield and the light yield MPV values with respect to the electric field density is illustrated in Fig.~\ref{fig:charge_light_separate_fits}.

The charge yield and light yield data points were fitted separately to the Birks' model, with results shown in Fig.~\ref{fig:charge_light_separate_fits} and Tab.~\ref{tab:fit_results}. We note that for the light yield fit (Fig.~\ref{fig:charge_light_separate_fits}, right), per Eq.~\ref{eq:LY_eq2}, the $A_\mathrm{Birks}$ and $\alpha_\mathrm{light}$ parameters are totally correlated and cannot be extracted independently.
The left panel of Fig.~\ref{fig:charge_light_separate_fits} also shows a comparison of the charge yield data (red points) to fits using a Birks' model (red curve) and Box model (green curve), alongside the results from the ICARUS experiment (blue curve), demonstrating good agreement between the results.

\begin{figure}
    \includegraphics[width = 0.75 \textwidth]{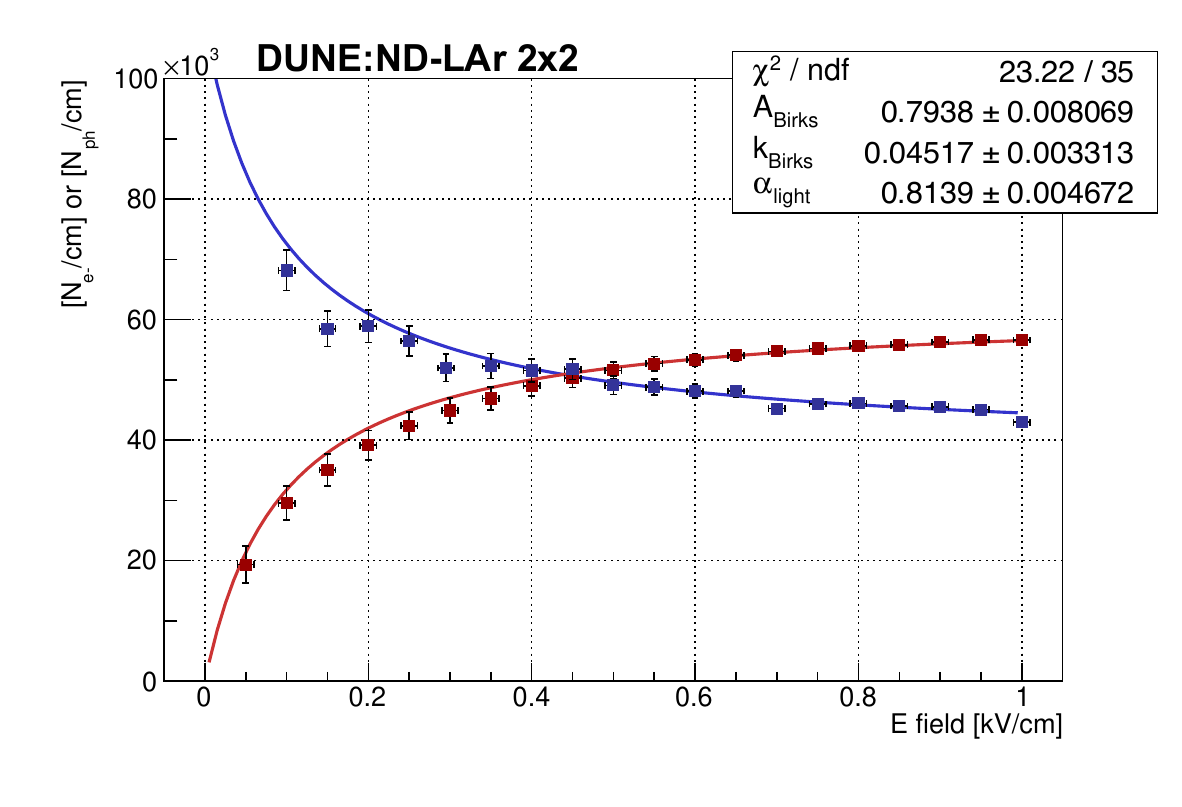}
    \caption{Light yield (blue) and charge yield (red) extracted from a simultaneous fit with the Birks' model.}
    \label{fig:combined_birk}
\end{figure}

\begin{table}
\begin{center}    
\begin{tabular}{cccc}
	\toprule
	Fit parameters & $A_{Birks} $ $[\SI{}{\kilo\volt\gram\per\cubic\centi\metre\per\mega\electronvolt}]$ & $k_{Birks   }$ $[\SI{}  {\kilo\volt\gram\per\cubic\centi\metre\per\mega\electronvolt}]$  \\
	\midrule
        Charge only fit     &  \SI{0.820 \pm 0.011}{}  & \SI{0.058 \pm 0.005}{} \\
        Light only fit  &  \SI{0.79 \pm 0.45}{}  &  \SI{0.037 \pm 0.004}{} \\
        Combined fit &  \SI{0.794 \pm 0.008}{}  &  \SI{0.045 \pm 0.003}{} \\
	\bottomrule
\end{tabular}
	\caption{The fitted parameters of the Birks' model using the Module-0 data.}
\label{tab:fit_results}
\end{center}
\end{table}
Next, a combined fit of the Birks' model to both charge and light yield data sets was performed. Fig.~\ref{fig:combined_birk} shows the final result of the correlation study. The best fit results for the Birks' model parameters are $A_{Birks}=0.794 \pm 0.008$ and $k_{Birks}=0.045 \pm 0.003$, with a $\chi^2/$ndf of $23.2/35$, where the number of degrees of freedom calculated based on 19 fit points per dataset (charge and light) included in the fit and three fit parameters.
Table~\ref{tab: ICARUS Model Parameter Comparison} summarizes the Birks' model parameters obtained with the \mbox{Module-0} detector and compares them with the parameters found in the ICARUS and the ArgoNeuT experiments.
The results of the simultaneous fit of the Birks' model to the light and charge distributions show reasonable agreement with previous experiments.

\begin{table}
\begin{tabular}{cccc}
	\toprule
	Experiment & $A_{Birks} $ $[\SI{}{\kilo\volt\gram\per\cubic\centi\metre\per\mega\electronvolt}]$ & $k_{Birks}$ $[\SI{}{\kilo\volt\gram\per\cubic\centi\metre\per\mega\electronvolt}]$ & Reference \\
	\midrule
        ICARUS     &  \SI{0.800 \pm 0.003}{}  & \SI{0.0486 \pm 0.0006}{} & \cite{Amoruso:2004dy} \\
        ArgoNeuT   &  \SI{0.806 \pm 0.010}{}  &  \SI{0.052 \pm 0.001}{} & \cite{ArgoNeut_2013} \\
        Module-0 &  \SI{0.794 \pm 0.008}{}  &  \SI{0.045 \pm 0.003}{} & This work \\
	\bottomrule
\end{tabular}
	\caption{Comparison of the ICARUS and ArgoNeuT results with the current study.}
\label{tab: ICARUS Model Parameter Comparison}
\end{table}

\subsection{Michel electrons}
\label{sec:michel}

\begin{figure}[ht!]
 \includegraphics[width=1.0\textwidth]{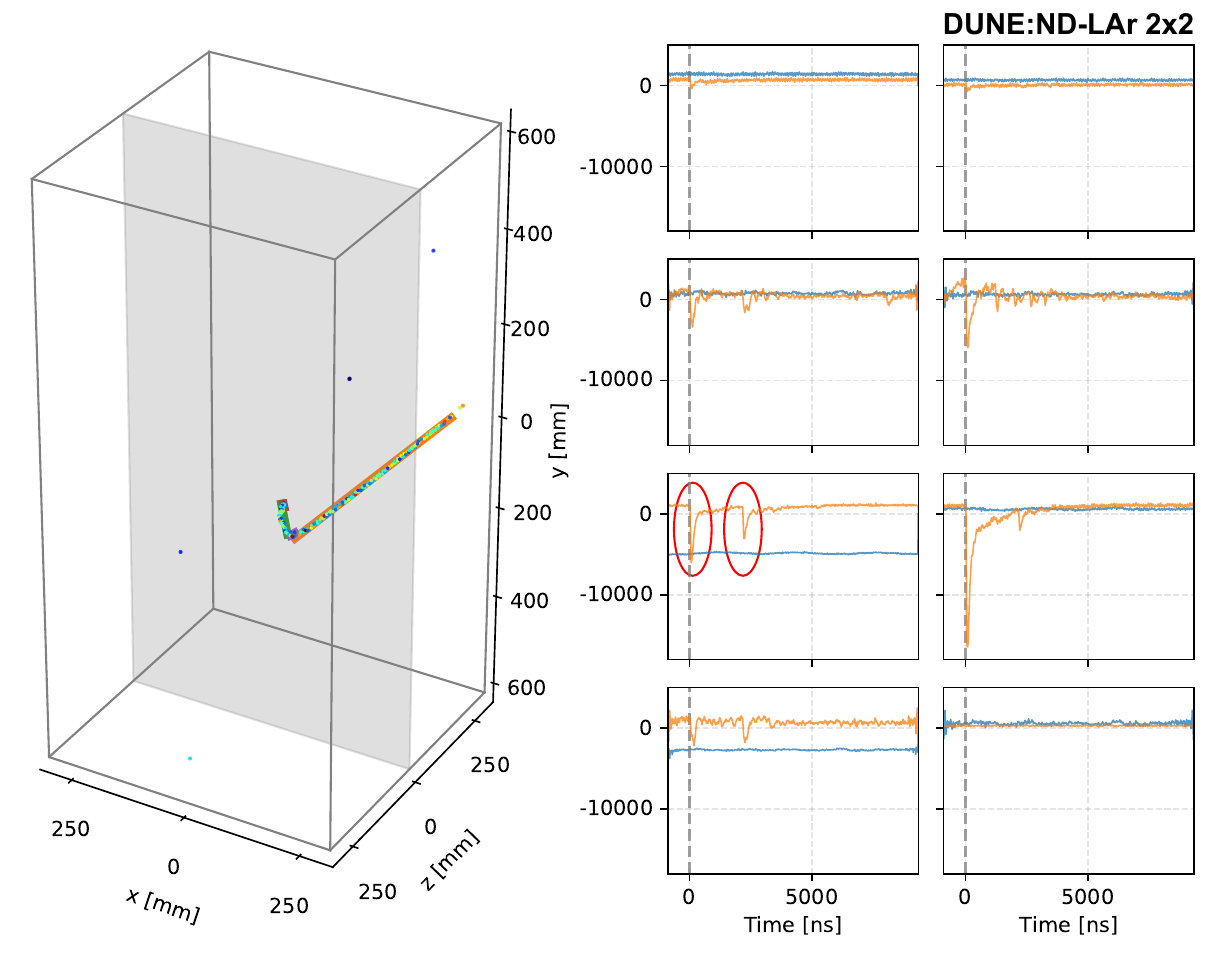}
 \caption{Event display of a Michel electron candidate shown in a 3D view (left) and with associated waveforms from photon detectors (right). In the right panel, orange and blue indicate the two optically isolated semi-TPCs. The red circles highlight an example the two pulses on the photon detectors correspond to the entering muon and the electron resulting from its decay.}
 \label{fig:michel_eventDisplayWithLight}
\end{figure}

Michel electrons, i.e. electrons from stopped muon decay, constitute a readily available and versatile tool for the study and characterisation of the performance of a LArTPC. They are abundant for surface-level detectors exposed to a large cosmic ray muon flux, and with $\mu \rightarrow e \overline{\nu}_{e} \nu_{\mu}$ as the almost exclusive decay channel, the number of events is given by the probability of the muon to come to rest in the detector. The electrons produced by the decay have a well-characterised energy spectrum with a cutoff at $\sim50$~MeV and their topology is relatively easy to tag: a long muon track ending with a Bragg peak followed by a short ionization track from the electron at a different angle with respect to the muon direction.
Fig.~\ref{fig:fig_eventgallery} includes one example of a stopping muon decaying with a Michel electron in \mbox{Module-0}.
The effective muon lifetime of $\sim2$~$\mu$s is short relative to the TPC drift speed, leading to minimal displacement of the muon track endpoint and electron track start. However, it is large relative to the time resolution of the light readout system, allowing the two signals to be tagged separately: the first light pulse corresponding to the muon ionization, and the second to the electron, can be easily separated for a large majority of events due to the excellent timing resolution of ArCLight and LCM detectors. Fig.~\ref{fig:michel_eventDisplayWithLight} shows the event display of a selected Michel electron candidate, with the two peaks showing the waveforms of the light detectors located in one of the two half-TPCs.

\begin{figure}[ht!]
 \includegraphics[width=0.7\textwidth]{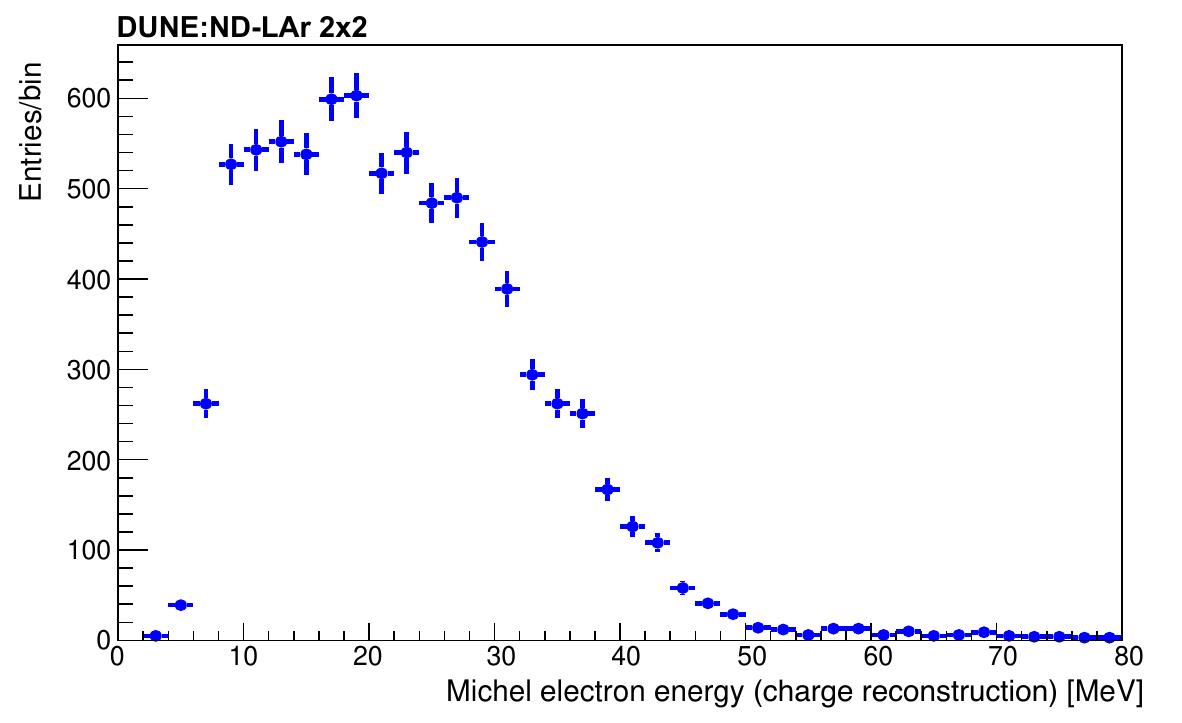}
 \caption{Charge-based energy spectrum of Michel electron candidates from a sample of reconstructed muon decays, using the full data set and automated event reconstruction.}
 \label{fig:michel_spectrum}
\end{figure}

The Michel electron candidates' topology is mainly characterised by a long ionisation trail left over by the crossing muon. An automatic selection algorithm based on the event topology and the presence of the Bragg peak at the end of the muon track was developed and applied to the subset of cosmic data. Visual event validation was performed on selected events to validate the analysis. The final distribution of the reconstructed Michel electron energy based on the automated charge reconstruction is shown in Fig.~\ref{fig:michel_spectrum}. The end point is near the expected true end point of 53~MeV. The spectrum peaks at lower energies mainly as a consequence of partial containment, imperfect clustering, and charge below threshold, particularly from electrons Compton-scattered by Bremmstrahlung photons radiated from the primary electron~\cite{DUNE:2022meu,LArIAT:2019gdz,MicroBooNE:2017kvv}. 

\subsection{Detector simulation validation with cosmic ray tracks}
\label{sec:cosmics}

\begin{figure}[htbp]
    \includegraphics[width=0.95\textwidth]{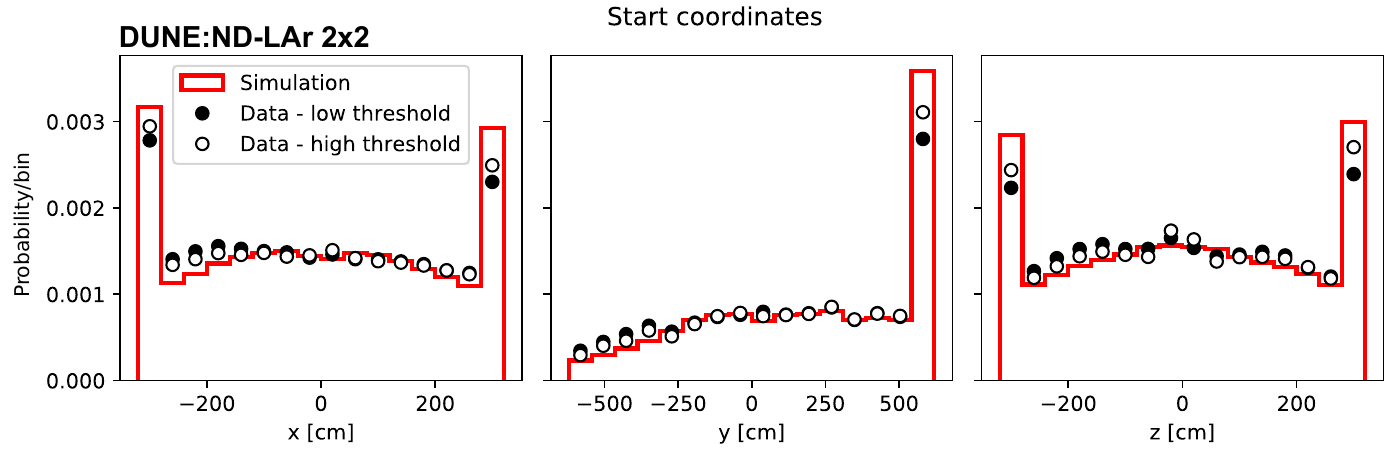}
    \includegraphics[width=0.95\textwidth]{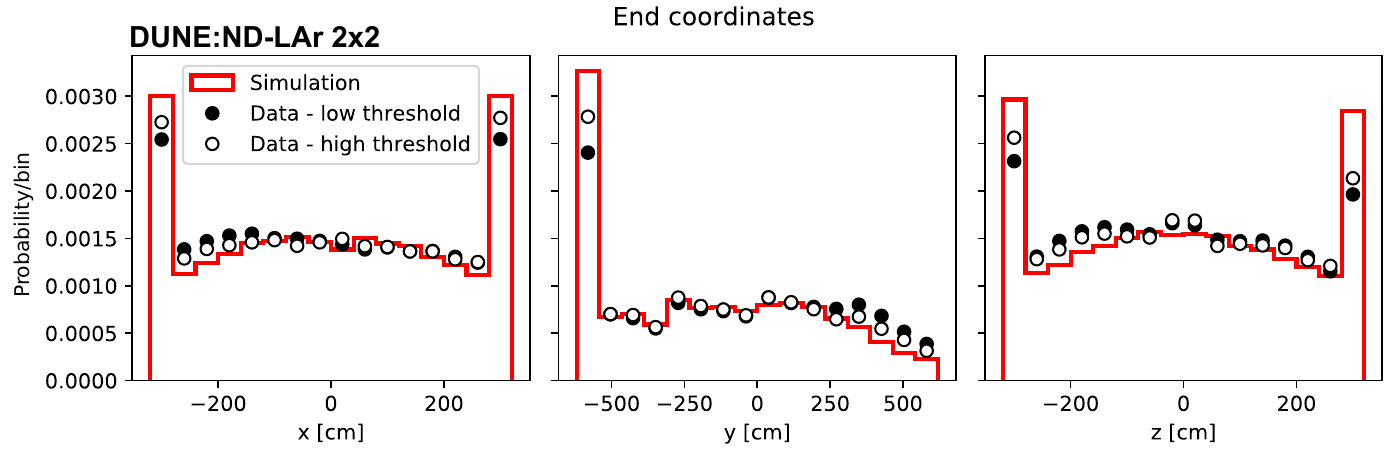}
    \caption{Start and end coordinates of stitched tracks in data (high and low threshold runs) and simulation.}
    \label{fig:xyz}
\end{figure}

Finally, selected samples of cosmic ray tracks are compared in detail to a cosmic ray simulation based on the CORSIKA event generator and the detailed microphysical detector simulation introduced in Section~\ref{sec:cosmic-analysis}.
Starting from the cosmic ray track reconstruction described there,
the track's start and end points are found by projecting the 3D points onto the 
cluster's principal  components. The DBSCAN+RANSAC fit is applied on outlying hits
until all are placed within a cluster or no hits remain.
This is sufficient for studies of low-level detector response, as it provides a local 
approximation of the track trajectory with minimal impact from $\delta$-rays
and hard scatters. Reconstructed tracks may show artificial gaps due to the 
presence of disabled channels. Also, cathode-piercing tracks will usually be 
reconstructed as separated tracks, due to the non-zero cathode thickness. Thus, 
tracks with an angle smaller than $20^{\circ}$ and closer than 10~cm are 
stitched together for the following studies.
A comparison between the spatial coordinates of the stitched tracks in data and 
simulation is shown in Fig.~\ref{fig:xyz}. 

\begin{figure}[htbp]
    \includegraphics[width=0.95\textwidth]{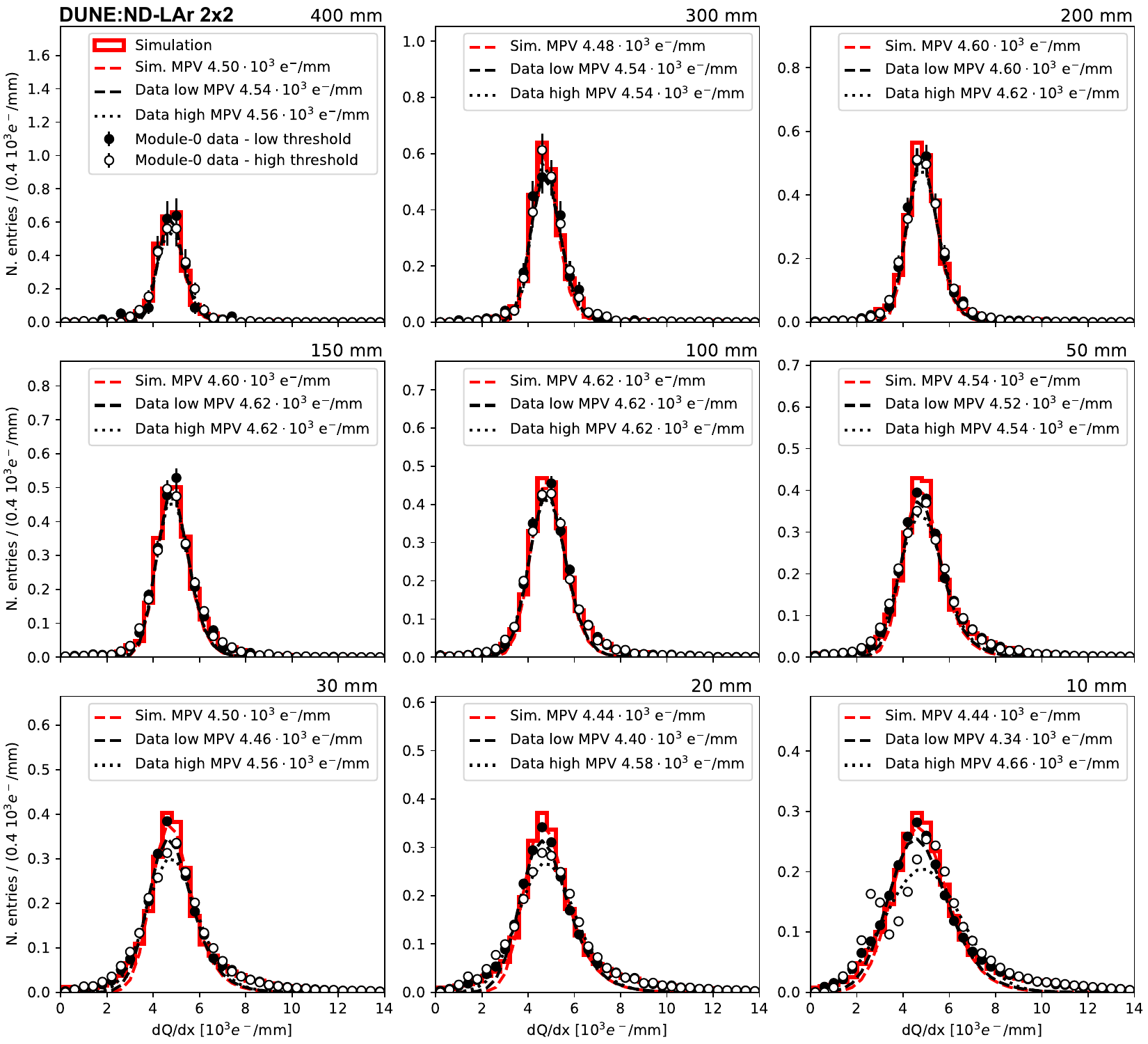}
    \caption{$dQ/dx$ measured for segments of different lengths for low threshold runs (black dots), high threshold runs (white dots) and a sample of simulated cosmic rays (red line). The distributions have been fitted with a Gaussian-convolved Moyal function (dashed lines).}
    \label{fig:dqdx_sim}
\end{figure}

Fig.~\ref{fig:dqdx_sim} shows a comparison of the $dQ/dx$ for low threshold and high threshold runs with a sample of simulated cosmic rays. The $dQ/dx$ has been measured for segments of different lengths, following the procedure described in Section~\ref{sec:cosmics}. The simulation assumes the Birks model for electron recombination and a gain of 4~mV/$10^3$~e$^{-}$ \cite{Birks:1951boa}. In the data, the amount of charge that reaches the anode is corrected by the electron lifetime factor calculated in Section \ref{sec:electron-lifetime}.

\begin{figure}[htbp]
 \includegraphics[width=0.95\textwidth]{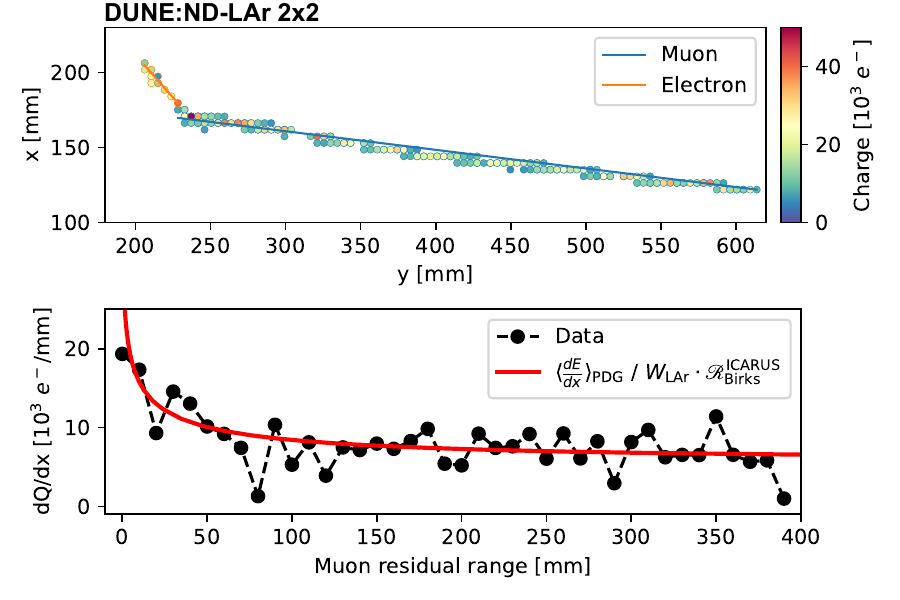}
 \caption{Top: event display of the anode plane for a selected stopping muon (blue) and subsequent Michel electron (orange). Bottom: $dQ/dx$ for the reconstructed muon track as a function of the residual range $dQ/dx$ and the theoretical curve for muons stopping in liquid argon (red line).}
 \label{fig:stopped_muon}
\end{figure}

Next, the $dQ/dx$ as a function of the reconstructed track residual range is considered. As noted in Section~\ref{sec:michel},
for a muon that stops in the detector the amount of deposited charge per unit length will increase as it approaches the end point, forming a Bragg peak. Fig.~\ref{fig:stopped_muon} shows an example of a stopping muon and the subsequent Michel electron. The $dQ/dx$ has been measured by subdividing the reconstructed track in 10~mm segments (our $dx$) and summing the charge contained in each segment (the $dQ$). The data show a Bragg peak near the end of the reconstructed track, where the residual range is close to zero. The theoretical prediction is obtained by taking the $\langle\frac{dE}{dx}\rangle$ values tabulated in Ref. \cite{Groom:2001kq} for muons in LAr, divided by the argon ionization energy (23.6~eV) and multiplied by the recombination factor ${R}_\mathrm{Birks}^{\mathrm{ICARUS}}$, calculated in Ref.~\cite{Amoruso:2004dy}.

The observed distributions indicate good overall agreement between data and simulations, in particular with the ability to correctly reproduce the position of the $dQ/dx$ peak. \mbox{Module-0} data provide input that can be used to further tune the detector simulation, including modeling of additional noise sources and details of the anode response. Meanwhile, the strong overall agreement in the vertex positioning and calorimetry indicates that the initial detector response model is able to capture the main features of the cosmic ray track samples.

\FloatBarrier
\section{Conclusions}
\label{sec:conclusions}

We have reported here the experimental results of exposing the Module-0 demonstrator, a tonne-scale LArTPC with pixel-based charge readout, to cosmic rays. This new type of neutrino detector is designed to meet the challenges of the near detector complex of the forthcoming DUNE experiment, which will be exposed to a very intense beam-related flux of particles. These challenges are expected to severely hamper the performance of a conventional, wire-readout, monolithic LArTPC, where reconstruction of complex 3D event topologies using a small number of 2D projections can lead to unsolvable ambiguities, particularly when multiple events overlap in the drift direction.
The novel Module-0 design features a combination of new technological solutions: a pixelated anode to read out the ionization electron signal that provides native three-dimensional charge imaging, a modular structure with relatively short drift length, high-performance scintillation light detection systems, and an innovative approach to field shaping using a low-profile resistive shell. Module-0 is one of four units that will comprise the $2\times2$ demonstrator (ProtoDUNE-ND) being installed at Fermilab to be exposed to the NuMI neutrino beam.

A detailed assessment of this technology has been performed by operating Module-0, as well as the associated cryogenics, data acquisition, trigger, and timing infrastructure, at the University of Bern. A large sample of 25 million self-triggered cosmic ray-induced events was collected and analyzed, along with an array of dedicated diagnostic data runs. 
The response of the 78,400-pixel readout system was studied, as well as the performance of the two independent and complementary light detection systems.
The data analysis demonstrated key physics requirements of this technology, such as the electron lifetime, the uniformity of the electric field, and the matching/correlation between the charge and light signals. The reconstruction of particle tracks and Michel electrons illustrates the physics capabilities, and the comparison with detailed, microphysical simulations has demonstrated a robust understanding of the workings of this new type of LArTPC detector.
Overall, these results demonstrate the key design features of the technique and provide a confirmation of the outstanding imaging capabilities of this next-generation LArTPC design.

\section{Acknowledgments}
\label{sec:acknowledgments}

This document was prepared by the DUNE collaboration using the
resources of the Fermi National Accelerator Laboratory 
(Fermilab), a U.S. Department of Energy, Office of Science, 
HEP User Facility. Fermilab is managed by Fermi Research Alliance, 
LLC (FRA), acting under Contract No. DE-AC02-07CH11359.
This work was supported by
CNPq,
FAPERJ,
FAPEG and 
FAPESP,                         Brazil;
CFI, 
IPP and 
NSERC,                          Canada;
CERN;
M\v{S}MT,                       Czech Republic;
ERDF, 
H2020-EU and 
MSCA,                           European Union;
CNRS/IN2P3 and
CEA,                            France;
INFN,                           Italy;
FCT,                            Portugal;
NRF,                            South Korea;
CAM, 
Fundaci\'{o}n ``La Caixa'',
Junta de Andaluc\'ia-FEDER,
MICINN, and
Xunta de Galicia,               Spain;
SERI and 
SNSF,                           Switzerland;
T\"UB\.ITAK,                    Turkey;
The Royal Society and 
UKRI/STFC,                      United Kingdom;
DOE and 
NSF,                            United States of America.
This research used resources of the 
National Energy Research Scientific Computing Center (NERSC), 
a U.S. Department of Energy Office of Science User Facility 
operated under Contract No. DE-AC02-05CH11231.

\bibliography{bibliography.bib}

\end{document}